\pdfoutput=1
\documentclass[12pt,a4paper]{article}

\usepackage{ifthen} 
\newboolean{pdflatex}
\setboolean{pdflatex}{true} 

\newboolean{articletitles}
\setboolean{articletitles}{true} 

\newboolean{uprightparticles}
\setboolean{uprightparticles}{false} 

\usepackage{afterpage}

\usepackage{float}

\def\paperauthors{LHCb collaboration} 

\def\paperasciititle{Search for massive long-lived particles decaying semileptonically at sqrt(s)=13TeV
} 
\def\papertitle{Search for massive long-lived particles decaying semileptonically at \sqs=13\tev
} 

\def\paperkeywords{{High Energy Physics}, {LHCb}} 
\def\papercopyright{\the\year\ CERN for the benefit of the LHCb collaboration} 
\def\paperlicence{CC BY 4.0 licence}
\def\paperlicenceurl{https://creativecommons.org/licenses/by/4.0/}

\usepackage[top=1in, bottom=1.25in, left=1in, right=1in]{geometry}

\usepackage{microtype}
\usepackage{lineno}  
\usepackage{xspace} 
\usepackage{caption} 

\usepackage{graphicx}  
\usepackage{color}
\usepackage{colortbl}
\graphicspath{{./figs/}} 

\usepackage{amsmath} 
\usepackage{amssymb}
\usepackage{amsfonts}
\usepackage{upgreek} 

\newcommand*\patchAmsMathEnvironmentForLineno[1]{%
\expandafter\let\csname old#1\expandafter\endcsname\csname #1\endcsname
\expandafter\let\csname oldend#1\expandafter\endcsname\csname
end#1\endcsname
 \renewenvironment{#1}%
   {\linenomath\csname old#1\endcsname}%
   {\csname oldend#1\endcsname\endlinenomath}%
}
\newcommand*\patchBothAmsMathEnvironmentsForLineno[1]{%
  \patchAmsMathEnvironmentForLineno{#1}%
  \patchAmsMathEnvironmentForLineno{#1*}%
}
\AtBeginDocument{%
\patchBothAmsMathEnvironmentsForLineno{equation}%
\patchBothAmsMathEnvironmentsForLineno{align}%
\patchBothAmsMathEnvironmentsForLineno{flalign}%
\patchBothAmsMathEnvironmentsForLineno{alignat}%
\patchBothAmsMathEnvironmentsForLineno{gather}%
\patchBothAmsMathEnvironmentsForLineno{multline}%
\patchBothAmsMathEnvironmentsForLineno{eqnarray}%
}

\usepackage{hyperxmp}

\usepackage[pdftex,
            pdfauthor={\paperauthors},
            pdftitle={\paperasciititle},
            pdfkeywords={\paperkeywords},
            pdfcopyright={Copyright (C) \papercopyright},
            pdflicenseurl={\paperlicenceurl}]{hyperref}

\usepackage[bottom,flushmargin,hang,multiple]{footmisc}

\usepackage[all]{hypcap} 

\usepackage{xspace} 
\usepackage{upgreek}

\def\lhcb   {\mbox{LHCb}\xspace}

\def\velo   {VELO\xspace}

\def\MagUp {\mbox{\em Mag\kern -0.05em Up}\xspace}

\ifthenelse{\boolean{uprightparticles}}%
{

 \def\Pmu         {\ensuremath{\upmu}\xspace}

 \def\Ppsi        {\ensuremath{\uppsi}\xspace}

 \def\PDelta      {\ensuremath{\Delta}\xspace}                 
 \def\PXi         {\ensuremath{\Xi}\xspace}                 
 \def\PLambda     {\ensuremath{\Lambda}\xspace}                 
 \def\PSigma      {\ensuremath{\Sigma}\xspace}                 
 \def\POmega      {\ensuremath{\Omega}\xspace}                 
 \def\PUpsilon    {\ensuremath{\Upsilon}\xspace}

 \def\PB      {\ensuremath{\mathrm{B}}\xspace}                 
                  
 \def\PD      {\ensuremath{\mathrm{D}}\xspace}

 \def\PJ      {\ensuremath{\mathrm{J}}\xspace}                 
 \def\PK      {\ensuremath{\mathrm{K}}\xspace}

 \def\PW      {\ensuremath{\mathrm{W}}\xspace}

 \def\PZ      {\ensuremath{\mathrm{Z}}\xspace}                 
                  
 \def\Pb      {\ensuremath{\mathrm{b}}\xspace}                 
 \def\Pc      {\ensuremath{\mathrm{c}}\xspace}

 \def\Pi      {\ensuremath{\mathrm{i}}\xspace}

 \def\Ps      {\ensuremath{\mathrm{s}}\xspace}

 \def\thebaroffset{0.0em}
}
{

 \def\Pmu         {\ensuremath{\mu}\xspace}

 \def\Ppsi        {\ensuremath{\psi}\xspace}                 
                  
 \mathchardef\PDelta="7101
 \mathchardef\PXi="7104
 \mathchardef\PLambda="7103
 \mathchardef\PSigma="7106
 \mathchardef\POmega="710A
 \mathchardef\PUpsilon="7107
                  
 \def\PB      {\ensuremath{B}\xspace}                 
                  
 \def\PD      {\ensuremath{D}\xspace}

 \def\PJ      {\ensuremath{J}\xspace}                 
 \def\PK      {\ensuremath{K}\xspace}

 \def\PW      {\ensuremath{W}\xspace}

 \def\PZ      {\ensuremath{Z}\xspace}                 
                  
 \def\Pb      {\ensuremath{b}\xspace}                 
 \def\Pc      {\ensuremath{c}\xspace}

 \def\Pi      {\ensuremath{i}\xspace}

 \def\Ps      {\ensuremath{s}\xspace}

 \def\thebaroffset{0.18em}
}
\newcommand{\offsetoverline}[2][\thebaroffset]{\kern #1\overline{\kern -#1 #2}}%

\makeatletter
\ifcase \@ptsize \relax
  \newcommand{\miniscule}{\@setfontsize\miniscule{4}{5}}
\or
  \newcommand{\miniscule}{\@setfontsize\miniscule{5}{6}}
\or
  \newcommand{\miniscule}{\@setfontsize\miniscule{5}{6}}
\fi
\makeatother

\DeclareRobustCommand{\optbar}[1]{\shortstack{{\miniscule (\rule[.5ex]{1.25em}{.18mm})}
  \\ [-.7ex] $#1$}}

\def\mumu       {{\ensuremath{\Pmu^+\Pmu^-}}\xspace}

\def\W      {{\ensuremath{\PW}}\xspace}

\def\Z      {{\ensuremath{\PZ}}\xspace}

\def\squark    {{\ensuremath{\Ps}}\xspace}

\def\cquark    {{\ensuremath{\Pc}}\xspace}
\def\cquarkbar {{\ensuremath{\overline \cquark}}\xspace}
\def\ccbar     {{\ensuremath{\cquark\cquarkbar}}\xspace}
\def\bquark    {{\ensuremath{\Pb}}\xspace}
\def\bquarkbar {{\ensuremath{\overline \bquark}}\xspace}
\def\bbbar     {{\ensuremath{\bquark\bquarkbar}}\xspace}

\def\kaon    {{\ensuremath{\PK}}\xspace}

\def\KorKbar {\kern \thebaroffset\optbar{\kern -\thebaroffset \PK}{}\xspace}

\def\Kstarz  {{\ensuremath{\kaon^{*0}}}\xspace}

\def\D       {{\ensuremath{\PD}}\xspace}

\def\DorDbar {\kern \thebaroffset\optbar{\kern -\thebaroffset \PD}\xspace}

\def\Dp      {{\ensuremath{\D^+}}\xspace}
\def\Dm      {{\ensuremath{\D^-}}\xspace}

\def\DpDm    {\ensuremath{\Dp {\kern -0.16em \Dm}}\xspace}

\def\B       {{\ensuremath{\PB}}\xspace}

\def\BorBbar {\kern \thebaroffset\optbar{\kern -\thebaroffset \PB}\xspace}

\def\Bd      {{\ensuremath{\B^0}}\xspace}

\def\BdorBdbar {\kern \thebaroffset\optbar{\kern -\thebaroffset \Bd}\xspace}

\def\Bs      {{\ensuremath{\B^0_\squark}}\xspace}

\def\BsorBsbar {\kern \thebaroffset\optbar{\kern -\thebaroffset \Bs}\xspace}

\def\jpsi     {{\ensuremath{{\PJ\mskip -3mu/\mskip -2mu\Ppsi}}}\xspace}

\def\Upsilonres  {{\ensuremath{\PUpsilon}}\xspace}
\def\Y#1S{\ensuremath{\PUpsilon{(#1S)}}\xspace}

\def\LorLbar     {\kern \thebaroffset\optbar{\kern -\thebaroffset \PLambda}\xspace}

\newcommand{\decay}[2]{\ensuremath{#1\!\to #2}\xspace} 

\def\to                 {\ensuremath{\rightarrow}\xspace}

\def\eps   {{\ensuremath{\varepsilon}}\xspace}

\def\AT#1     {\ensuremath{A_{\mathrm{T}}^{#1}}\xspace}           

\def\C#1      {\ensuremath{\mathcal{C}_{#1}}\xspace}                       
\def\Cp#1     {\ensuremath{\mathcal{C}_{#1}^{'}}\xspace}                    
\def\Ceff#1   {\ensuremath{\mathcal{C}_{#1}^{\mathrm{(eff)}}}\xspace}        
\def\Cpeff#1  {\ensuremath{\mathcal{C}_{#1}^{'\mathrm{(eff)}}}\xspace}       
\def\Ope#1    {\ensuremath{\mathcal{O}_{#1}}\xspace}                       
\def\Opep#1   {\ensuremath{\mathcal{O}_{#1}^{'}}\xspace}                    

\newcommand{\nospaceunit}[1]{\ensuremath{\text{#1}}}       
\newcommand{\aunit}[1]{\ensuremath{\text{\,#1}}}       

\newcommand{\tev}{\aunit{Te\kern -0.1em V}\xspace}
\newcommand{\gev}{\aunit{Ge\kern -0.1em V}\xspace}
\newcommand{\mev}{\aunit{Me\kern -0.1em V}\xspace}
\newcommand{\kev}{\aunit{ke\kern -0.1em V}\xspace}
\newcommand{\ev}{\aunit{e\kern -0.1em V}\xspace}
 
\newcommand{\mevc}{\ensuremath{\aunit{Me\kern -0.1em V\!/}c}\xspace}
\newcommand{\gevc}{\ensuremath{\aunit{Ge\kern -0.1em V\!/}c}\xspace}
\newcommand{\mevcc}{\ensuremath{\aunit{Me\kern -0.1em V\!/}c^2}\xspace}
\newcommand{\gevcc}{\ensuremath{\aunit{Ge\kern -0.1em V\!/}c^2}\xspace}

\def\mm   {\aunit{mm}\xspace}

\def\mum  {\ensuremath{\,\upmu\nospaceunit{m}}\xspace}

\def\pb {\aunit{pb}\xspace}

\def\fb   {\ensuremath{\aunit{fb}}\xspace}
\def\invfb   {\ensuremath{\fb^{-1}}\xspace}

\def\ps   {\ensuremath{\aunit{ps}}\xspace}

\newcommand{\chisq}{\ensuremath{\chi^2}\xspace}
\newcommand{\chisqndf}{\ensuremath{\chi^2/\mathrm{ndf}}\xspace}

\def\gsim{{~\raise.15em\hbox{$>$}\kern-.85em
          \lower.35em\hbox{$\sim$}~}\xspace}
\def\lsim{{~\raise.15em\hbox{$<$}\kern-.85em
          \lower.35em\hbox{$\sim$}~}\xspace}

\def\sqs   {\ensuremath{\protect\sqrt{s}}\xspace}

\def\pt         {\ensuremath{p_{\mathrm{T}}}\xspace}

\def\ptot       {\ensuremath{p}\xspace}

\def\geant      {\mbox{\textsc{Geant4}}\xspace}

\def\pythia     {\mbox{\textsc{Pythia}}\xspace}

\def\tell1  {TELL1\xspace}
\def\ukl1   {UKL1\xspace}

\newcommand{\eg}{\mbox{\itshape e.g.}\xspace}

\usepackage{cite} 
\usepackage{mciteplus}

\usepackage{longtable} 

\usepackage[version-1-compatibility]{siunitx}

\newcommand{\chino}{\tilde{\chi}}

\newcommand{\khi}{\ensuremath{\chio_{1}}\xspace}
\newcommand{\chio}{\chino^{0}}

\newcommand{\tauchi}{\ensuremath{\tau_{\khi}}\xspace}

\newcommand{\RXY}{\ensuremath{R_{\rm xy}}\xspace}

\newcommand{\IPMU}{\ensuremath{{\rm IP^{\mu}}}\xspace}
\newcommand{\PTMU}{\ensuremath{{p^{\mu}_{\mathrm{T}}}}\xspace}

\newcommand{\mhzero}{\ensuremath{m_{h^0}}\xspace}

\newcommand{\lhcbX}{ \scalebox{0.7}{{\makebox[1.1\width]{\lhcb}}} }
\newcommand{\lhcbL}{ \scalebox{0.7}{{\makebox[1.1\width]{5.4\invfb}}} }

\newcolumntype{M}[1]{>{\raggedright}m{#1}}

\begin{document}

\renewcommand{\thefootnote}{\fnsymbol{footnote}}
\setcounter{footnote}{1}


\begin{titlepage}
\pagenumbering{roman}

\vspace*{-1.5cm}
\centerline{\large EUROPEAN ORGANIZATION FOR NUCLEAR RESEARCH (CERN)}
\vspace*{1.5cm}
\noindent
\begin{tabular*}{\linewidth}{lc@{\extracolsep{\fill}}r@{\extracolsep{0pt}}}
\ifthenelse{\boolean{pdflatex}}
{\vspace*{-1.5cm}\mbox{\!\!\!\includegraphics[width=.14\textwidth]{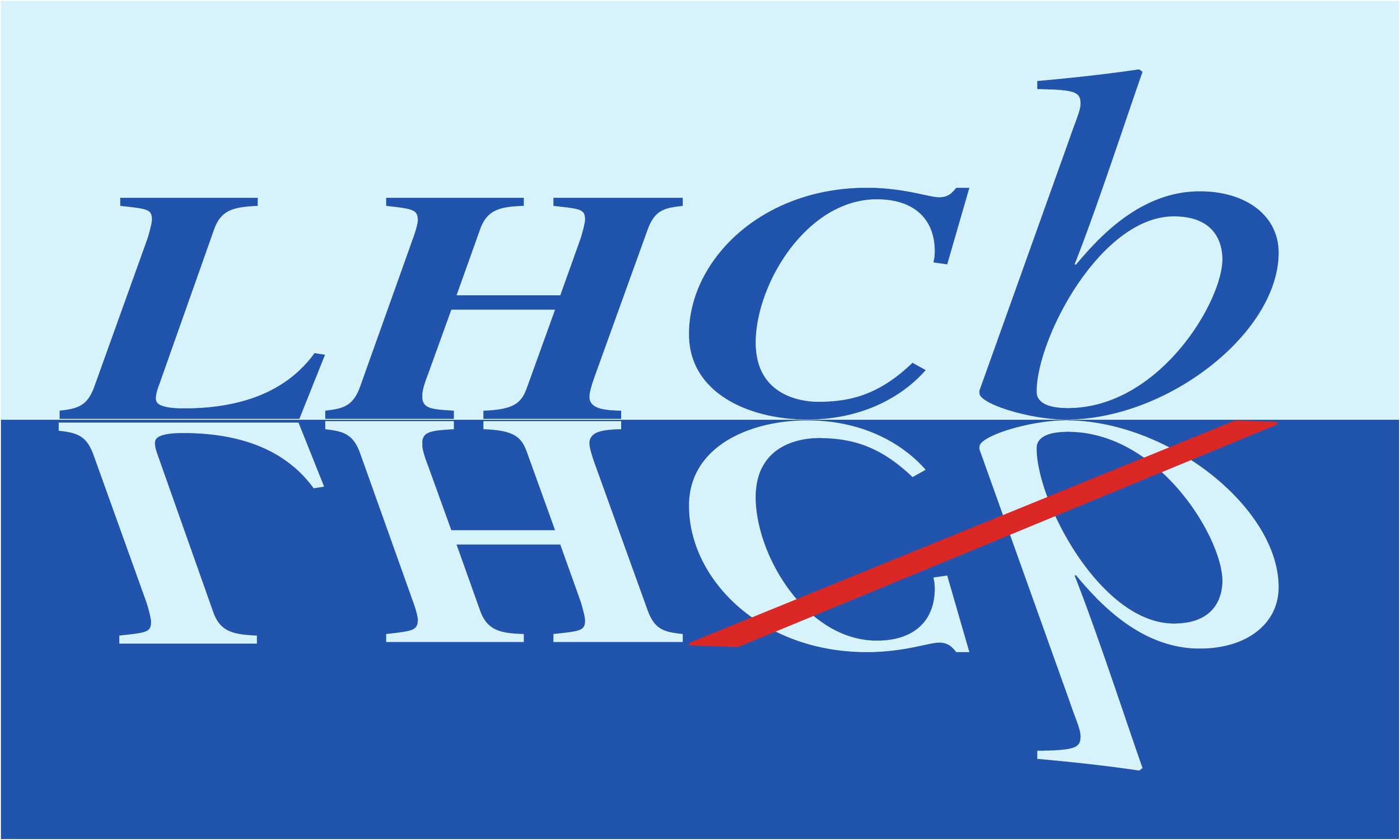}} & &}%
{\vspace*{-1.2cm}\mbox{\!\!\!\includegraphics[width=.12\textwidth]{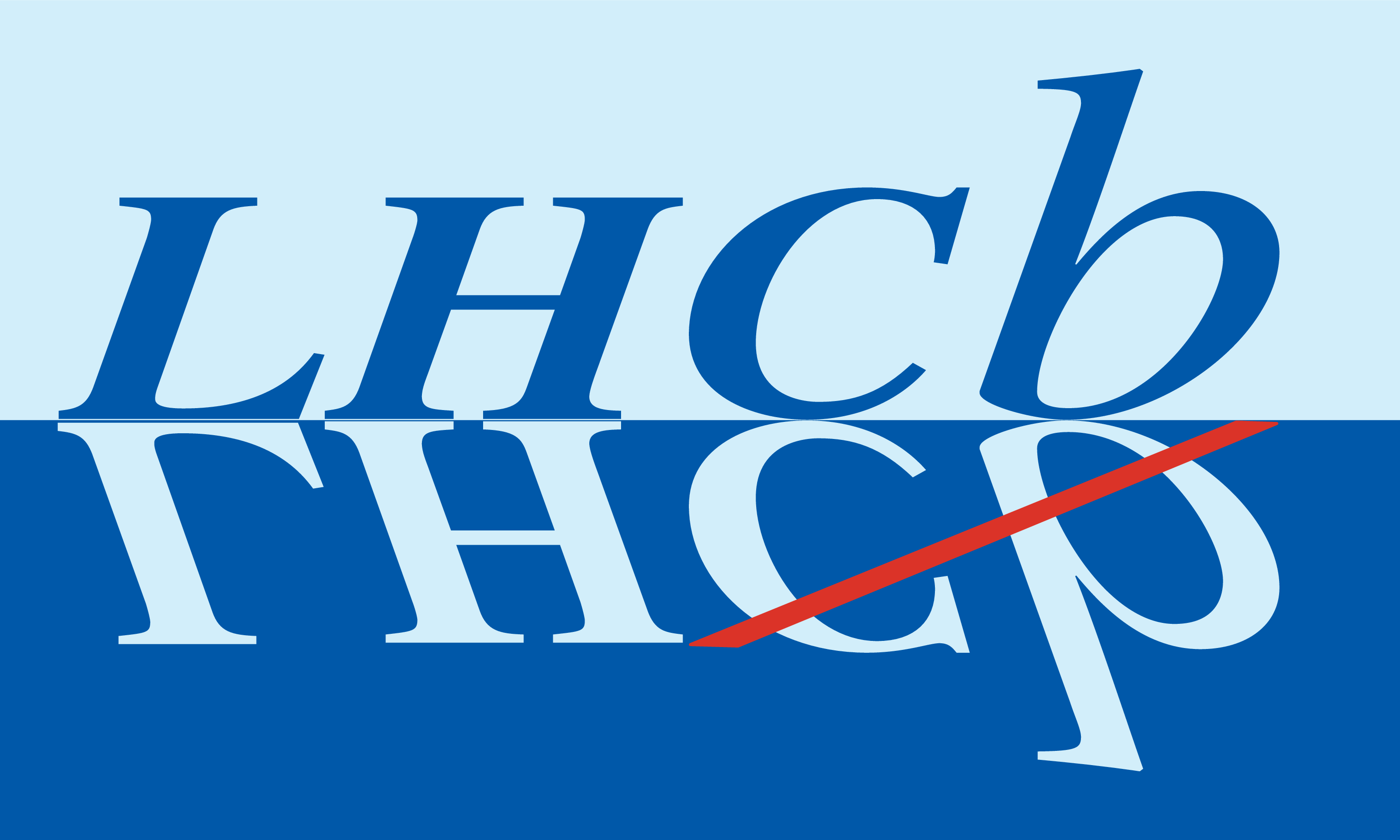}} & &}%
\\
 & & CERN-EP-2021-186 \\  
 & & LHCb-PAPER-2021-028 \\  
 & & 28 April 2022 \\ 
\end{tabular*}

\vspace*{4.0cm}

{\normalfont\bfseries\boldmath\huge
\begin{center}
  \papertitle 
\end{center}
}

\vspace*{1.0cm}

\begin{center}
\paperauthors\footnote{Authors are listed at the end of this paper.}
\end{center}

\vspace{\fill}

\begin{abstract}
\noindent
  A search is performed for massive long-lived particles (LLPs) decaying semileptonically into a muon and two quarks. 
  Two kinds of LLP production processes were considered.
  In the first, a Higgs-like boson with mass from 30 to 200\gevcc is produced by gluon fusion
  and decays into two LLPs. The analysis covers LLP mass values from
  10\gevcc up to about one half the Higgs-like boson mass.
  The second LLP production mode is directly from quark interactions, with
  LLP masses from 10 to 90\gevcc.
  The LLP lifetimes considered range from 5 to 200 ps.
  This study uses LHCb data collected from proton-proton collisions at $\sqrt{s} = 13\tev$,
  corresponding to an integrated luminosity of 5.4\invfb.
  No evidence of these long-lived states has been observed,
  and upper limits on the production cross-section times branching ratio have been set
  for each model considered.
  
\end{abstract}

\vspace*{1.0cm}

\begin{center}
  Published in Eur. Phys. J C82 (2022) 82:373
 \end{center}

\vspace{\fill}

{\footnotesize 
\centerline{\copyright~\papercopyright. \href{\paperlicenceurl}{\paperlicence}.}}
\vspace*{2mm}

\end{titlepage}


\newpage
\setcounter{page}{2}
\mbox{~}
%
%
%
%


\renewcommand{\thefootnote}{\arabic{footnote}}
\setcounter{footnote}{0}

\cleardoublepage


\pagestyle{plain} 
\setcounter{page}{1}
\pagenumbering{arabic}


\section{Introduction}

Supersymmetry (SUSY) is one of the most popular extensions of the Standard Model (SM), which can solve
the hierarchy problem, can unify the gauge couplings at the Planck scale and proposes dark matter candidates.
The minimal supersymmetric extension of the Standard Model (MSSM) is the simplest phenomenologically viable
realisation of SUSY~\cite{MSSM,Martin:1997ns}.
The present study addresses a subset of models featuring massive
long-lived particles (LLPs) with a measurable flight distance~\cite{KaplanDisplaced2012,hv2}, decaying semileptonically.
%
Long-lived particles decaying semileptonically with displaced jets composed of SM particles have been studied by the experiments at the
LHC~\cite{ATLAS-neutralino-lepton-2012,ATLAS-LLP-8TEV-2015,CMS-LLPinclusive-2018,ATLAS-2020,LHCb-PAPER-2016-047}
%
Additional information on searches for LLPs at collider experiments can be found in Refs.~\cite{art:graham,Lee:2018pag,LLP-LHC-2020}.

\begin{figure}[b]
  \begin{center}
    \includegraphics[width=0.6\linewidth]{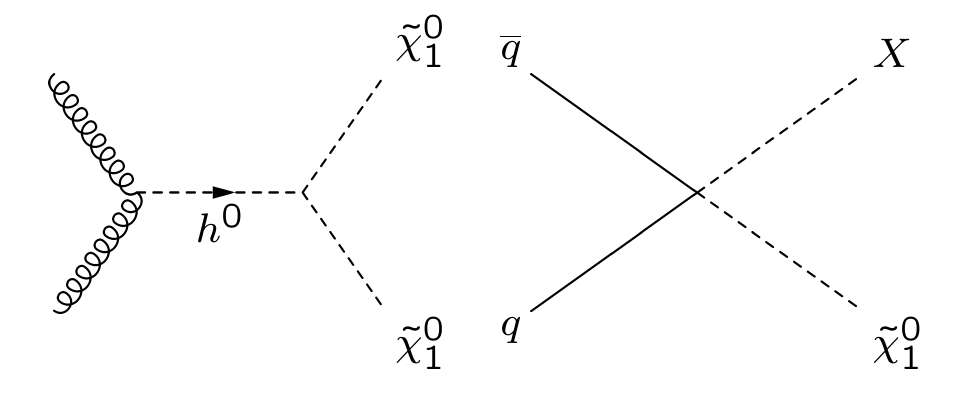}
    \put(-220,10){(a)}
      \put(-80,10){(b)}
    \vspace*{-0.5cm}
  \end{center}
  \caption{
    LLP production processes considered in this paper, where the \khi represents the LLP:
    (a) di-LLP production via a scalar particle $h^0$;
    (b) non-resonant, direct LLP production from quark interactions, where $X$ is a stable particle, with mass identical to the LLP.
    The LLP decays into a muon and two quarks:
    $\khi \rightarrow \mu^{+} q_{i} q_{j}( \mu^{-} \bar{q_{i}} \bar{q_{j}})$.
  }
  \label{fig:evttopo}  
\end{figure}

This analysis uses proton-proton ($pp$) collision data at a centre-of-mass energy \sqs=13\tev collected by
the \lhcb experiment at the LHC, corresponding to a total integrated luminosity of 5.4\invfb.
It extends the analysis of Ref.~\cite{LHCb-PAPER-2016-047} on data collected at $\sqs=7$ and 8\tev.
%
The adopted theoretical framework is inspired by the SUper GRAvity (mSUGRA)
with  R-parity violation (RPV)~\cite{art:msugra2}, in which the neutralino can decay into a
muon and two quarks:  $\khi \rightarrow \mu^{+} q_{i} q_{j}( \mu^{-} \bar{q_{i}} \bar{q_{j}})$.
Neutralinos can be produced by a variety of processes. In this paper
the analysis has been performed assuming the two mechanisms depicted in Fig.~\ref{fig:evttopo}.
In the first process,
a Higgs-like particle, $h^0$, is produced by gluon fusion and decays into two LLPs.
The analysis covers $h^0$ masses from 30 to 200\gevcc,
LLP lifetimes from 5 to 200 ps and LLP mass values from 10\gevcc
up to about one half the $h^0$ mass.
The second mode is a direct LLP production from quark interactions.
The  LLP lifetime range considered is from 5 to 200 ps and the
mass range from 10 to 90\gevcc. 
The LLP lifetime range begins at 5 ps, well above the typical b-hadron lifetime, and extends up to 200 ps,
where most of the vertices are still within the \lhcb vertex locator (\velo).
The mass range avoids the region of the SM b-quark states, but also takes
into account the forward acceptance of the \lhcb detector within which the decay products of relatively light LLPs can be efficiently detected.

The LLP signature is a displaced vertex made of charged particle tracks accompanied by an
isolated muon with high transverse momentum with respect to the proton beam direction, \pt.
This study benefits from  the excellent vertex reconstruction provided by the
\velo, and by the low \pt threshold of the muon trigger, compared to the other LHC experiments.
In addition, the \lhcb experiment is probing a rapidity region only partially accessible by other LHC experiments.
These properties allow the \lhcb experiment to be
complementary to similar analyses performed by the two central detectors at the LHC
and even explore regions of the theoretical parameter space where these
experiments are limited by their low efficiency to reconstruct highly boosted LLPs.

\section{Detector description and simulation}
\label{sec:Detector}

The \lhcb detector~\cite{LHCb-DP-2008-001,LHCb-DP-2014-002} is a single-arm forward
spectrometer covering the \mbox{pseudorapidity} range $2<\eta <5$,
designed for the study of particles containing \bquark or \cquark
quarks. The detector includes a high-precision tracking system
consisting of the \velo which is a silicon-strip detector surrounding the $pp$
interaction region~\cite{LHCb-DP-2014-001}, a large-area silicon-strip detector located
upstream of a dipole magnet with a bending power of about
$4{\mathrm{\,Tm}}$, and three stations of silicon-strip detectors and straw
drift tubes~\cite{LHCb-DP-2013-003,LHCb-DP-2017-001} placed downstream of the magnet.
The tracking system provides a measurement of the momentum, \ptot, of charged particles with
a relative uncertainty that varies from 0.5\% at low momentum to 1.0\% at 200\gevc.
The minimum distance of a track to a primary $pp$ collision vertex (PV), the impact parameter, 
is measured with a resolution of $(15+29/\pt)\mum$,
where \pt is in \,\gevc.
Different types of charged hadrons are distinguished using information
from two ring-imaging Cherenkov detectors~\cite{LHCb-DP-2012-003}. 
Photons, electrons and hadrons are identified by a calorimeter system consisting of
scintillating-pad and preshower detectors, an electromagnetic (ECAL)
and a hadronic calorimeter (HCAL)~\cite{LHCb-DP-2020-001}. Muons are identified by a
system composed of alternating layers of iron and multiwire
proportional chambers~\cite{LHCb-DP-2012-002}.
The online event selection is performed by a trigger~\cite{LHCb-DP-2012-004}, 
which consists of a hardware stage, based on information from the calorimeter and muon
systems, followed by a software stage, which applies a full event
reconstruction.
During data taking an alignment and calibration of the detector is performed in near real-time and used
in the software trigger~\cite{LHCb-DP-2019-001}.
The same alignment and calibration information is propagated to the offline reconstruction.

Simulation is used to model the effects of the detector acceptance and the imposed selection requirements.
In the simulation, $pp$ collisions are generated using
\pythia8~\cite{Sjostrand:2007gs,*Sjostrand:2006za} 
with a specific \lhcb configuration~\cite{LHCb-PROC-2010-056} and
with parton density functions taken from CTEQ6L~\cite{cteq6l}.
The interaction of the generated particles with the detector, and its response, are implemented using the \geant
toolkit~\cite{Allison:2006ve, *Agostinelli:2002hh} as described in Ref.~\cite{LHCb-PROC-2011-006}.
The simulation includes pileup events with an average of 1.1 $pp$ visible interactions per bunch crossing.

Several sets of signal events have been produced
assuming the processes illustrated in Fig.~\ref{fig:evttopo},
where the \khi plays the role of a long-lived particle. 
For the first process considered, two \khi particles are obtained from the decay of the Higgs-like boson
produced by gluon fusion, $ gg \rightarrow h^0  \rightarrow \khi\khi $.
For the second process, the LLP is produced in a non-resonant mode, $ q\bar{q} \rightarrow   \khi X $.
Here $X$ is a stable neutral particle with the same mass as that of the \khi state.
This production of a LLP in association with a stable particle X is included, which enables 
probing the sensitivity to this topology, with the signal LLP recoiling against such a particle. 

The LLP decays into a muon and two quarks; the branching ratio of
\mbox{$\khi \rightarrow \mu^{+} q_{i} q_{j}( \mu^{-} \bar{q_{i}} \bar{q_{j}})$}
is set to be equal for each quark combination ($q_{i} = u,c$ and $q_{j} = \bar{d},\bar{s},\bar{b}$),
with an equal proportion of $\mu^{+}$ and $\mu^{-}$.

In the following, the model name is indicated by the values of \mhzero, $m_{\khi}$ and $\tau_{\khi}$;
\mbox{h125-chi40-10ps}, for example, corresponds to $\mhzero=125\gevcc$, $m_{\khi}=40\gevcc$, $\tau_{\khi}=10\ps$.
For the direct production, the Higgs mass is omitted from this notation, such as for example in chi30-10ps. 
  
The most relevant background in this analysis is from events containing heavy quarks.
The background from heavy quarks directly produced in $pp$ collisions,
as well as from $\W$, $\Z$, Higgs boson and top quark decays, is studied using the simulation.
The simulation of inclusive \bbbar and \ccbar events is not efficient to produce a
large enough sample to cover the relevant high-\pt muon kinematic region.
Hence, a dedicated sample of
$20 \times 10^6$ ($1 \times 10^6)$ simulated \bbbar (\ccbar) events has been produced with a 
minimum parton  $\hat{p}_{\rm T}$ of 20\gevc and requiring a muon with $\pt>12\gevc$ and $1.5<\eta<5.0$.
All the simulated background species
are suppressed by the multivariate analysis presented in the next section.
Therefore, a data-driven approach is employed for the final background estimation.

\section{Signal selection}\label{sec:evtsel}

Signal events are selected by requiring a
vertex displaced from any PV in the event and containing one isolated, high-\pt muon.
Due to the relatively high LLP mass, the muons from the LLP decay are expected to be
more isolated than muons from hadron decays.
The events from $pp$ collisions are selected online by a trigger requiring muons with $\pt>10\gevc$.
The offline analysis requires that the triggering muon has an impact parameter, \IPMU, with respect to any PV,
larger than $0.25\mm$ and a transverse momentum, \PTMU,   larger than $12\gevc$.
Primary and displaced vertices are reconstructed offline from charged particle tracks~\cite{LHCb-PUB-2014-044}.
Genuine PVs are identified by a small radial distance from the beam axis, $\RXY<0.3$~mm.
Once the set of PVs is identified, all the other vertices are candidates for the decay position of LLPs.
An LLP candidate is formed by requiring three or more tracks including the muon
and having an invariant mass above 4.5\gevcc.
There is no requirement for the reconstructed momentum to point to a specific PV.
Particles interacting with the detector material are an important source of background.
Therefore, a geometric veto is used to reject candidates with vertices in regions
occupied by detector material~\cite{LHCb-PAPER-2012-023}.
The event preselection requires at least one PV in the event and at least one LLP candidate.

\begin{figure}[H]
\centering
\mbox{
 \includegraphics[height=4cm,width=6.2cm]{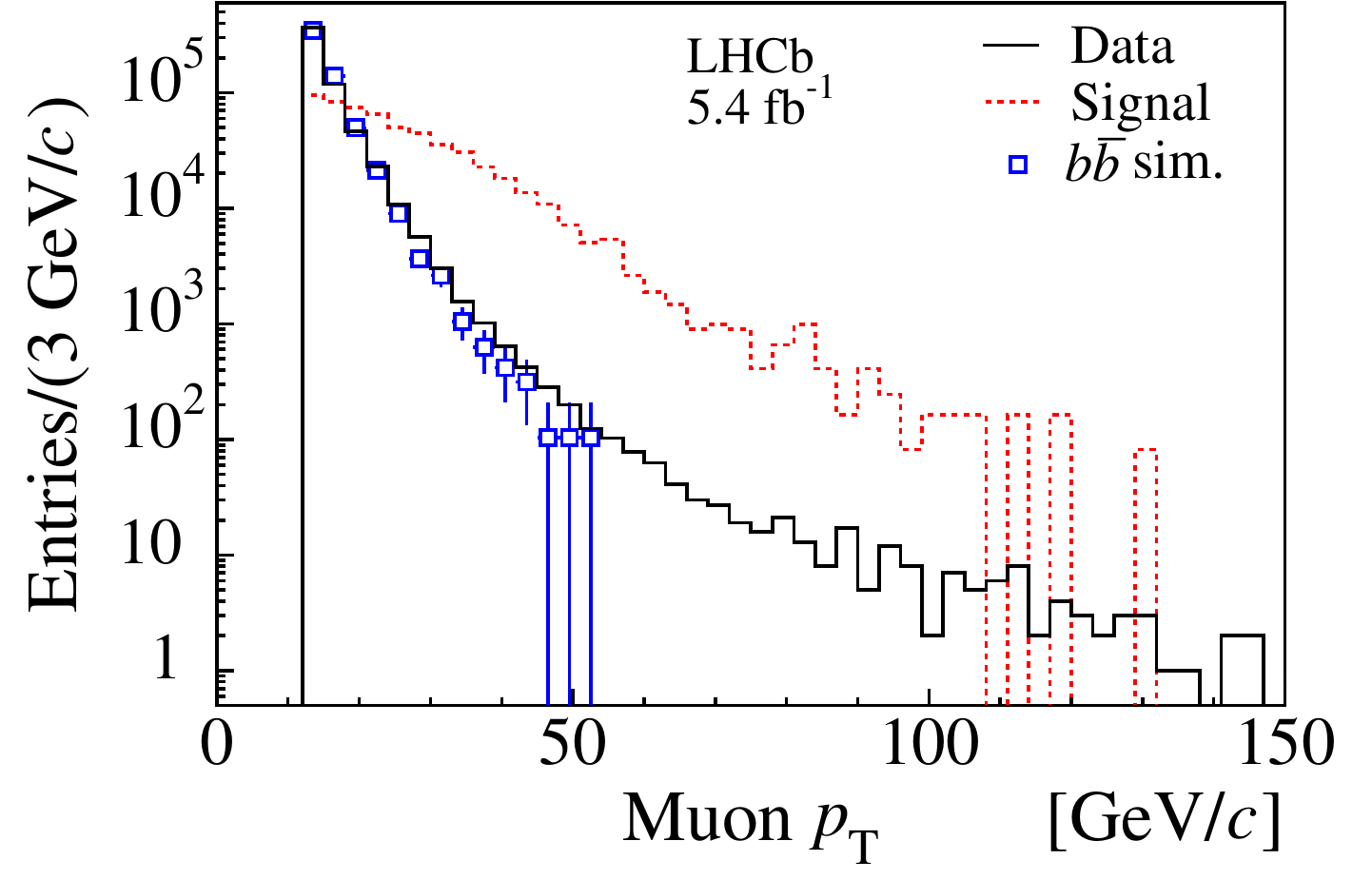}
 \includegraphics[height=4cm,width=6.2cm]{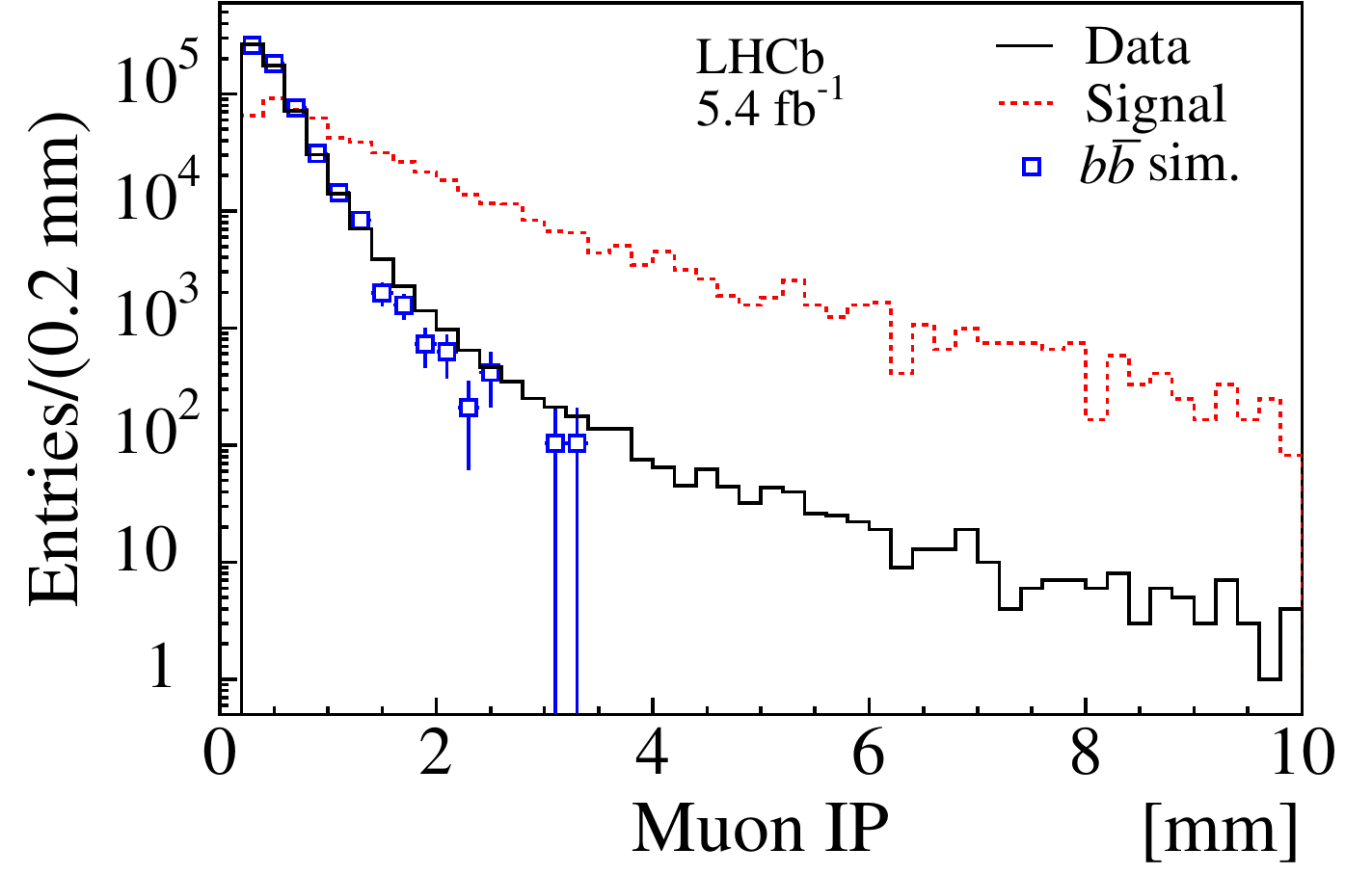}
 \put(-314,30){\small (a)} \put(-142,30){\small (b)}
}
\mbox{
\includegraphics[height=4cm,width=6.2cm]{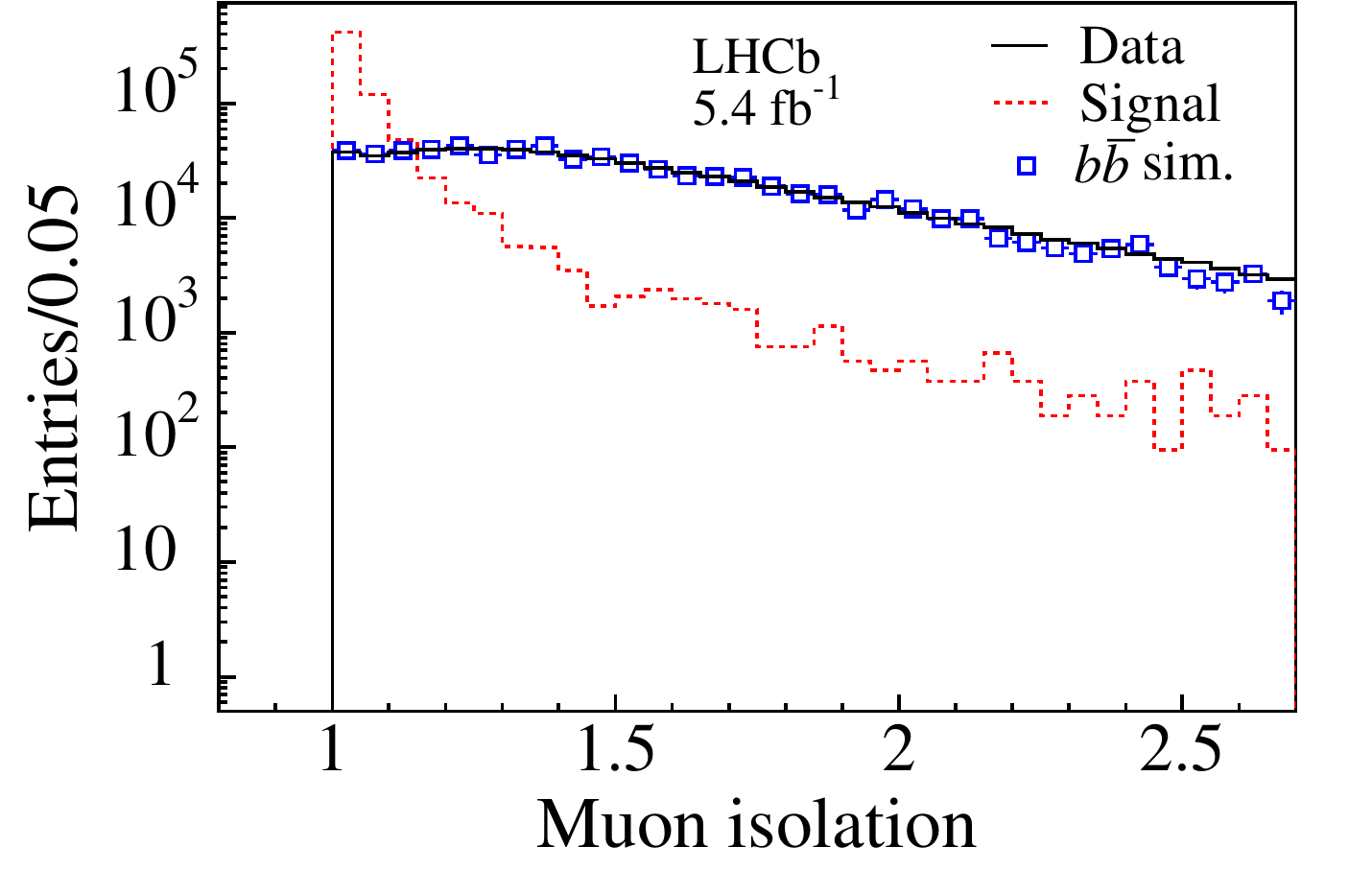}
\includegraphics[height=4cm,width=6.2cm]{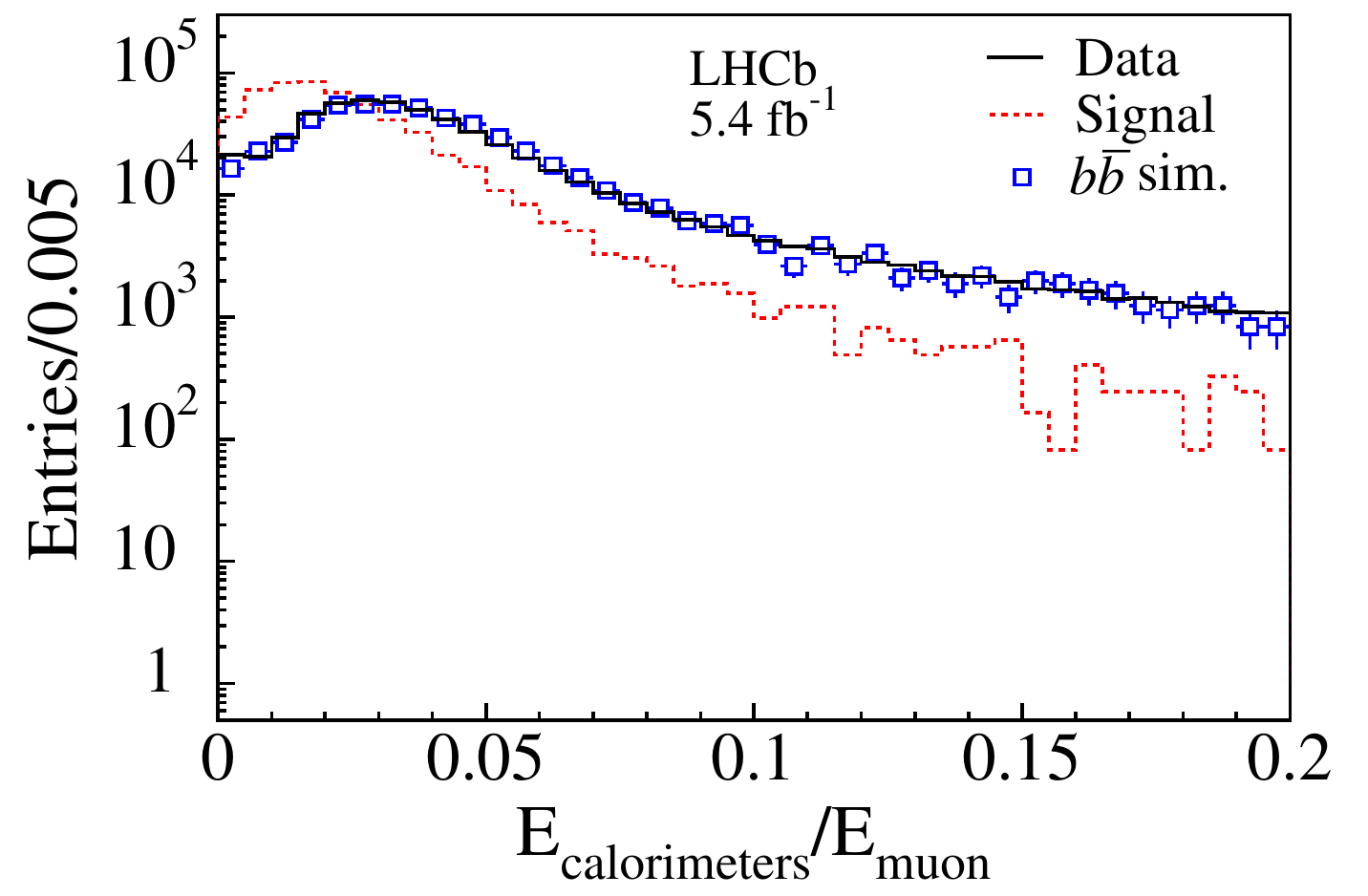}
 \put(-314,30){\small (c)} \put(-142,30){\small (d)}
}
\mbox{
\centering
\includegraphics[height=4cm,width=6.2cm]{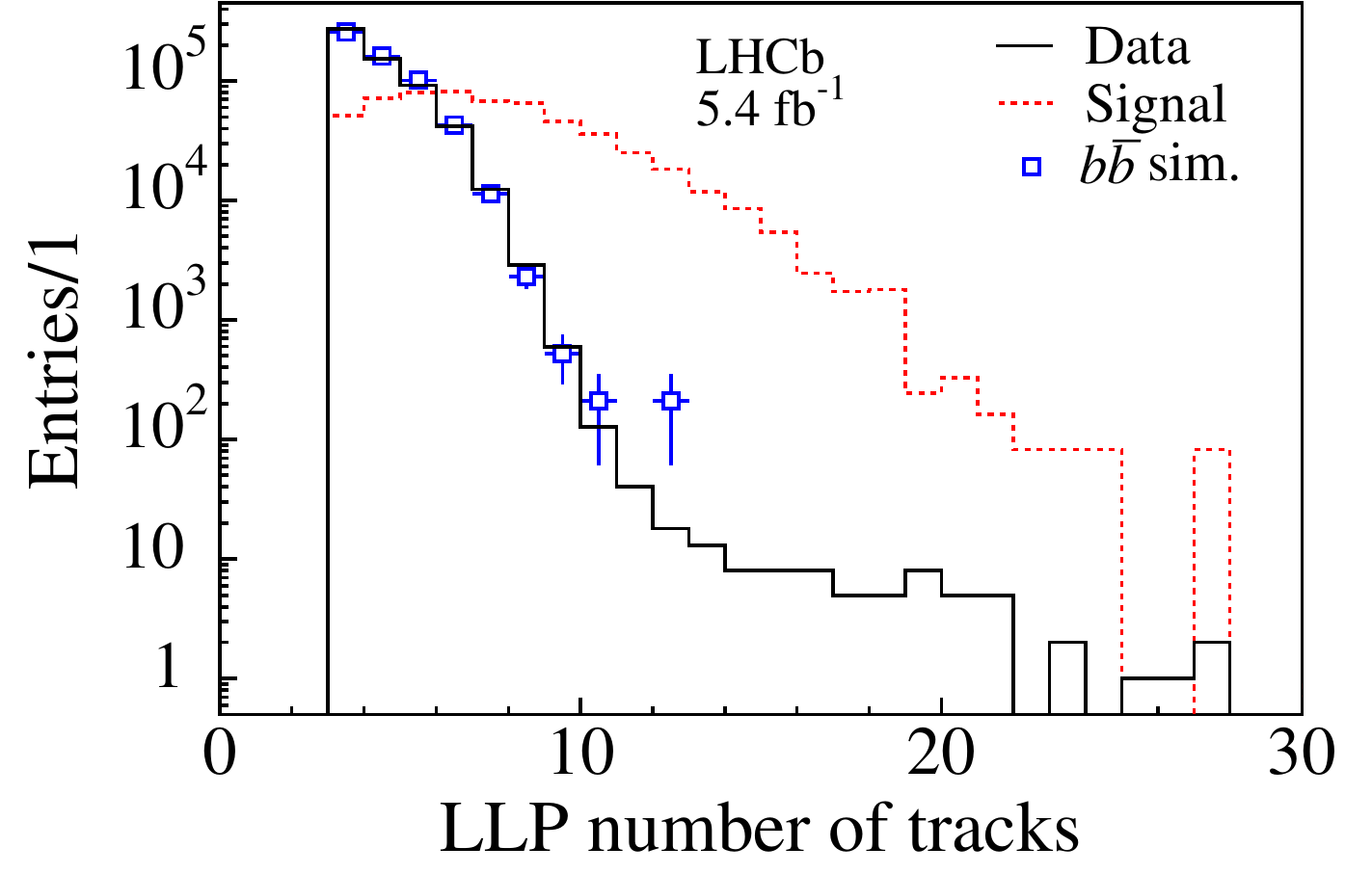}
\includegraphics[height=4cm,width=6.2cm]{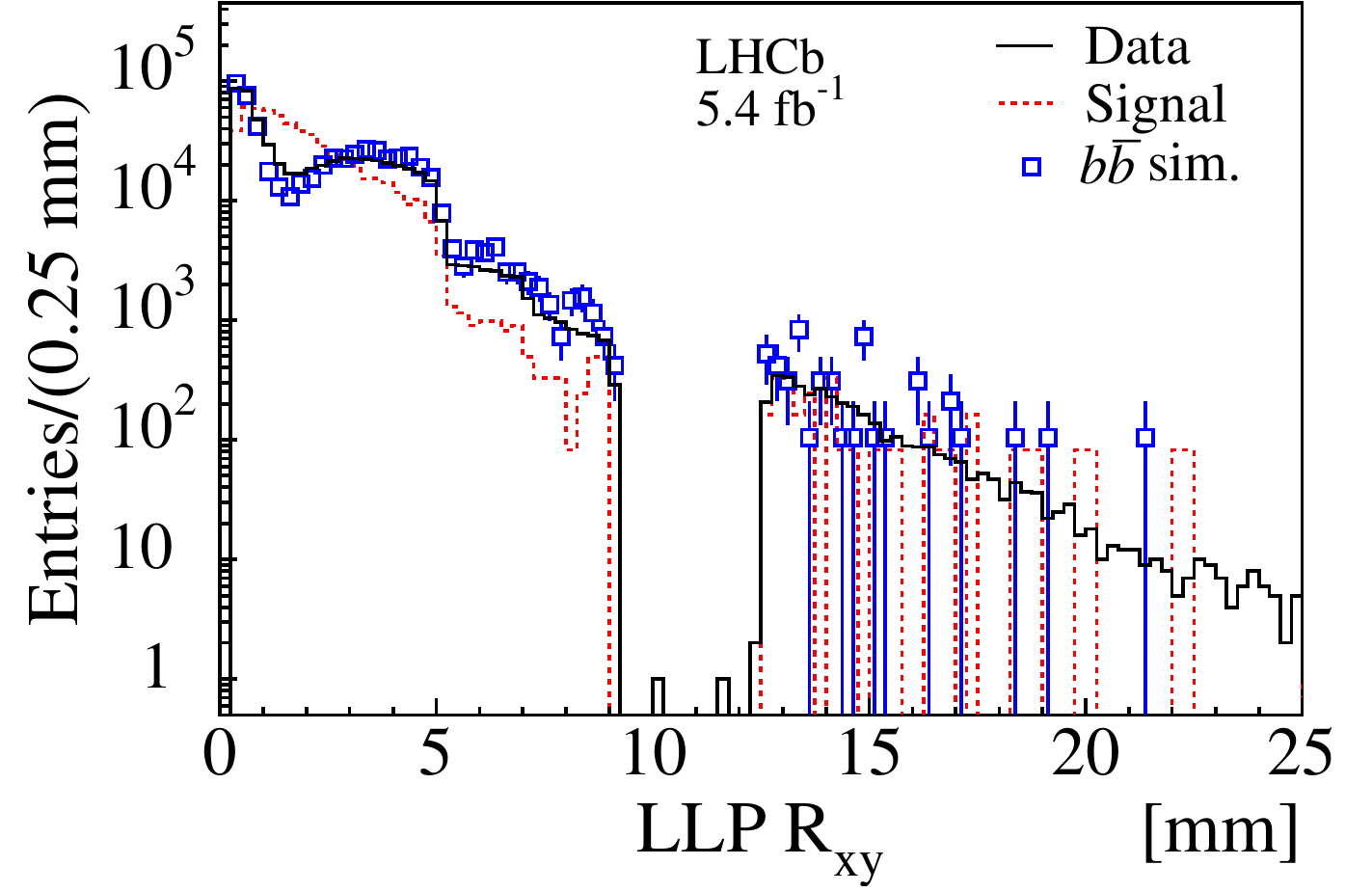}
 \put(-314,30){\small (e)} \put(-142,30){\small (f)}
}
\mbox{
\centering
\includegraphics[height=4cm,width=6.2cm]{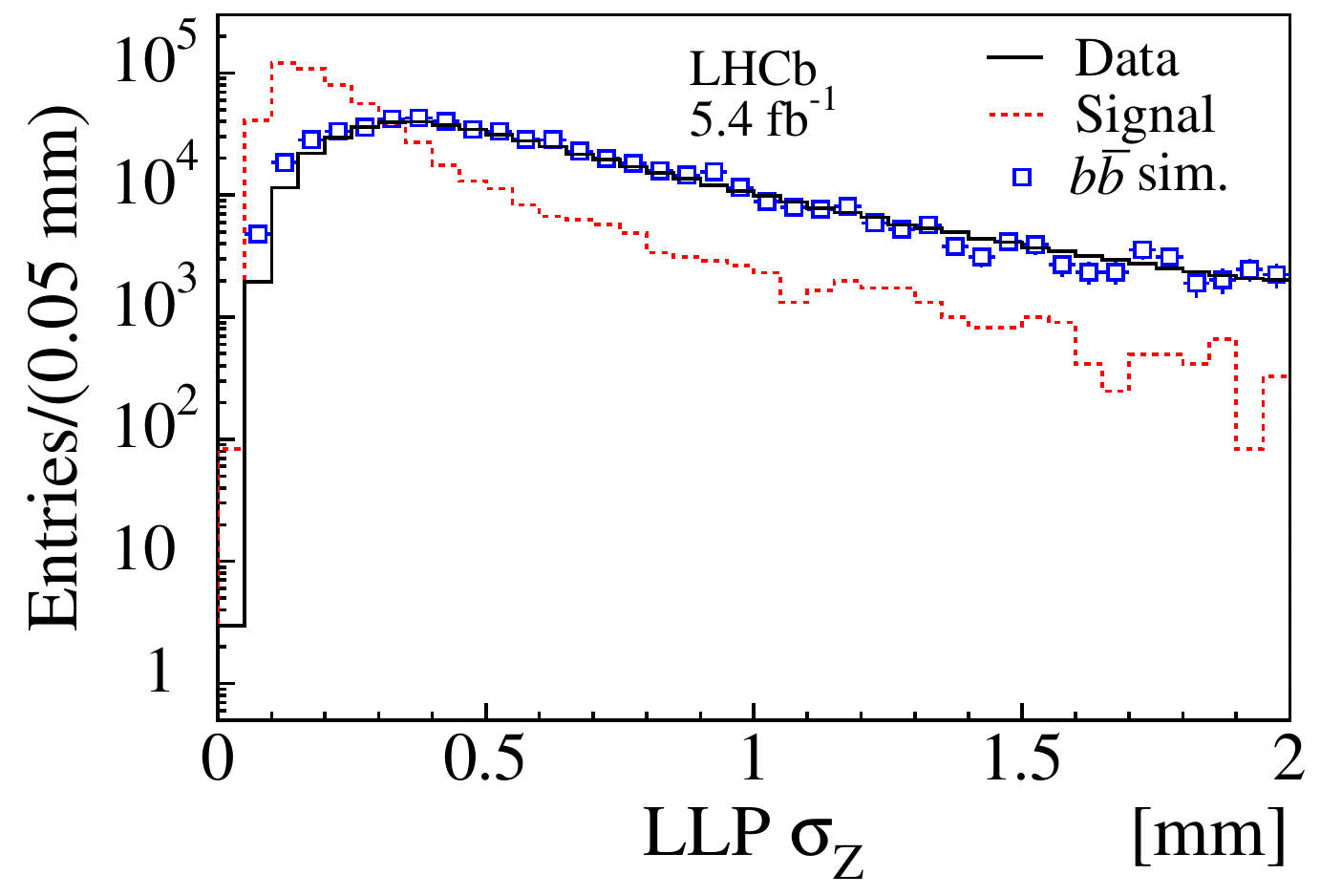}
\includegraphics[height=4cm,width=6.2cm]{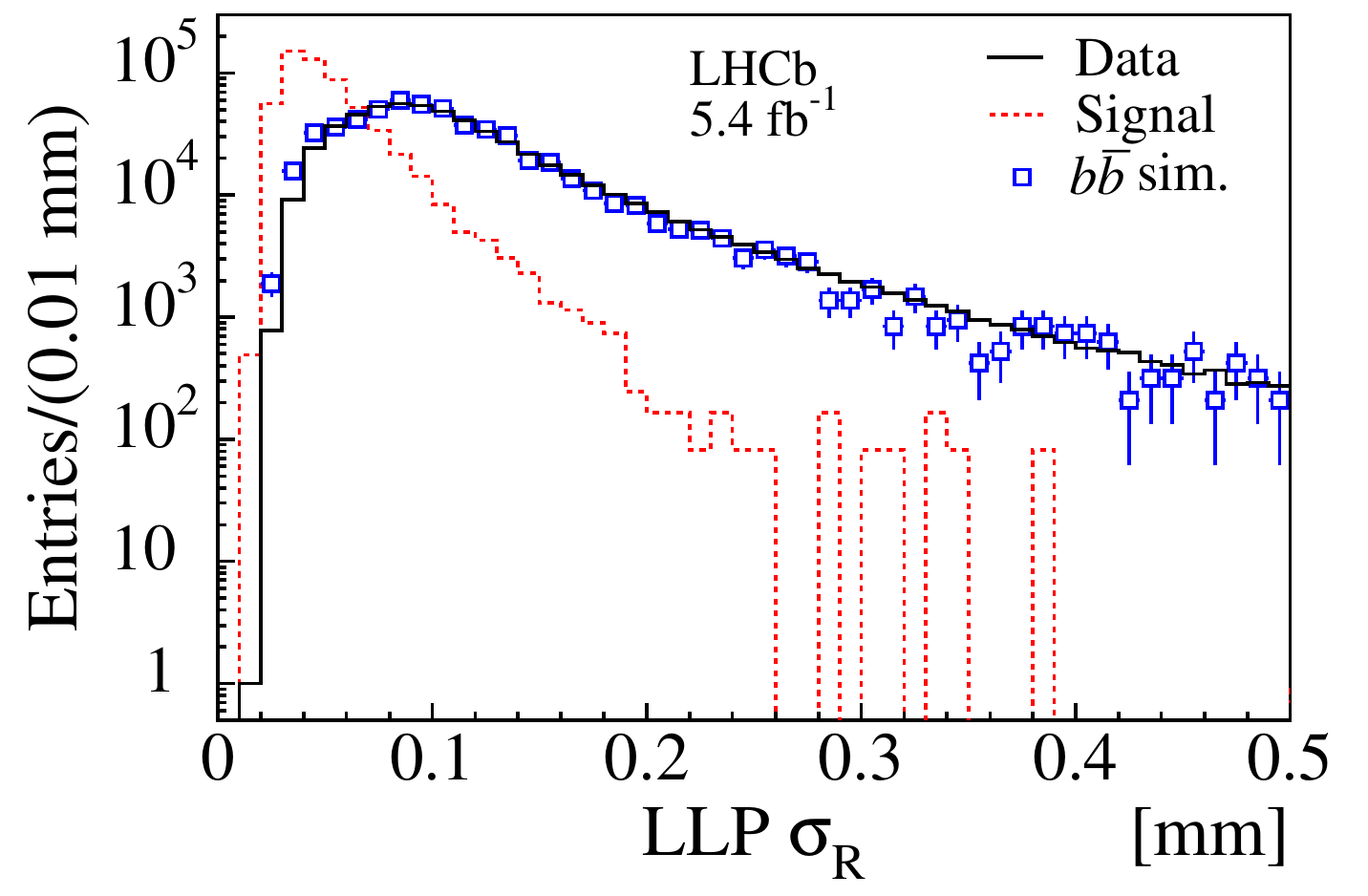}
 \put(-314,30){\small (g)} \put(-140,30){\small (h)}
}
\mbox{
  \includegraphics[height=4cm,width=6.2cm]{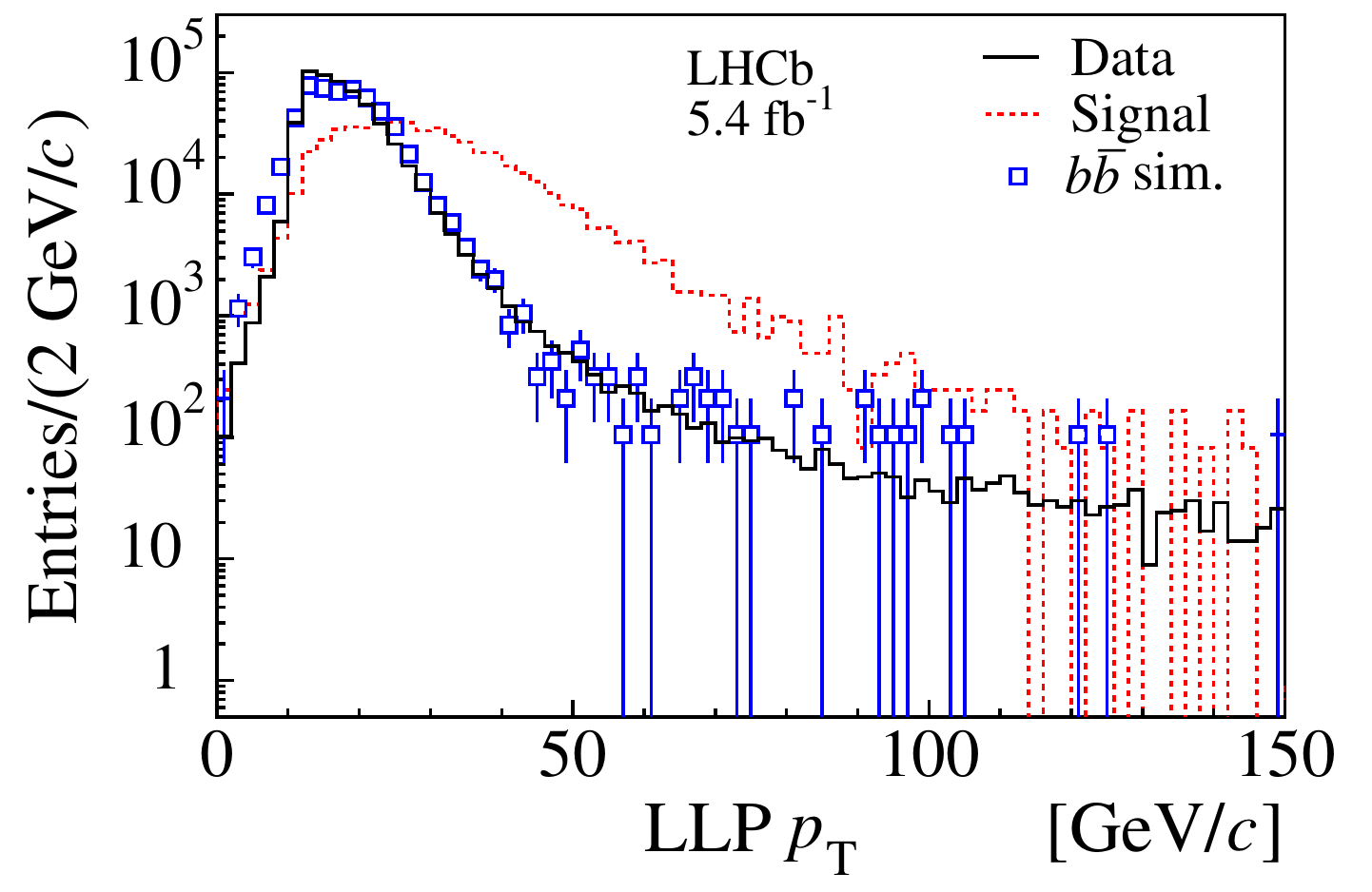}
  \includegraphics[height=4cm,width=6.2cm]{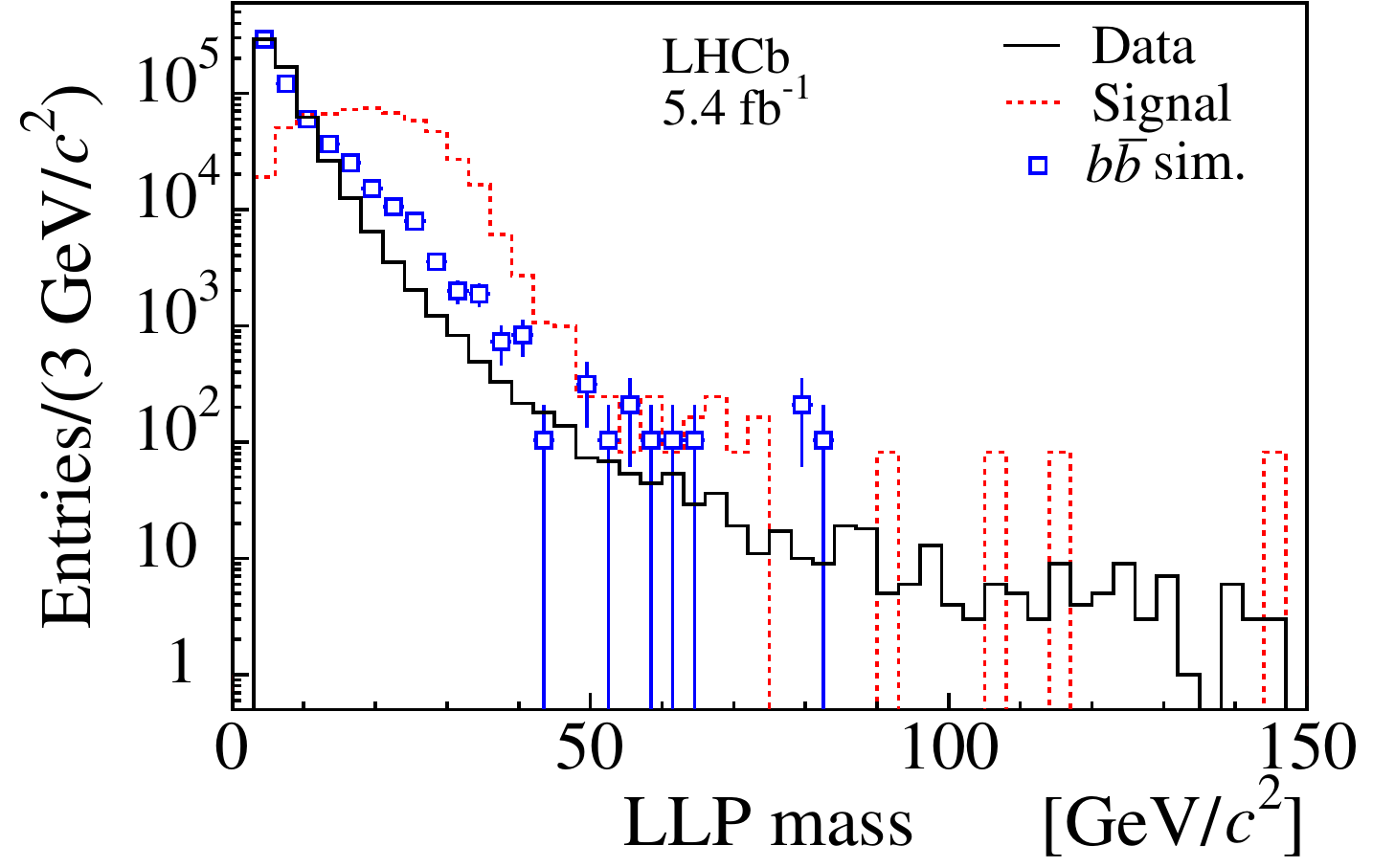}
  \put(-314,30){\small (i)} \put(-140,30){\small (j)}
}
\caption{
\small
Distributions from data compared to simulated \bbbar events (blue)
and the simulated signal h125-chi40-10p (red),
after preselection.
  From (a) to (j):
  muon transverse momentum;
  muon impact parameter;
  muon isolation; the calorimetric energy, $\rm E_{calorimeters}$, associated with the muon normalised by the muon energy,
  $\rm E_{muon}$;
  the number of tracks used to reconstruct the LLP vertex including the muon;
  the radial distance to the beam line of the reconstructed vertex;
  longitudinal and radial vertex fit errors, $\sigma_Z$ and $\sigma_R$;
  reconstructed transverse momentum and mass of the LLP candidate.
  The distributions from simulated events are normalised to the data.
  }
\label{fig:presel}
\end{figure}

Figure~\ref{fig:presel} compares the distributions from data and from the simulated \bbbar events
for the relevant observables, after preselection.
For illustration the shapes of simulated h125-chi40-10ps events are also superimposed.
The effect of the geometric veto is visible in the \RXY distribution, for candidates with \RXY above 5\mm.
From simulation, the veto introduces a loss of efficiency of 3\% (27\%)
for the detection of LLPs with a 50\gevcc mass and a 10 ps (200 ps) lifetime, \mhzero=125\gevcc. 
The muon-isolation variable is defined as the sum of the energy of tracks
surrounding the muon direction, including the muon itself, in a cone of radius $R_{\eta \phi}= 0.3$ in the  pseudorapidity-azimuthal $(\eta, \phi)$ space, divided by the energy of the muon track.
The radius is reduced to  \,$R_{\eta \phi}= 0.2$\, when the theoretical hypothesis assumes a LLP mass of 10\gevcc,
to account for the reduced aperture of the jet of particles produced by the LLP decay.
A muon-isolation value of unity denotes a fully isolated muon.
In simulation the muon from the signal is found to be more isolated than the hadronic background.
The variables  $\sigma_R$ and $\sigma_Z$ are the vertex uncertainties in the radial direction and in the z direction respectively.

The reconstructed vertex mass is very broad and does not peak at the neutralino mass values, because it misses some charged particle tracks, and any neutrals produced in the LLP decay.

The shapes of the distributions in Fig.~\ref{fig:presel} are all consistent with a dominant
\bbbar composition of the background. 
This is confirmed by comparing  the yields in data and simulation:
after preselection and requiring the isolation parameter below 1.2,
the total number of LLP candidates in data is $148\times 10^3$.
The predicted background yields from \bbbar and \ccbar events are
$(120 \pm 20)\times 10^3$ and $(14 \pm 4)\times10^3$, respectively.
Small contributions are expected from processes with \W, \Z bosons plus jets,
top and Standard Model Higgs events:
260, 20, 2, and 1 candidates, respectively.
The \bbbar and \ccbar prediction uses the cross-sections measured by the LHCb experiment at
13 TeV~\cite{LHCb-PAPER-2016-031,LHCb-PROC-2016-001}.
The acceptance of this analysis is computed with {\mbox{\textsc{MadGraph5-aMC@NLO}}\xspace}~\cite{Alwall_2014}
and the detection efficiency is obtained from simulated events.
As already stated, these background estimations are only used for cross-checks.

A multivariate analysis (MVA) based on a boosted decision tree~\cite{Breiman,AdaBoost}
is used to further purify the data sample.
Ten MVA input variables are selected to optimise the signal--background separation.
They are: \PTMU and \IPMU,
the ratio of the energies associated with the muon measured in ECAL and HCAL normalised to the muon energy,
the LLP candidate \pt, its pseudorapidity, the number of tracks forming the LLP,
the vertex uncertainties  $\sigma_R$ and $\sigma_Z$,
and the vertex \RXY distance.

Larger vertex uncertainties are expected on the vertices of candidates from \bbbar events
compared to signal LLPs. The former are more boosted and produce more collimated tracks,
while the relatively heavier signal LLPs decay into more divergent tracks.
This effect decreases when the mass of the LLP approaches the mass of \bquark-quark hadrons.
The selection based on the energy deposit in the calorimeters is efficient 
to suppress the background due to kaons or pions punching through the calorimeters
and being misidentified as muons.
The muon-isolation variable and the reconstructed mass of the long-lived particles are not included in the
classifier; the discrimination power of these two variables is
subsequently exploited for the signal determination.

The signal MVA training samples are provided by simulation.
The background training sample is obtained from data,
based on the hypothesis that the fraction of signal in the data after preselection is small.
This automatically includes all possible background sources, with the correct relative abundance.

The training is performed independently for each simulated model.
The MVA classifier is subsequently applied to the data and to the simulated signal.
For each model, the optimal MVA cut value is chosen by an iterative minimization procedure to give the
best expected cross-section upper limit, but keeping at least ten candidates to allow the invariant-mass fit to work properly.

The classifier can be biased by the presence of signal in data used as
background training set. To quantify the potential bias,
the MVA training is performed adding a fraction of simulated signal events (up to 5\%)
to the background set.
This test demonstrates a negligible effect on the MVA performance for all the signal models.

\section{Determination of the signal yield}\label{sec:signal-extraction}

The signal yield is determined with an unbinned extended maximum-likelihood fit
to the distribution of the reconstructed LLP mass.
The shape of the signal component is taken from the simulated models, and a background component is added.
After the MVA selection, no simulated background survives,
therefore the background shape is determined by a data-driven method,
which also avoids potential simulation mismodeling of the reconstructed mass.
The data candidates are separated into a signal region with muon isolation below 1.2
and a background region with isolation values from 1.4 to 2.0.
The signal-region selection accepts more than 80\% of the signal for all the models considered
(see \eg Fig.~\ref{fig:presel}). Any potential signal yield in the background region is considered negligible. 
The reconstructed mass distribution obtained from the background candidates is used to constrain
an empirical  probability density function (PDF) consisting of the sum of two negative-slope exponential functions, one
of them convolved with a Gaussian function. Shape parameters and amplitudes are left to vary in the fit.
It is possible that the mass distribution obtained after selection of the background
region does not represent exactly the background component in the signal region. Hence, a correction is applied before 
performing the fit: the mass distribution selected in the background region is weighted with
weights deduced from the comparison of the candidate mass distributions of signal and background regions
obtained from data with a relaxed MVA selection.
This relaxed selection is required to have sufficiently populated samples
and to minimise the correlation with the final distributions from which signal yields are obtained.
The consistency of this procedure is tested on \bbbar simulated events.
%
%
%
 
\begin{center}
\begin{footnotesize}
\begin{longtable}[t]
{cccccS[tabnumalign=centre,tabformat=4.2]@{\:$\pm$\:}S[tabnumalign=centre,tabformat=3.2]S[tabnumalign=centre,tabformat=4.2]@{\:$\pm$\:}Sc}
\caption {
 Signal detection efficiency, in percent, after preselection, 
 $\eps_{presel}$, including the geometrical acceptance, and after MVA selection,  $\eps$,
 and numbers of fitted signal and background events in the signal region, $N_{\rm s}$ and $N_{\rm b}$,
 for the different  signal hypotheses with resonant LLP production  (masses units are \gevcc, lifetimes in \ps).
 The last column gives the value of  \chisq per degree of freedom, ndf, from the fit.}
\label{tab:eff-fit-h} \\
\hline
 $m_{h0}$ &  m$_{\khi}$ & $\tau_{khi}$  & $\eps_{presel}$ & $\eps$ &  \multicolumn{2}{c}{$N_{\rm b}$} & \multicolumn{2}{c}{$N_{\rm s}$} & {\chisqndf} \\
\hline
\endfirsthead
\multicolumn{9}{c} {\tablename\ \thetable{} -- continued from previous page} \\
\hline
 $\mhzero$ &  $m_{\khi}$ & $\tau_{khi}$    & $\eps_{presel}$  & $\eps$ &  \multicolumn{2}{c}{$N_{\rm b}$} & \multicolumn{2}{c}{$N_{\rm s}$} & {\chisqndf} \\
\hline
\endhead
\hline \multicolumn{9}{r}{\textit{Continued on next page}}
\endfoot
\hline
\endlastfoot
    30 &  10 &    10 &    0.19 &    0.03 &  1025.7 &  159.4 &   -89.7 &   39.1 &  2.75 \\
    41 &  10 &    10 &    0.29 &    0.01 &    61.3 &   51.7 &    -5.3 &    5.2 &  0.67 \\
    41 &  20 &    10 &    0.53 &    0.08 &    73.2 &   14.2 &    13.8 &   12.8 &  1.22 \\
    50 &  10 &    10 &    0.46 &    0.09 &  1421.3 &   84.4 &   -71.4 &   75.1 &  1.03 \\
    50 &  20 &    10 &    0.95 &    0.12 &    18.3 &    6.9 &     6.7 &    6.4 &  1.12 \\
    80 &  10 &    10 &    0.74 &    0.05 &   127.0 &   17.8 &     1.0 &   14.5 &  1.35 \\
    80 &  20 &    10 &    1.86 &    0.42 &    53.9 &   45.3 &    -5.4 &    5.0 &  1.07 \\
    80 &  30 &    10 &    2.34 &    0.72 &    22.4 &    4.2 &    -0.4 &    4.2 &  0.67 \\
    80 &  40 &    10 &    1.83 &    0.43 &    12.1 &    4.1 &     1.9 &    4.5 &  0.53 \\
   125 &  10 &     5 &    0.63 &    0.04 &   333.7 &   28.0 &   -10.6 &   21.7 &  0.77 \\
   125 &  10 &    10 &    0.68 &    0.06 &   198.3 &   19.2 &     4.7 &   14.7 &  1.72 \\
   125 &  10 &    20 &    0.73 &    0.09 &  1680.8 &   42.9 &  -172.2 &   34.0 &  0.96 \\
   125 &  10 &    30 &    0.64 &    0.08 &   419.0 &   45.1 &   -29.9 &   34.8 &  1.43 \\
   125 &  10 &    40 &    0.49 &    0.06 &   184.7 &   30.4 &     8.3 &   28.2 &  0.77 \\
   125 &  10 &    50 &    0.51 &    0.11 &   667.0 &   46.9 &    -2.0 &   39.5 &  1.20 \\
   125 &  10 &   100 &    0.32 &    0.04 &    53.7 &   12.4 &    10.3 &   11.7 &  1.14 \\
   125 &  10 &   200 &    0.19 &    0.03 &    71.5 &    8.4 &    -0.5 &    4.1 &  0.57 \\
   125 &  20 &     5 &    1.86 &    0.36 &    56.1 &   43.3 &    -5.8 &    5.6 &  1.27 \\
   125 &  20 &    10 &    2.28 &    0.67 &    39.6 &   36.8 &    -3.9 &    5.4 &  1.06 \\
   125 &  20 &    20 &    2.24 &    0.47 &    18.1 &   15.4 &    -0.1 &    5.5 &  0.97 \\
   125 &  20 &    30 &    2.10 &    0.35 &     9.5 &    4.0 &     7.5 &    4.2 &  0.93 \\
   125 &  20 &    50 &    1.73 &    0.44 &    28.9 &    9.4 &     8.1 &    8.7 &  1.67 \\
   125 &  20 &    80 &    1.36 &    0.21 &    22.9 &    5.8 &     2.1 &    5.5 &  0.24 \\
   125 &  20 &   100 &    1.18 &    0.25 &    15.0 &   14.9 &    -1.0 &    4.1 &  0.80 \\
   125 &  20 &   200 &    0.71 &    0.09 &    11.0 &    2.0 &    -3.5 &    2.2 &  0.40 \\
   125 &  30 &     5 &    2.24 &    0.68 &    93.5 &    8.8 &   -15.0 &    5.5 &  2.26 \\
   125 &  30 &    10 &    3.13 &    0.72 &     7.0 &    4.0 &     3.9 &    4.0 &  0.84 \\
   125 &  30 &    20 &    3.24 &    0.93 &     6.6 &    3.2 &     3.4 &    3.1 &  0.35 \\
   125 &  30 &    30 &    2.92 &    0.87 &    14.3 &    2.9 &    -1.6 &    2.8 &  0.84 \\
   125 &  30 &    50 &    2.51 &    1.00 &    22.7 &   22.6 &    -1.7 &    6.2 &  0.83 \\
   125 &  30 &   100 &    1.84 &    0.58 &     8.8 &    5.5 &    -2.7 &    2.1 &  0.96 \\
   125 &  30 &   200 &    1.09 &    0.45 &    25.3 &    4.8 &    -4.8 &    4.1 &  1.16 \\
   125 &  40 &     5 &    2.41 &    0.59 &     9.9 &    6.3 &    -2.5 &    1.9 &  0.77 \\
   125 &  40 &    10 &    3.44 &    0.84 &    11.0 &    1.6 &    -2.2 &    2.1 &  0.94 \\
   125 &  40 &    20 &    3.79 &    1.90 &    10.5 &    3.2 &     4.5 &    2.6 &  0.72 \\
   125 &  40 &    30 &    3.71 &    1.37 &     6.3 &    3.4 &     4.7 &    3.4 &  1.31 \\
   125 &  40 &    50 &    3.24 &    1.39 &    28.6 &    1.7 &   -10.6 &    4.8 &  0.87 \\
   125 &  40 &   100 &    2.30 &    0.64 &     9.9 &    8.3 &    -1.4 &    2.2 &  0.56 \\
   125 &  40 &   200 &    1.49 &    0.61 &     9.9 &    8.1 &    -1.5 &    2.2 &  1.28 \\
   125 &  50 &     5 &    2.25 &    0.48 &     8.3 &    1.5 &    -0.3 &    2.4 &  0.60 \\
   125 &  50 &    10 &    3.41 &    1.60 &    20.9 &    4.0 &     1.1 &    1.8 &  0.58 \\
   125 &  50 &    20 &    4.10 &    2.01 &    16.5 &    0.5 &   -14.9 &   11.9 &  0.57 \\
   125 &  50 &    30 &    4.13 &    2.04 &     4.2 &    2.4 &     3.9 &    2.5 &  0.69 \\
   125 &  50 &    50 &    3.86 &    2.30 &    17.5 &    0.5 &    -0.5 &    0.3 &  0.64 \\
   125 &  50 &   100 &    3.02 &    1.74 &    15.4 &    0.8 &    -9.3 &    2.9 &  0.95 \\
   125 &  50 &   200 &    2.00 &    0.90 &     4.4 &    1.9 &     3.5 &    2.1 &  0.76 \\
   125 &  60 &     5 &    1.64 &    0.48 &    20.9 &   15.8 &    -2.6 &    3.1 &  0.64 \\
   125 &  60 &    10 &    2.85 &    1.22 &     9.8 &    2.1 &    -0.8 &    2.2 &  1.10 \\
   125 &  60 &    20 &    3.98 &    2.04 &    22.8 &    4.8 &     1.2 &    1.9 &  0.88 \\
   125 &  60 &    30 &    4.20 &    2.39 &    16.5 &    0.3 &   -10.0 &    0.9 &  0.73 \\
   125 &  60 &    50 &    4.57 &    2.87 &    20.9 &    2.9 &     3.1 &    1.4 &  1.13 \\
   125 &  60 &   100 &    3.68 &    2.30 &     9.9 &    1.9 &    -1.3 &    2.4 &  1.73 \\
   125 &  60 &   200 &    2.75 &    1.90 &    10.7 &    3.8 &     3.3 &    3.0 &  1.58 \\
   150 &  10 &    10 &    0.62 &    0.12 &  1514.7 &   52.5 &  -152.0 &   30.8 &  2.01 \\
   150 &  20 &    10 &    2.19 &    0.67 &    66.0 &   13.6 &    -8.2 &    6.0 &  1.47 \\
   150 &  30 &    10 &    3.20 &    0.60 &     6.9 &    2.0 &     1.1 &    1.1 &  0.67 \\
   150 &  40 &    10 &    3.55 &    1.44 &    15.4 &   12.1 &    -2.1 &    2.8 &  1.35 \\
   150 &  50 &    10 &    3.79 &    1.06 &     6.9 &    2.0 &     1.1 &    1.1 &  0.87 \\
   150 &  60 &    10 &    3.68 &    1.79 &    25.3 &    1.5 &   -15.0 &    1.7 &  1.41 \\
   200 &  10 &    10 &    0.46 &    0.05 &   433.1 &   34.2 &     7.0 &   27.6 &  1.59 \\
   200 &  20 &    10 &    1.93 &    0.45 &   136.2 &   26.2 &    21.8 &   24.5 &  0.95 \\
   200 &  30 &    10 &    2.94 &    0.48 &     8.6 &    3.9 &     4.4 &    4.2 &  0.79 \\
   200 &  40 &    10 &    3.43 &    0.84 &     9.3 &    2.8 &     3.7 &    2.2 &  1.30 \\
   200 &  50 &    10 &    3.84 &    1.48 &    12.1 &    2.1 &    -1.8 &    2.3 &  0.50 \\
   200 &  60 &    10 &    3.97 &    1.81 &    12.1 &    2.1 &    -1.8 &    2.5 &  0.56 \\
   200 &  50 &     5 &    2.90 &    1.20 &    11.8 &    2.1 &    -0.8 &    1.2 &  0.57 \\
   200 &  50 &    20 &    3.80 &    1.79 &     7.7 &    2.7 &     1.3 &    2.8 &  0.70 \\
   200 &  50 &    30 &    3.45 &    0.98 &    17.6 &    1.7 &    -6.0 &    2.7 &  0.62 \\
   200 &  50 &    50 &    2.91 &    1.14 &     7.7 &    0.8 &    -3.4 &    1.9 &  1.07 \\
   200 &  50 &   100 &    1.95 &    1.04 &    10.0 &    3.0 &     1.0 &    3.0 &  1.07 \\
   200 &  50 &   200 &    1.17 &    0.72 &    17.6 &    3.5 &    -4.0 &    3.6 &  0.50 \\
 \hline
\end{longtable}
\end{footnotesize}
 \end{center}

\begin{table}[ht]
  \caption {
Signal detection efficiency, in percent, after preselection, 
 $\eps_{presel}$, including the geometrical acceptance, and after MVA selection,  $\eps$,
 and numbers of fitted signal and background events in the signal region, $N_{\rm s}$ and $N_{\rm b}$,
 for the different signal hypotheses  for the LLP non-resonant production models (masses units are \gevcc, lifetimes in \ps).
 The last column gives the value of  \chisq per degree of freedom, ndf, from the fit.
} \label{tab:eff-fit-p216}
\centering
 \begin{footnotesize}
\begin{tabular}
 {ccccS[tabnumalign=centre,tabformat=4.2]@{\:$\pm$\:}S[tabnumalign=centre,tabformat=3.2]S[tabnumalign=centre,tabformat=4.2]@{\:$\pm$\:}Sc}
\hline
 $m_{\khi}$ & $\tau_{\khi}$    & $\eps_{presel}$  & $\eps$ &  \multicolumn{2}{c}{$N_{\rm b}$} & \multicolumn{2}{c}{$N_{\rm s}$} & {\chisqndf} \\
\hline
    10 &    10 &    0.61 &    0.13 &  2767.9 &   88.2 &  -141.8 &   69.7 &  1.69 \\
    20 &    10 &    0.66 &    0.23 &    43.9 &   40.1 &    -4.2 &    5.0 &  0.67 \\
    30 &    10 &    2.29 &    0.47 &    15.7 &    5.8 &     3.3 &    5.2 &  0.90 \\
    40 &    10 &    2.49 &    0.52 &     1.1 &    1.4 &     5.9 &    2.8 &  0.96 \\
    60 &    10 &    3.81 &    1.97 &    45.1 &    5.6 &    -8.0 &    4.3 &  0.80 \\
    90 &    10 &    2.52 &    1.68 &    30.8 &    2.2 &    -9.8 &    5.0 &  1.04 \\
    30 &     5 &    1.44 &    0.21 &    11.0 &    2.5 &    -1.0 &    2.7 &  0.67 \\
    30 &    20 &    2.64 &    0.66 &    13.8 &    4.4 &     3.2 &    4.2 &  0.65 \\
    30 &    30 &    2.52 &    0.74 &     5.6 &    2.2 &     2.4 &    2.1 &  0.41 \\
    30 &    50 &    2.25 &    0.81 &    16.5 &   16.1 &    -1.8 &    3.2 &  0.69 \\
    30 &   100 &    1.68 &    0.61 &     9.9 &    7.4 &    -1.7 &    3.1 &  1.10 \\
    30 &   200 &    1.06 &    0.29 &    38.0 &    6.3 &    -0.0 &    2.3 &  0.79 \\
 \hline
 \end{tabular}   
 \end{footnotesize}
 \end{table}

\begin{figure}[t]
  \centering
  \hspace{-2mm}
  {\includegraphics[clip, trim=0.5mm 0.5mm 0.5mm 0.7mm, height=4.5cm]{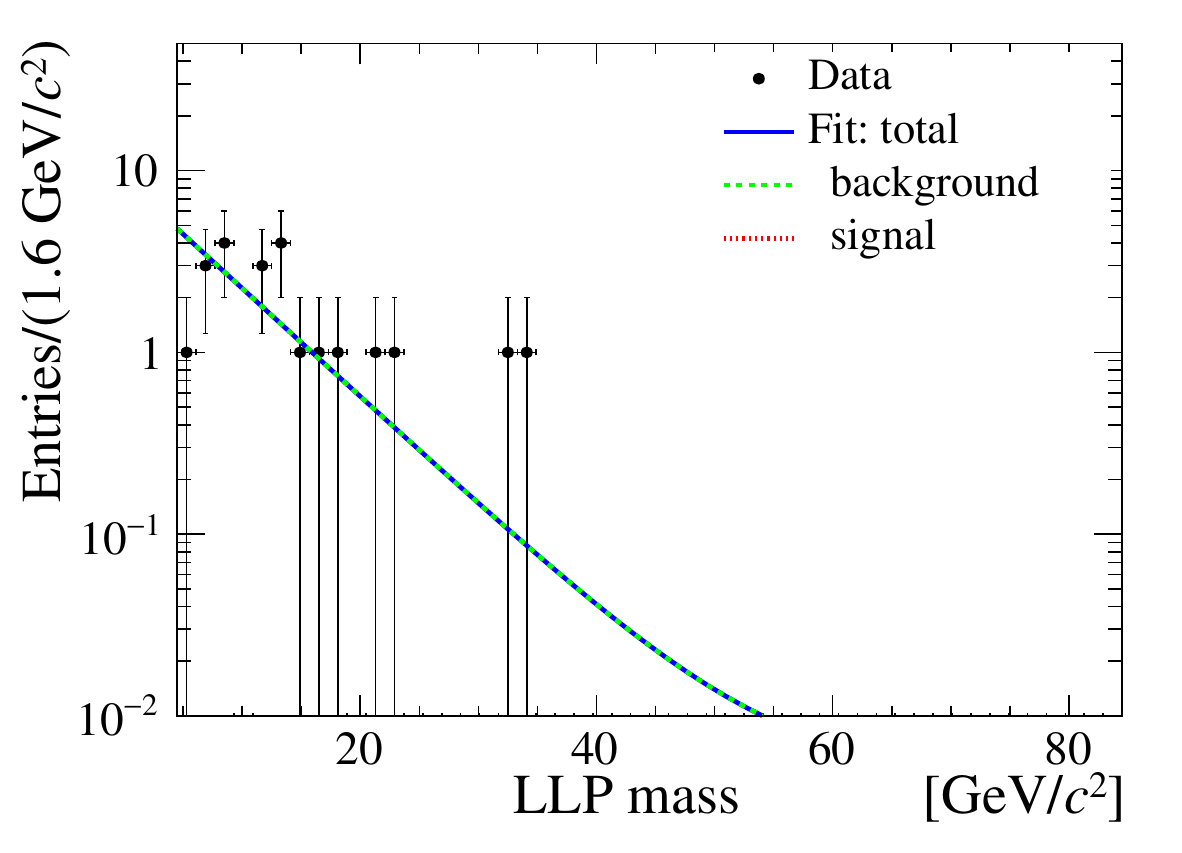}
    \put(-30,80){(a)} \put(-115,110){\lhcbX}\put(-115,100){\lhcbL}   }
  {\includegraphics[clip, trim=0.5mm 0.5mm 0.5mm 0.5mm, height=4.5cm]{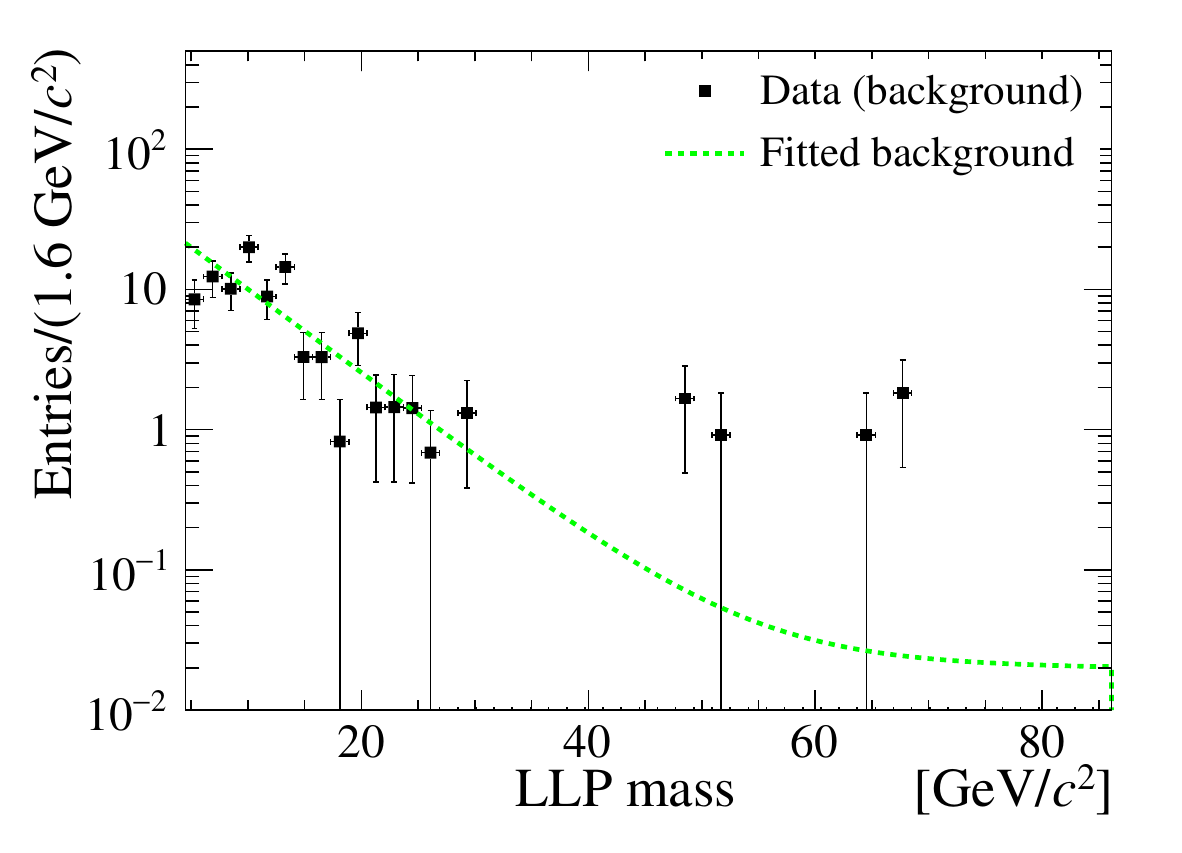}
    \put(-33,90){(b)} \put(-120,110){\lhcbX}\put(-120,100){\lhcbL} }

    {\includegraphics[clip, trim=0.5mm 0.2mm 0.5mm 0.2mm, height=4.5cm]{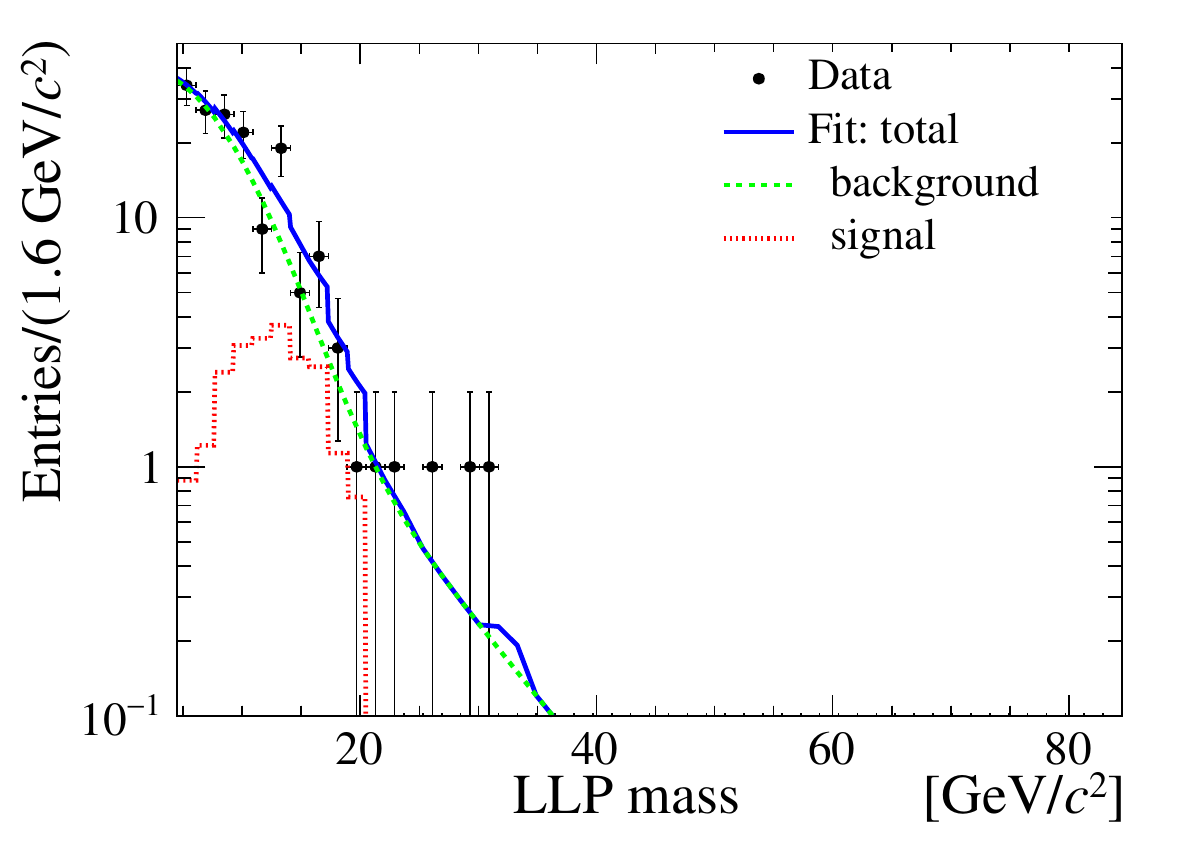}
  \put(-33,80){(c)} \put(-120,110){\lhcbX}\put(-120,100){\lhcbL} }
  {\includegraphics[clip, trim=0.5mm 0.5mm 0.5mm 0.1mm, height=4.5cm ]{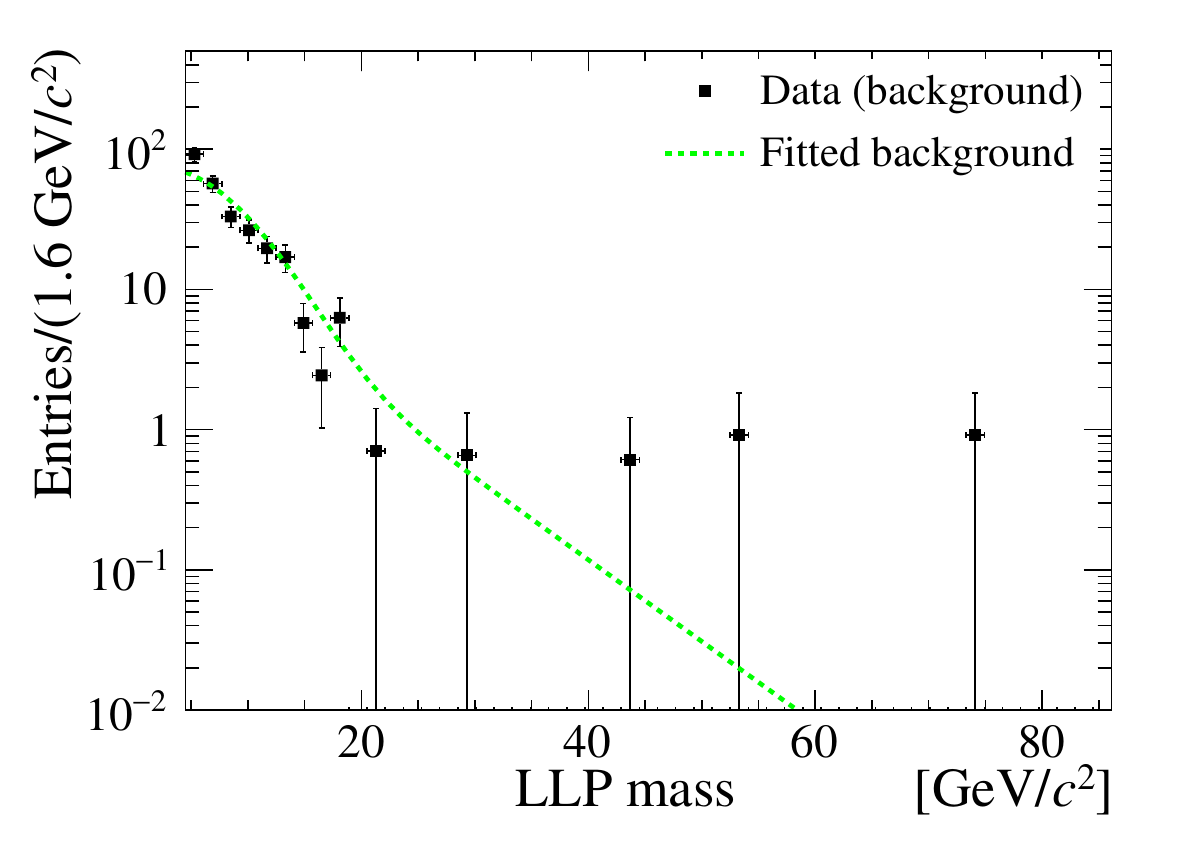}
    \put(-30,80){(d)}          \put(-115,110){\lhcbX}\put(-115,100){\lhcbL}  }

    {\includegraphics[clip, trim=0.5mm 0.5mm 0.5mm 0.1mm, height=4.5cm ]{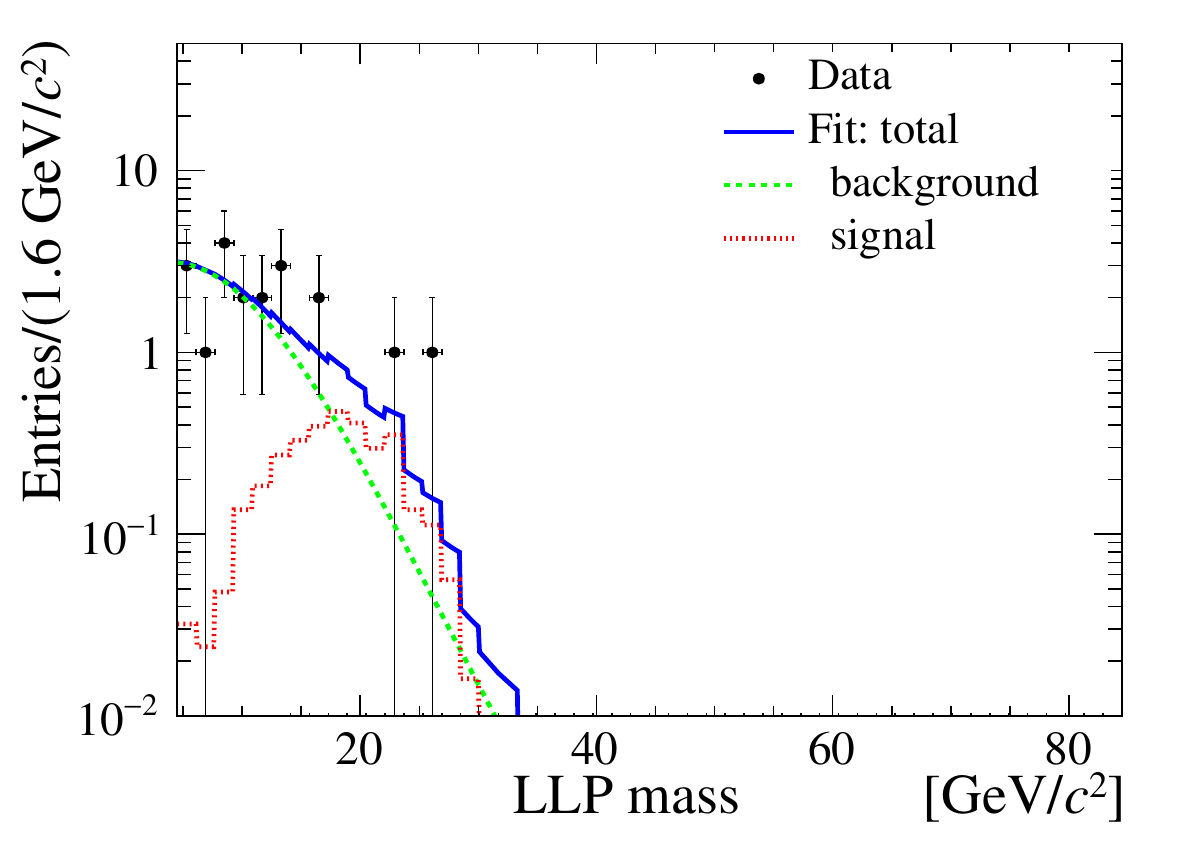}
    \put(-30,80){(e)}  \put(-115,110){\lhcbX}\put(-115,100){\lhcbL}   }
     {\includegraphics[clip, trim=0.5mm 0.2mm 0.5mm 0.2mm, height=4.5cm]{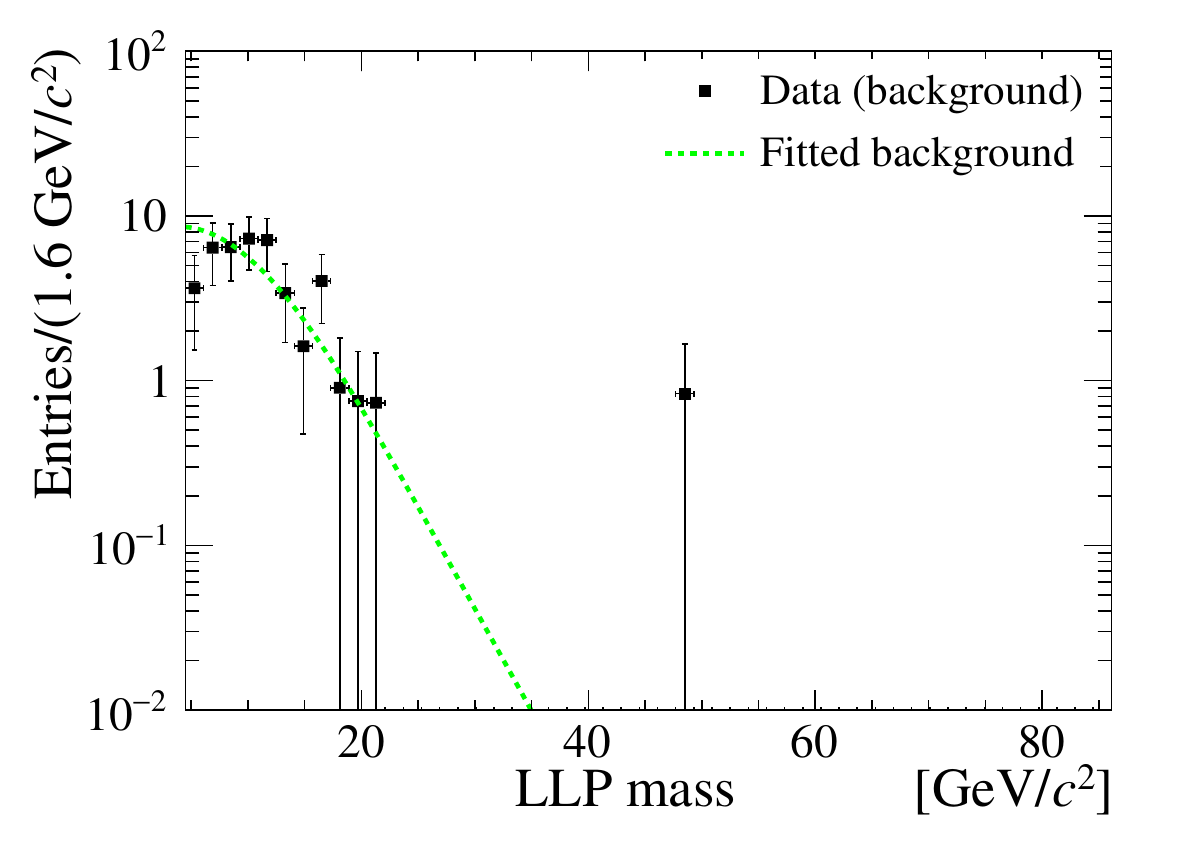}
    \put(-33,90){(f)} \put(-120,110){\lhcbX}\put(-120,100){\lhcbL} }
\caption{\small
  Reconstructed invariant mass of the LLP candidates.
  Subfigures (a), (c), and (e) correspond to the signal selections which assume the models h80-chi30-10ps,  
  h200-chi20-10ps, and the non-resonant model chi30-10ps, respectively.
  Subfigures (b), (d), and (f) are the corresponding distributions for candidates selected in the background region.
  The results of the fits are superimposed.
}
\label{fig:fit}
\end{figure}
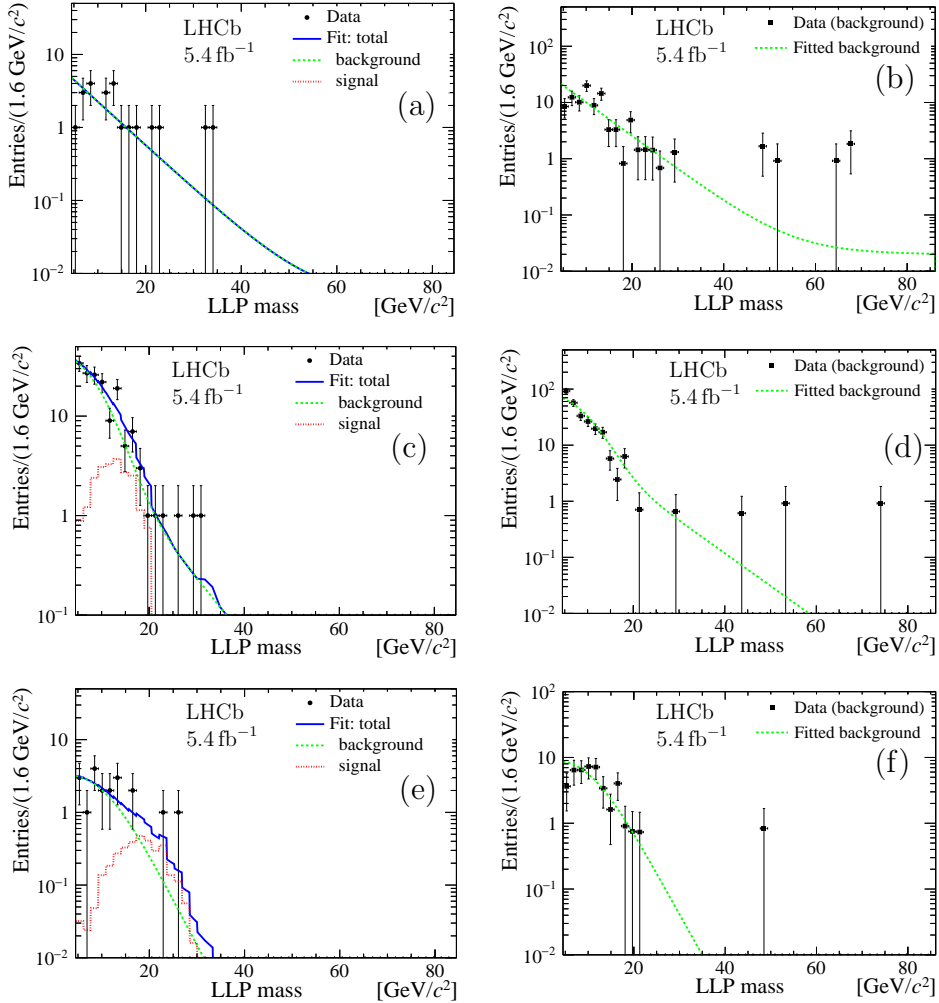
Examples of the invariant mass of the selected LLP candidates are shown in  Fig.~\ref{fig:fit}
for the signal and background regions.
The invariant-mass fit is performed simultaneously on LLP candidates from the signal and from the background regions.
In the former, the numbers of signal and background
events are free parameters of the fit.
The results of the fit are shown in the figure.
The sensitivity of the fit procedure is studied by adding a small number of simulated signal
events to the data according to a given signal model. The fitted yields are on average
consistent with the numbers of added events.
The fitted signal yields, given in Tables~\ref{tab:eff-fit-h} and~\ref{tab:eff-fit-p216} are compatible
with the background-only hypothesis for all the theoretical models.
%
%

\section{Detection efficiency and systematic uncertainties}\label{sec:eff}

The detection efficiency required in the calculation of the signal yield is estimated from the simulated signal events.
The efficiencies after preselection and after MVA selection are shown
in Table~\ref{tab:eff-fit-h} and \ref{tab:eff-fit-p216}, 
for the considered models of resonant and non-resonant LLP productions, respectively.
The values include the geometrical acceptance.
Several phenomena compete to determine the detection efficiency. 
In general the efficiency after preselection increases with the LLP mass because more particles are produced
in the decay of heavier LLPs. There is a loss of particles outside the spectrometer acceptance,
especially when the LLPs are produced from the decay of heavier states,
such as the Higgs-like particle.
In addition, the lower boost of heavier LLPs results in a shorter average flight length, 
which is disfavoured by the requirement of a minimum \RXY value.
With increasing LLP lifetimes  a larger portion of the decays falls into the material region and is vetoed.
Finally, a drop of sensitivity is expected for LLPs with a
lifetime close to the \bquark-hadron lifetimes,
where the contamination from \bbbar events becomes even more important, especially for low-mass LLPs.
The detection efficiency is reduced by up to one order of magnitude after the optimised MVA selection while the background is
reduced by 3--4 orders of magnitude.

A breakdown of the relative systematic uncertainties is shown in Table~\ref{tab:syst}.
The uncertainties of the partonic luminosity depend upon the process considered;
they are estimated following the procedure explained in Ref.~\cite{PDF4LHC,Butterworth:2015oua} and vary
from 3\% up to  6\%,  which is found for the gluon fusion process.
The integrated luminosity~\cite{LHCB-PAPER-2014-047} contributes with an uncertainty of 2\%.
The statistical precision of the efficiencies determined from simulation
is in the range 2--4\% for the different models.
Different sources of systematic uncertainty arising from discrepancies between data and simulation have
been considered. The size of those discrepancies for the relevant observables are inferred from a comparison of
the distributions obtained from data and from \bbbar simulated events, which describes the data quite completely,
or from other calibration processes. 

The muon detection efficiency, including trigger, tracking, and muon identification efficiencies,
is studied by a tag-and-probe technique applied to muons
from \mbox{$J/\psi \rightarrow \mumu$},  \mbox{$\Upsilonres(1S) \rightarrow \mumu$}  and  $\Z \rightarrow \mumu$ decays.
The corresponding systematic effects    
due to differences between data and simulation are estimated to be between 2\% and 3.7\%,
depending on the theoretical model considered.

A comparison of the simulated and observed \pt distributions of muons from \mbox{$\Z \rightarrow \mumu$} decays
shows a maximum difference of 0.2\gevc in the selected region;
this difference is propagated to the LLP analysis by shifting the muon \pt threshold by the same amount.
The corresponding systematic uncertainty is below 1\% for all models under consideration.

The muon impact-parameter distribution is also studied from $\Z$ decays and shows a discrepancy between
data and simulation of about 10\mum close to the \PTMU threshold.
By changing the minimum \IPMU requirement by this amount,
the change in the detection efficiency is below 1\% for all the models.

The vertex reconstruction efficiency has a complicated spatial structure due to the geometry of the VELO and the material veto.
Uncertainties in the estimated vertex-finding efficiency are due to the per-track efficiency,
track resolution, and differences in the contribution from background tracks due to the underlying interaction and pile-up.
In the material-free region, $\RXY < 4.5$ mm, the efficiency as a function of the flight distance has been studied
in the context of lifetime measurements~\cite{LHCb-PAPER-2013-065}, showing that the simulation reproduces the
data within 1\%. 
In the region $\RXY > 4.5$ mm a deviation of less than 6\% is inferred from the study of
inclusive \bbbar events in data and simulation.
By altering the efficiency in the simulation program as a function of the true vertex position, the effect
on the LLP detection efficiency is estimated to be 1--2\%.
A second method to determine this contribution uses vertices from
$B^0 \rightarrow \jpsi \Kstarz$ decays with $\decay{\jpsi}{\mumu}$ and $\Kstarz \rightarrow K^+ \pi^-$.
For this process the vertex detection efficiencies in data and simulation agree within 10\%.
This result, obtained from a process with four final-state particles,
is propagated to the LLP decay into a larger number of charged particle tracks and a detection threshold of three tracks.
A discrepancy of at most 2\% between the LLP efficiency in data and simulation is found, which is adopted as a contribution to
the systematic detection uncertainty.

The uncertainty on the position of the beam line in the transverse plane is less than 20\mum~\cite{LHCb-DP-2014-001}.
It can affect the secondary-vertex selection, mainly via the requirement on \RXY.
By altering the PV position in simulated signal events, the effect is estimated to be below 1\%.

The effect of the imperfect modelling on the observables used in the MVA training is estimated with pseudoexperiments.
As previously stated, the bias on each input variable is determined by comparing simulated and experimental
distributions of muons and LLP candidates from \Z and \W events, as well as from \bbbar events.
At the MVA test stage, each input variable is modified by a scale factor randomly selected from a Gaussian
distribution of width equal to the corresponding bias.
The standard deviation of the signal efficiency distribution is taken as a systematic uncertainty.

The signal and background samples are obtained through a selection on the muon isolation parameter.
By a comparison of data and muons from simulated \bbbar events,
the maximum uncertainty on this variable is estimated to be $\pm 0.015$ in the proximity of the thresholds,
with a maximal effect on the efficiency of 1.7\%.

Comparing the mass distributions of \bbbar and $Z\to\bbbar$ events, a maximum mass-scale
discrepancy between data and simulated events of 10\% is estimated in the proximity of the threshold,
which translates into a 1.4\% contribution to the detection efficiency uncertainty.

 \begin{table}[ht]
\begin{center}
\caption{\small
  Contributions to the relative systematic uncertainties.
  The indicated ranges cover the theoretical models considered.
  The contributions from the signal and background models used in the LLP mass fit are
  treated separately.
    }
\label{tab:syst}
\scalebox{0.9}{
\begin{tabular}{lc}
  \hline
Source & Contribution [$\%$] \\ 
\hline
Integrated luminosity       &    2.0       \\
Parton luminosity  gluons fusion (quarks)  &   6.0 (3.0)    \\
Simulation statistics            &    2.0--4.0     \\
Muon reconstruction         &    2.0--3.7     \\
\PTMU                       &    1.0        \\
\IPMU                       &    1.0        \\  
Vertex reconstruction       &    2.0      \\  
Beam line uncertainty (\RXY)&    0.9      \\  
Muon isolation              &    1.7      \\
MVA                         &    1.7-16   \\
Mass calibration            &    1.4      \\
\hline
Total                       &   7.3--18.9   \\
\hline
\end{tabular}
} 
\end{center}
 \end{table}

Finally, the total systematic uncertainty is obtained as the sum in quadrature of all contributions,
where the different components of the detection efficiency are assumed to be fully correlated.
 
The choice of the signal and background invariant-mass templates can affect the results of the LLP mass fits.
The uncertainty due to the signal model accounts for the mass scale and the mass resolution.
The mass scale and resolution discrepancies between data and simulation are
below 1\% and 1.5\% respectively, as obtained from \bbbar and $Z \rightarrow \bbbar$ events. 
Pseudoexperiments are used to estimate the effect on the cross-section calculation.
For each theoretical model, ten simulated signal events are added to the selected data
after a Gaussian smearing or after changing the mass scale.
The average deviation of the observed upper limits with respect to the one obtained from the
default signal and background distributions is below 2\%.

The background shape is deduced from data selected in the poorly isolated region after reweighting,
with weights inferred from the data distributions obtained with relaxed selection criteria.
The overall uncertainty is estimated by reducing by half the weights and running pseudoexperiments as before.
The average deviation of the observed upper limits is below 14\%.

\section{Results}
  
The 95\% confidence level (CL) upper limits, expected and observed, on the production cross-sections
times branching fraction are computed for each model using the CLs approach~\cite{art:cls}.
Statistical and systematic uncertainties on the signal efficiencies are included as
nuisance parameters of the likelihood function, assuming Gaussian distributions.
Finally, the upper limit values are corrected by the factors which account for the
imperfect modelling of signal and background templates.

\begin{figure}[ht]
\vspace{-6 mm}
\centering
\mbox{

\centering
 {\includegraphics[width=0.5\textwidth]{ 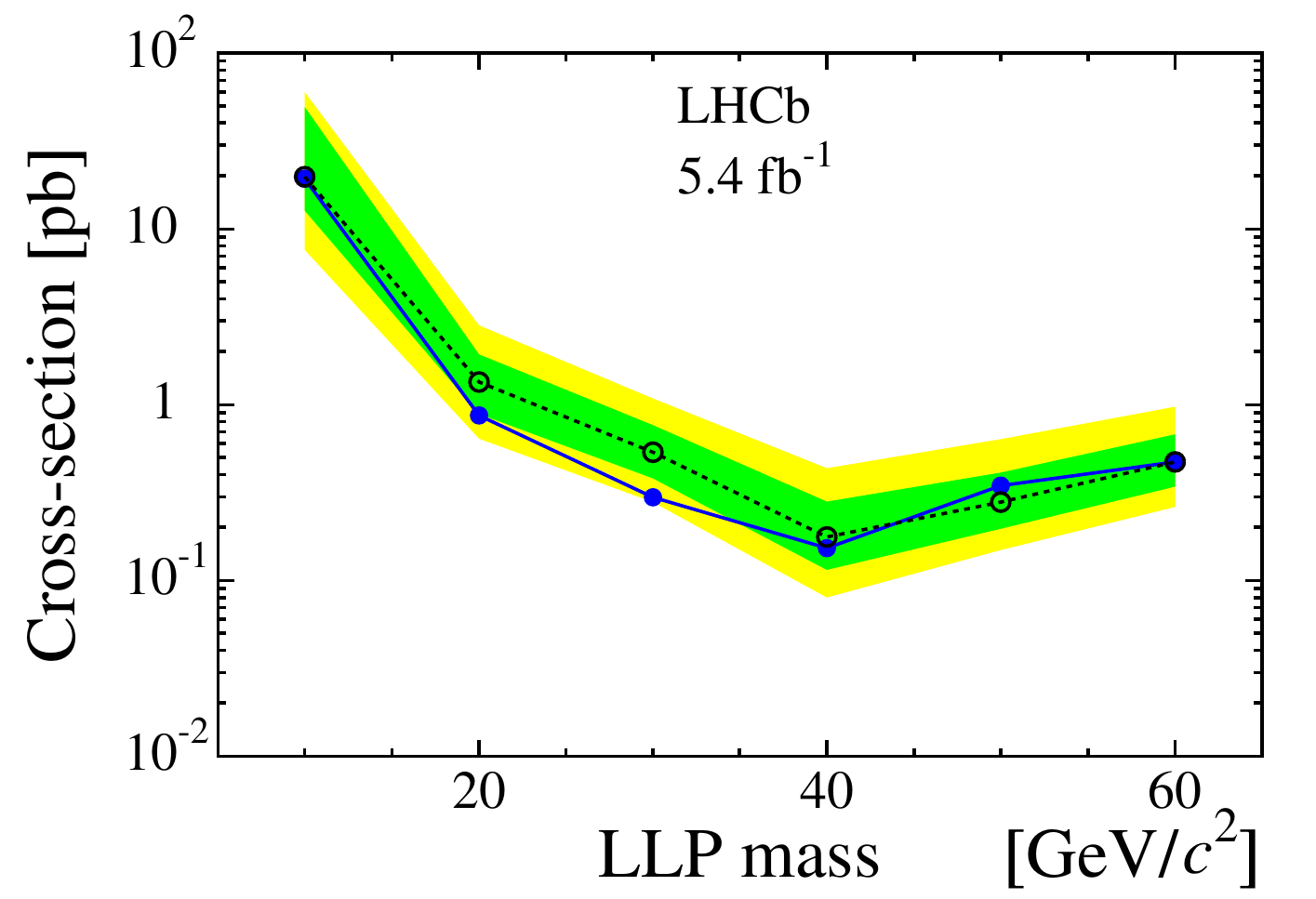}}
 {\includegraphics[width=0.5\textwidth]{ 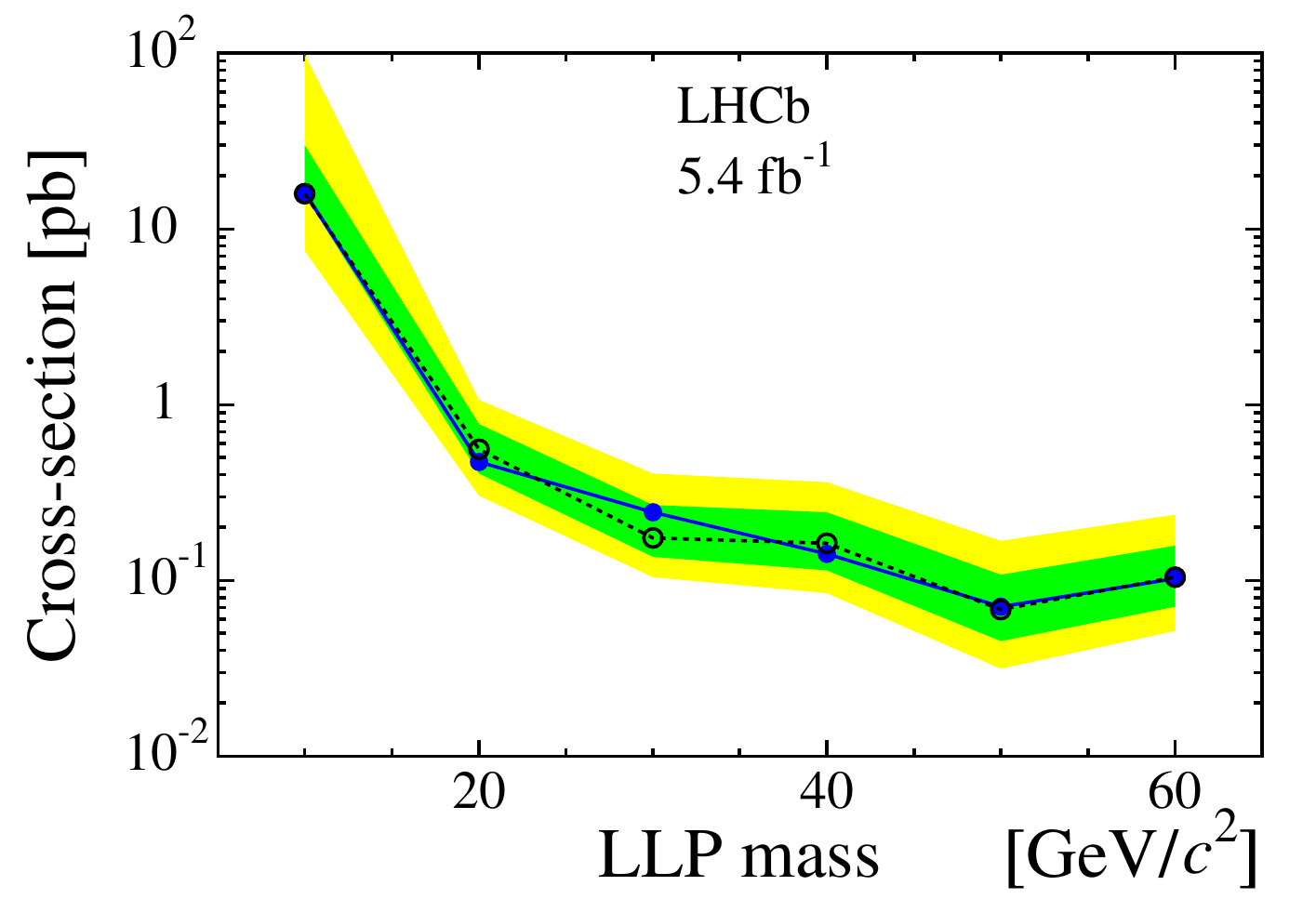}}
 \put(-270,120){(a)} \put(-35,120){(b)}
 \put(-300,100){\tauchi=5\ps} \put(-70,100){\tauchi=10\ps}
}
\mbox{\vspace{4 mm}}
\mbox{
  \centering
   {\includegraphics[width=0.5\textwidth]{ 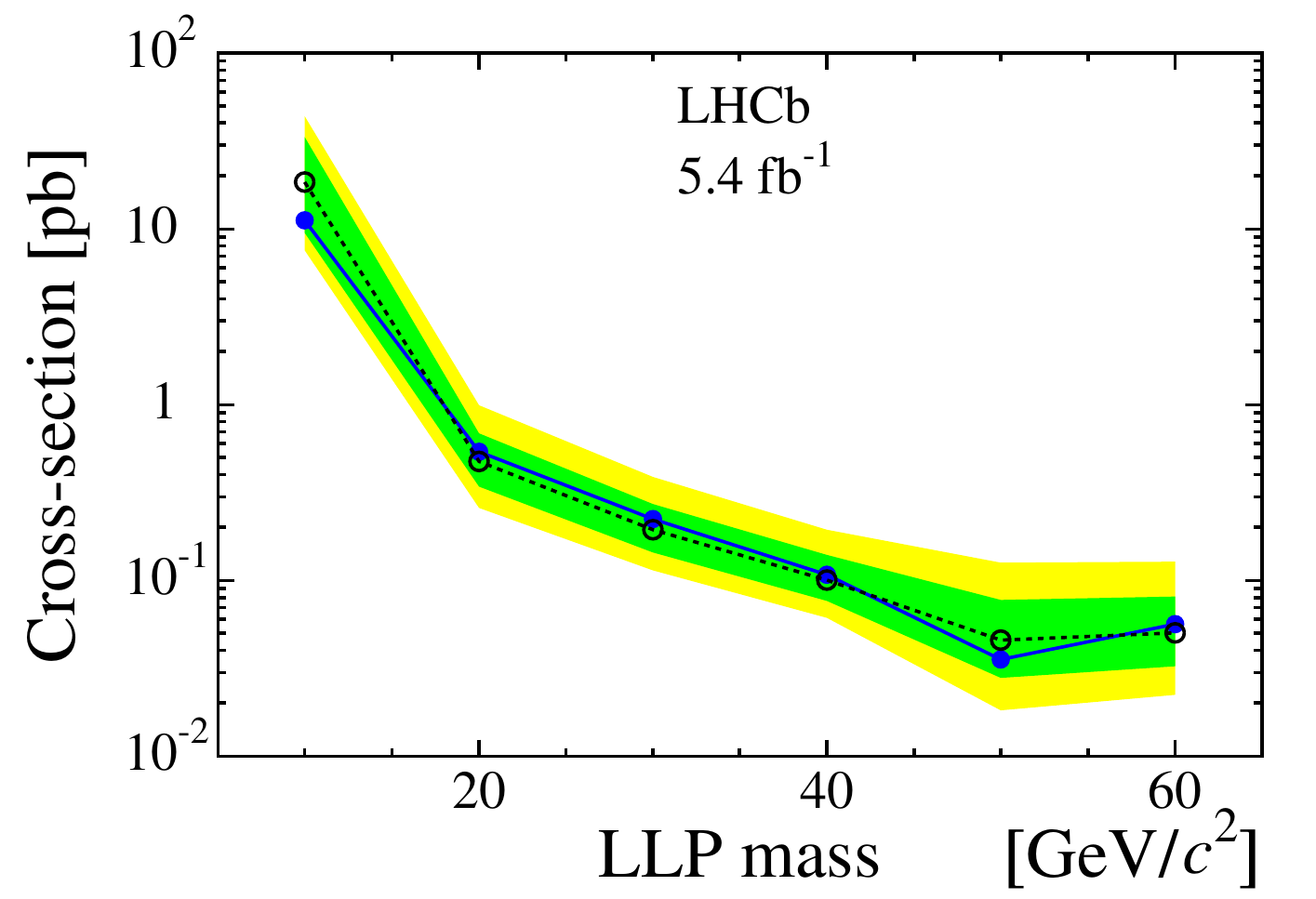}}
   {\includegraphics[width=0.5\textwidth]{ 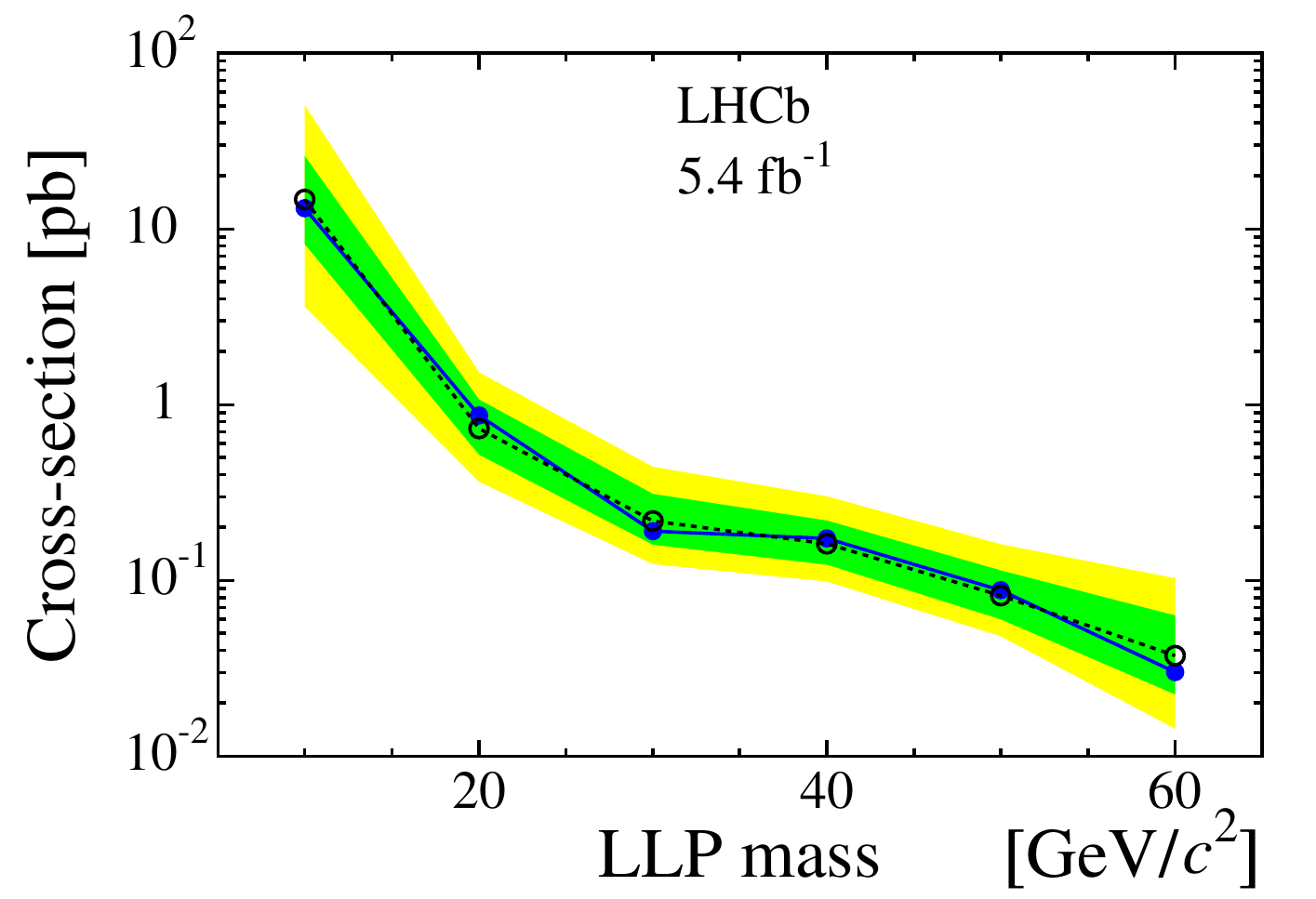}}
   \put(-270,120){(c)} \put(-35,120){(d)}
   \put(-300,100){\tauchi=20\ps} \put(-70,100){\tauchi=30\ps}
  }
\mbox{\vspace{4 mm}}
\mbox{
\vspace{0 mm}
  \centering
   {\includegraphics[width=0.5\textwidth]{ 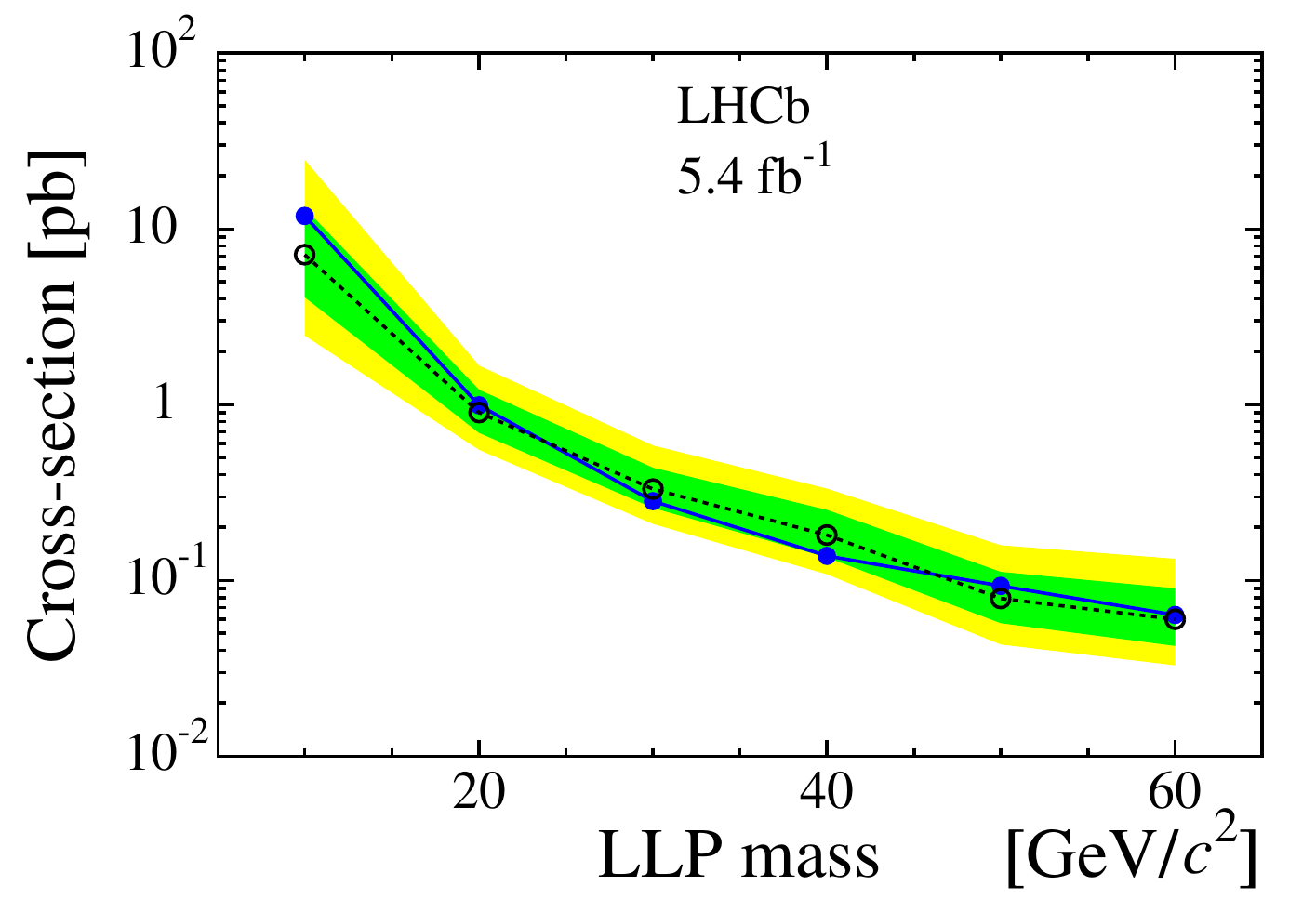}}
   {\includegraphics[width=0.5\textwidth]{ 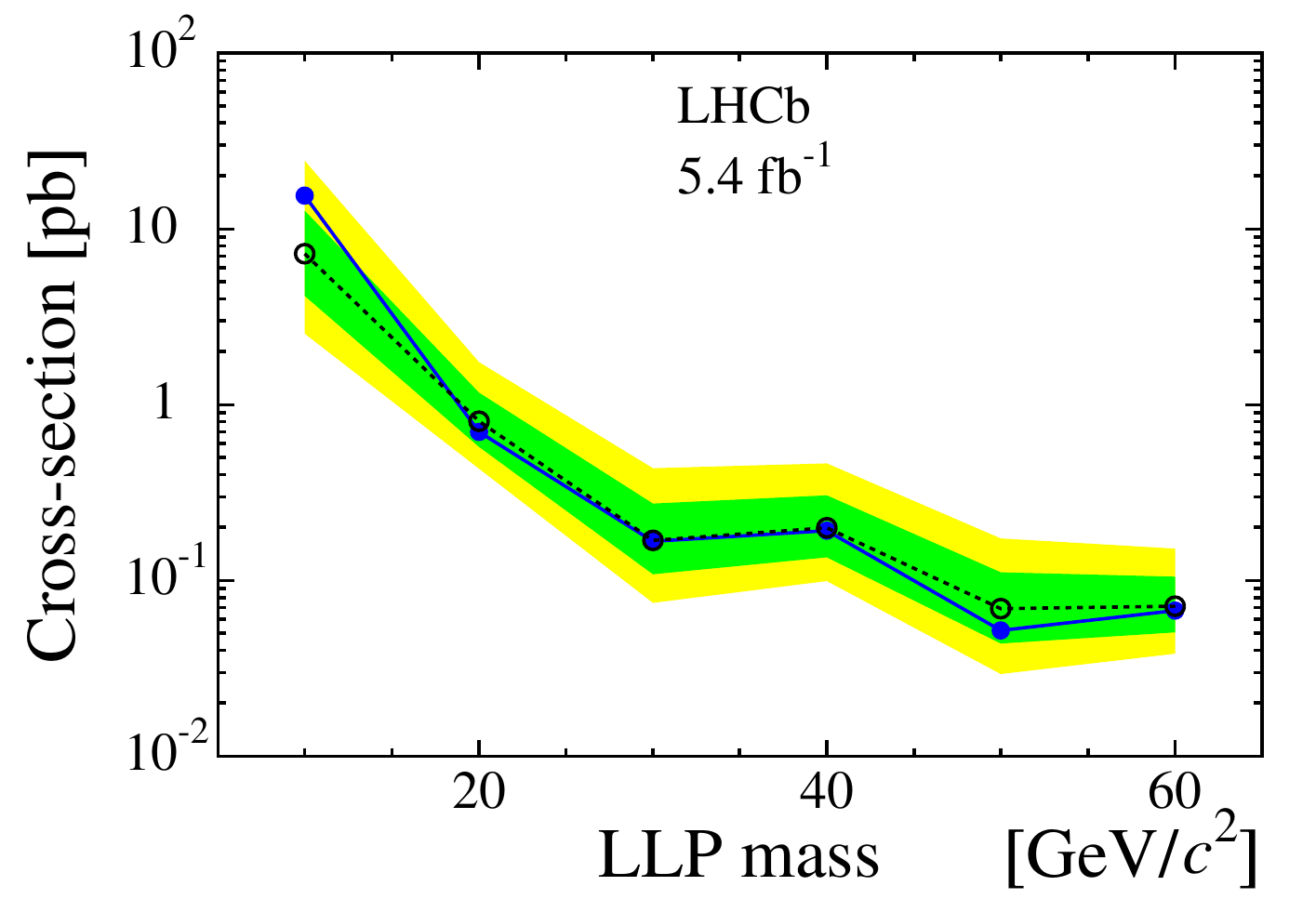}}
   \put(-270,120){(e)} \put(-35,120){(f)}
   \put(-300,100){\tauchi=50\ps} \put(-75,100){\tauchi=100\ps}
}

\vspace{4 mm}
  \centering
      {\includegraphics[width=0.5\textwidth]{ 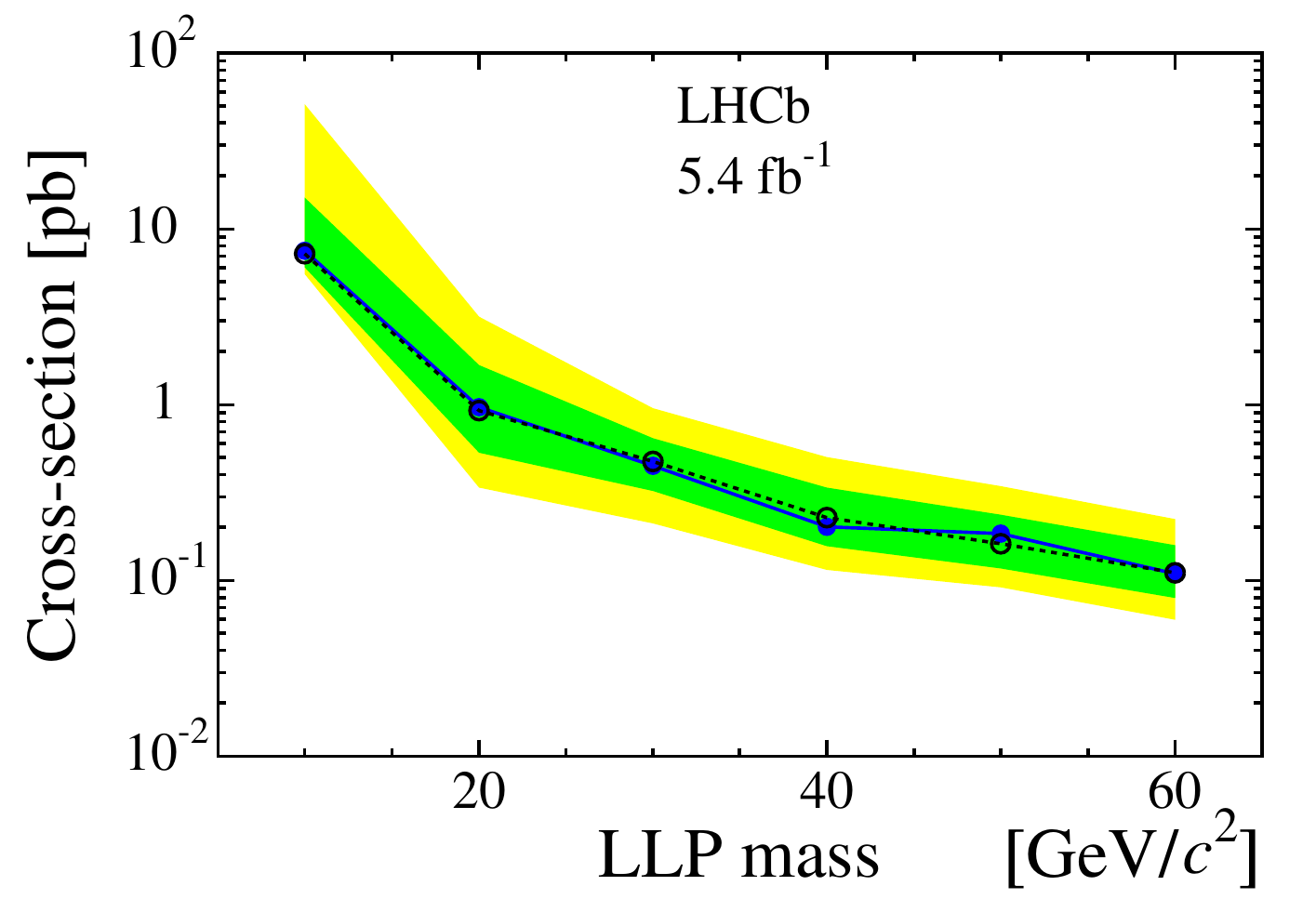}}
      \put(-40,120){(g)}
      \put(-75,100){\tauchi=200\ps} 
\caption{
  \small
  Expected (open dots and 1$\sigma$ and 2$\sigma$ bands) and observed (full dots)
  cross-section times branching fraction upper limits (95\% CL)
  as a function of m$_{\khi}$ for
  the resonant production processes with $\mhzero=125\gevcc$, and, from (a) to (g),
  $\tau_{\khi}$ of 5, 10, 20, 30, 50, 100, and 200\ps.
}
\label{fig:ul-1}
\end{figure}

\begin{figure}[ht]
\centering
\mbox{
\centering
 {\includegraphics[width=0.5\textwidth]{ 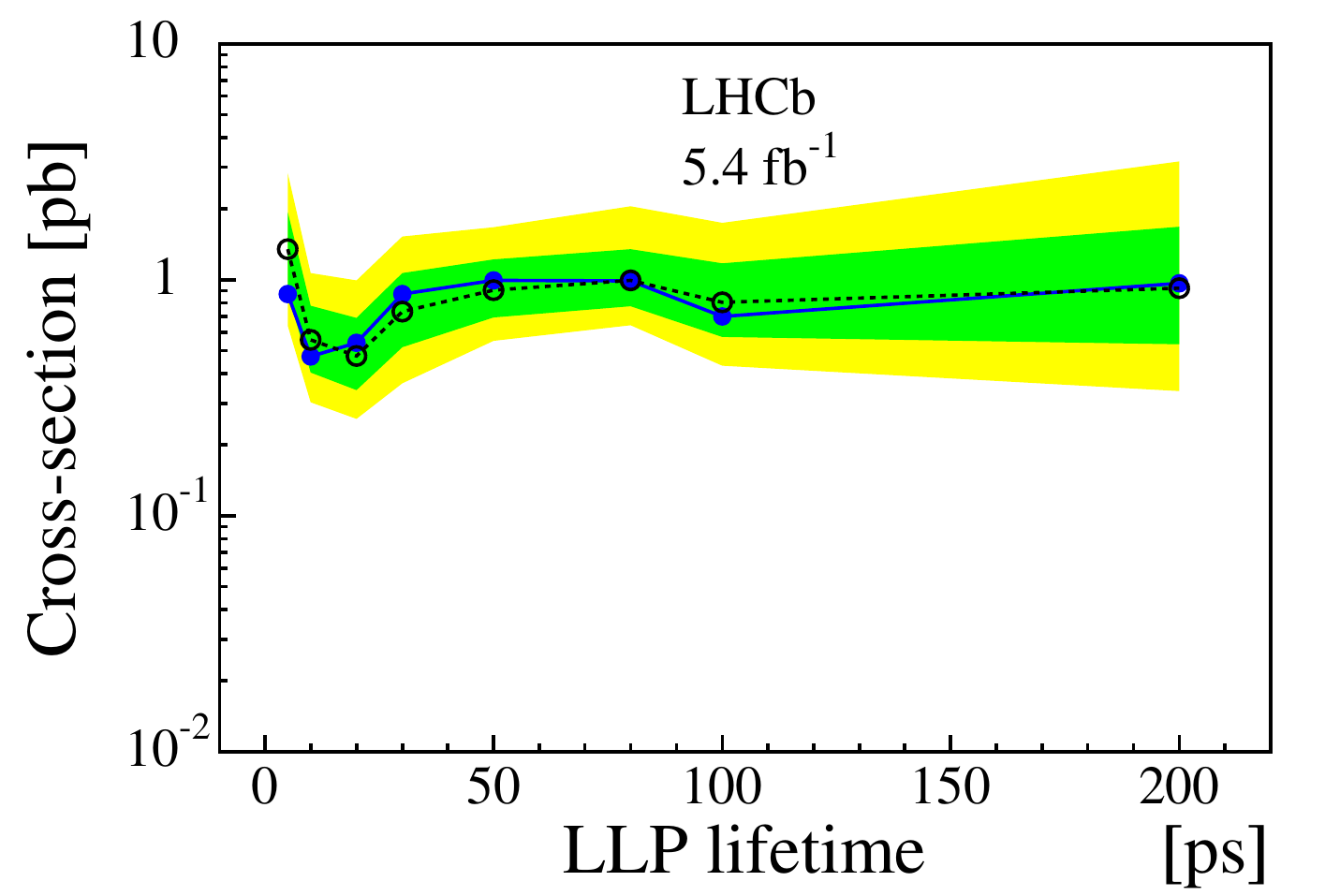}}
 {\includegraphics[width=0.5\textwidth]{ 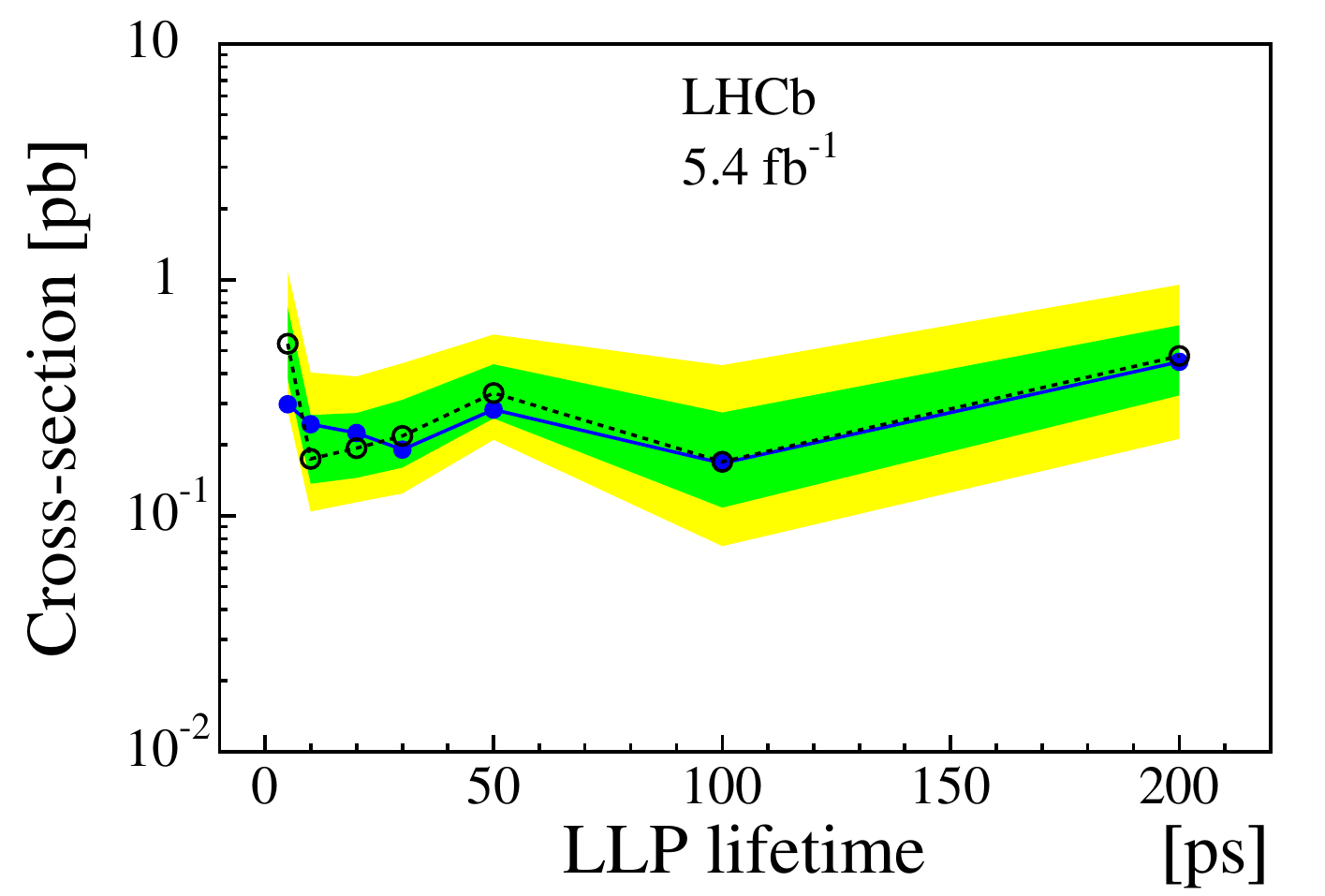}}
 \put(-270,120){(a)} \put(-35,120){(b)}
 \put(-370,50){ $m_{\khi}$=20\gevcc} \put(-135,50){ $m_{\khi}$=30\gevcc}
}
\mbox{\vspace{4 mm}}
\mbox{
  \centering
   {\includegraphics[width=0.5\textwidth]{ 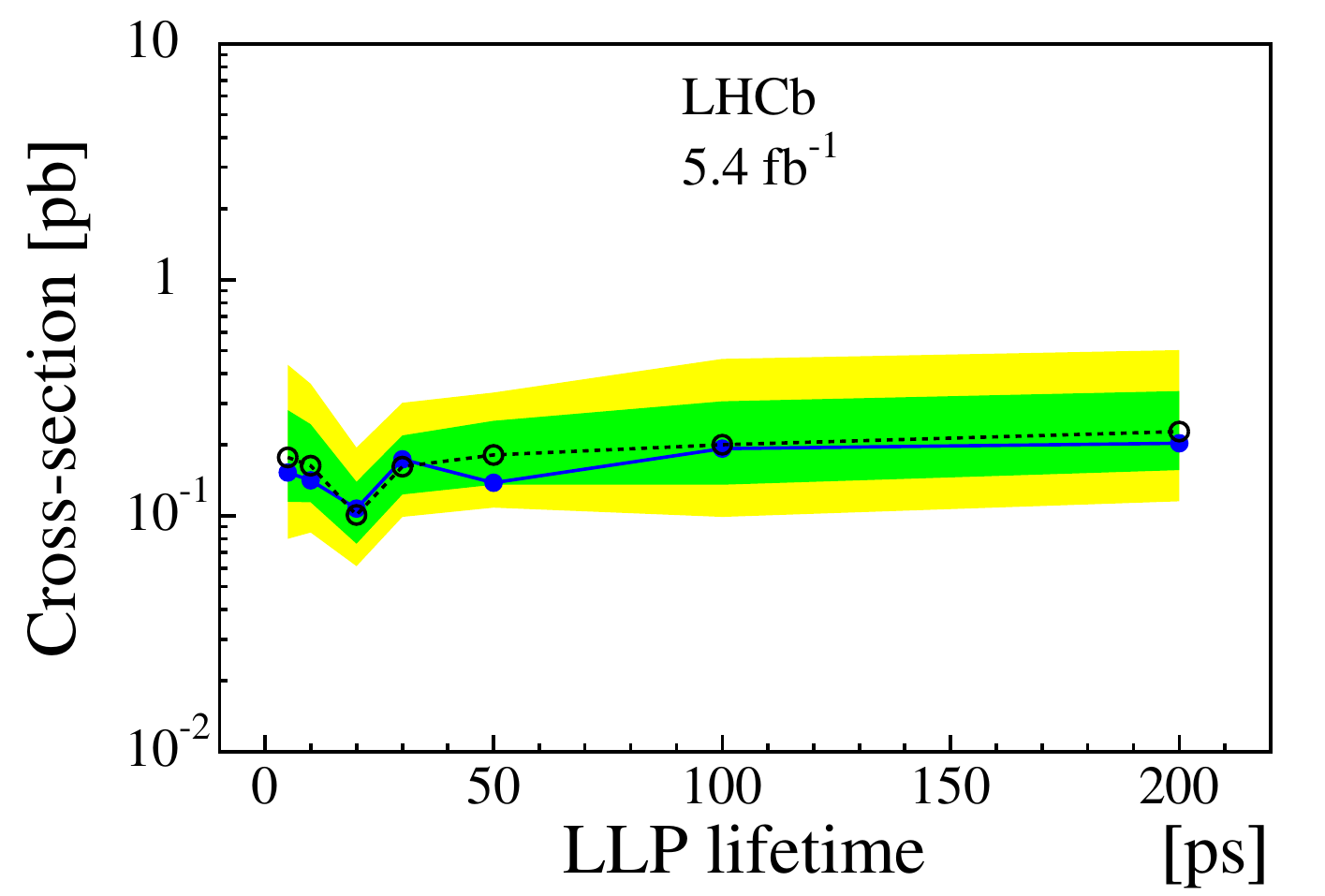}}
   {\includegraphics[width=0.5\textwidth]{ 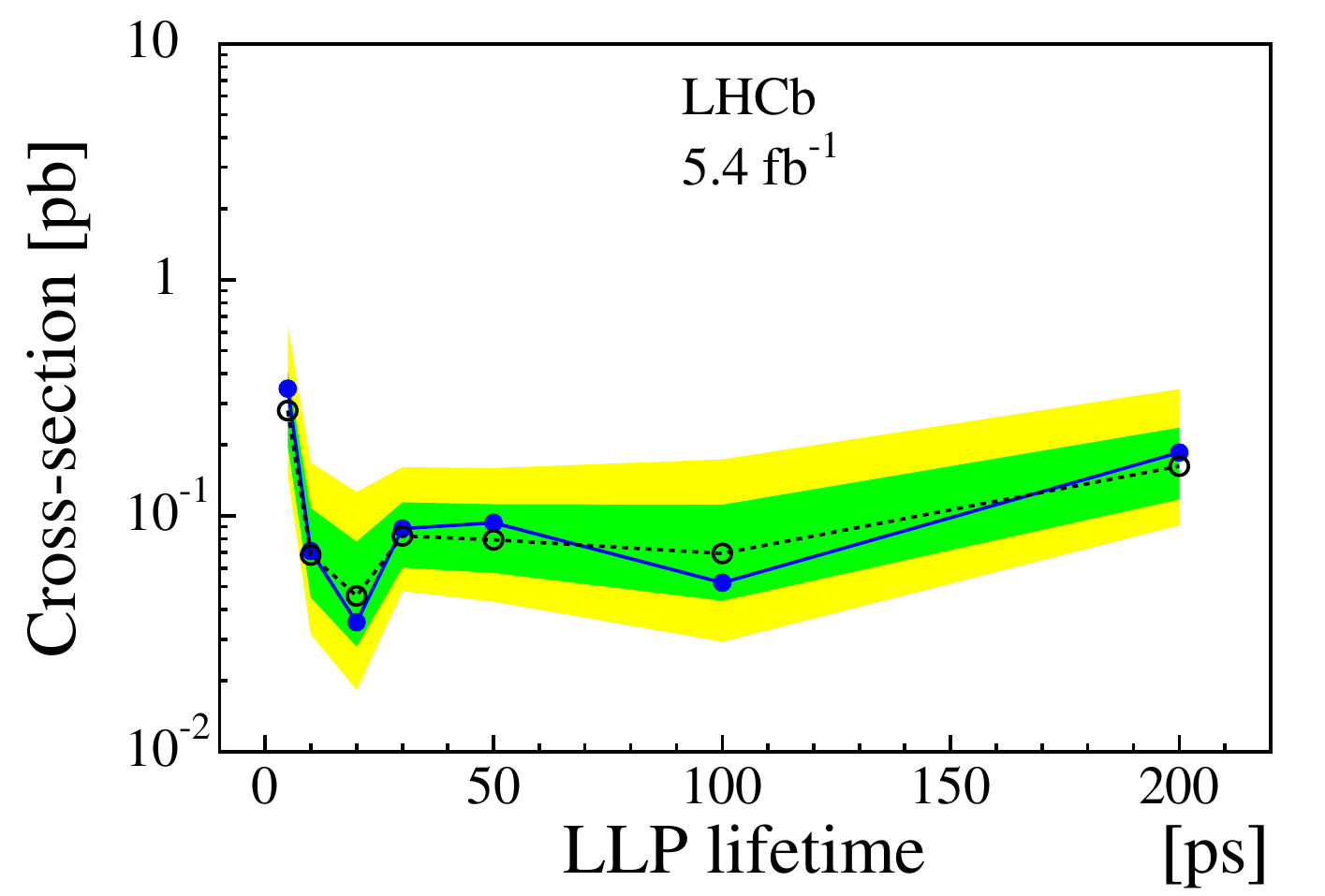}}
   \put(-270,120){(c)} \put(-35,120){(d)}
   \put(-370,105){ $m_{\khi}$=40\gevcc} \put(-135,105){ $m_{\khi}$=50\gevcc}
}
  
\vspace{5 mm}
\mbox{
  \centering
      {\includegraphics[width=0.5\textwidth]{ 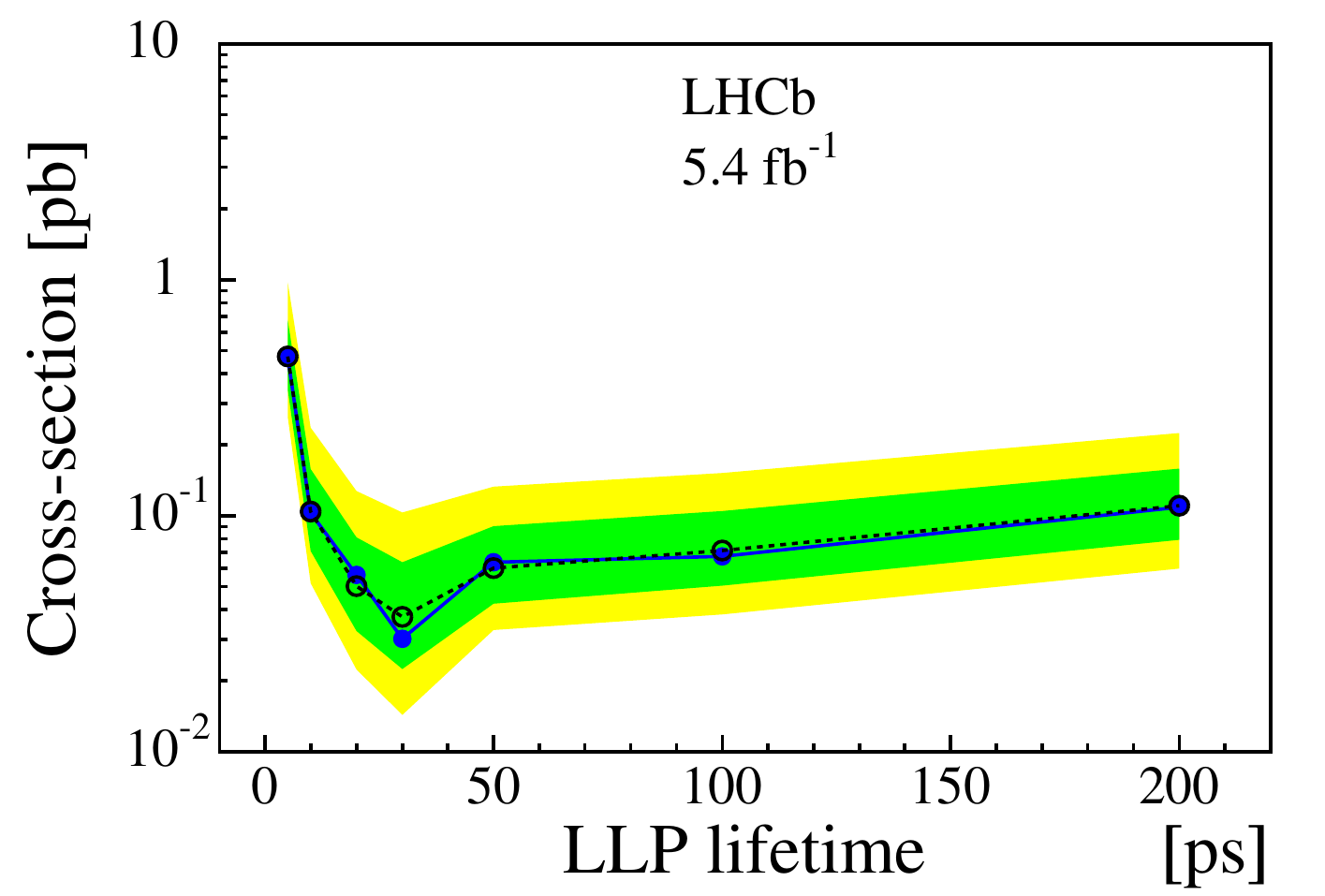}}
      \put(-40,120){(e)}
      \put(-140,105){ $m_{\khi}$=60\gevcc}
}
\caption{
  \small
  Expected (open dots and 1$\sigma$ and 2$\sigma$ bands) and observed (full dots)
  cross-section times branching fraction upper limits (95\% CL)
  as a function of $\tau_{\khi}$ for the resonant production with $\mhzero=125\gevcc$,
  and, from (a) to (e), $m_{\khi}$ of 20, 30, 40, 50, and 60\gevcc.
}
\label{fig:ul-2}
\end{figure}

\begin{figure}[ht]
\centering
\mbox{
\centering
 {\includegraphics[width=0.5\textwidth]{ 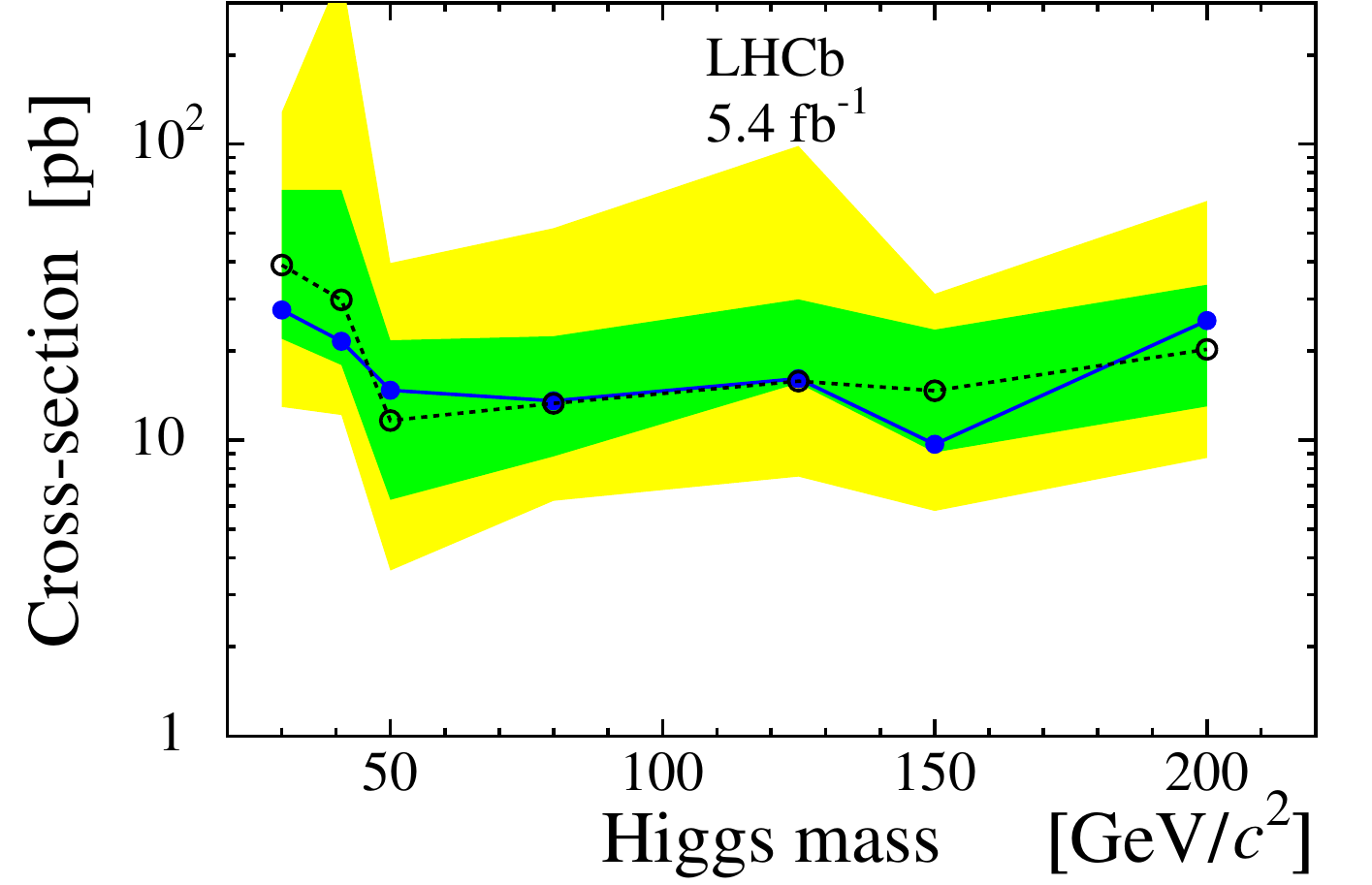}}
 {\includegraphics[width=0.5\textwidth]{ 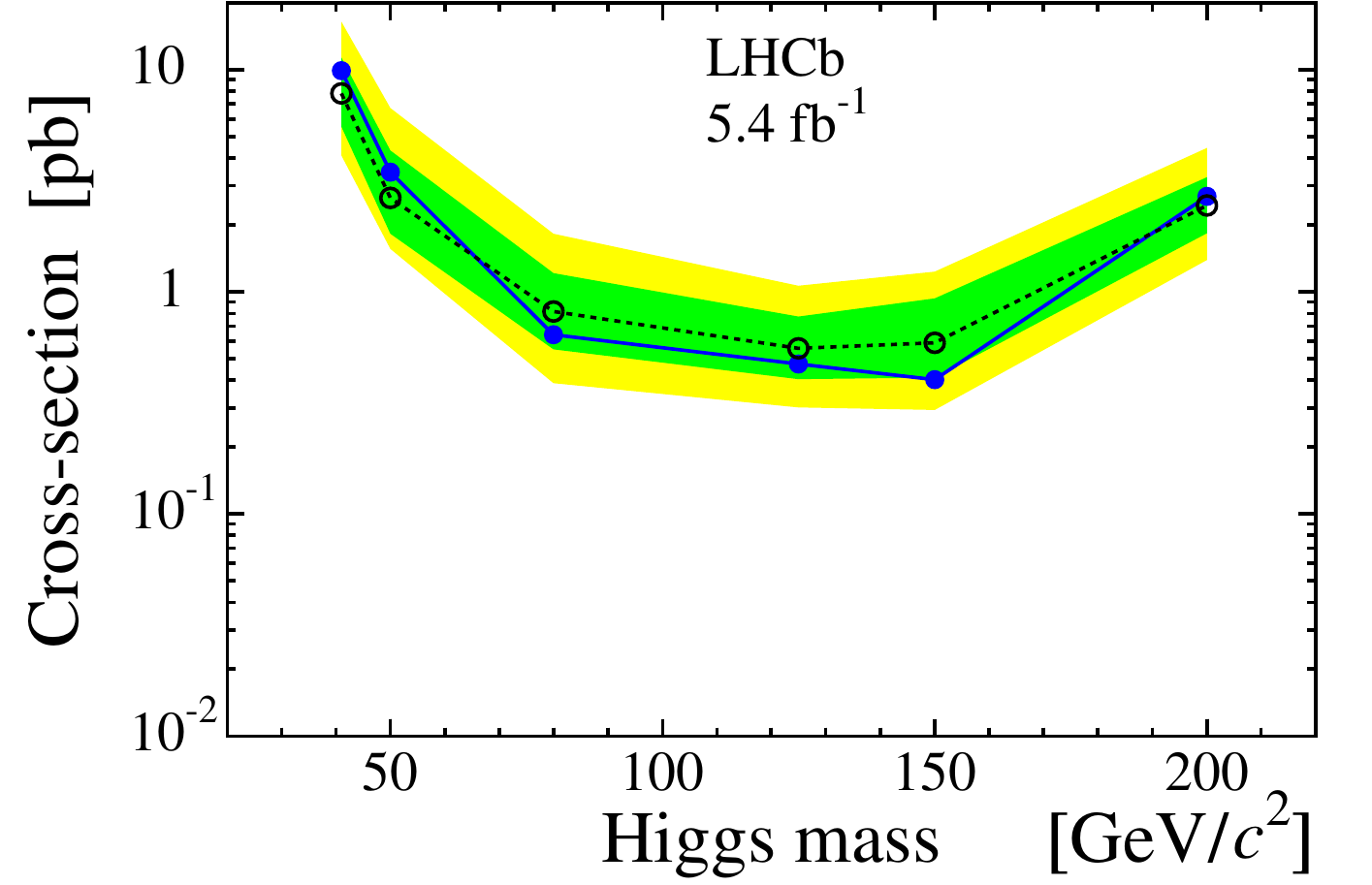}}
 \put(-270,120){(a)} \put(-35,120){(b)} 
 \put(-370,50){ $m_{\khi}$=10\gevcc} \put(-135,50){ $m_{\khi}$=20\gevcc}
}
\mbox{\vspace{4 mm}}
\mbox{
  \centering
   {\includegraphics[width=0.5\textwidth]{ 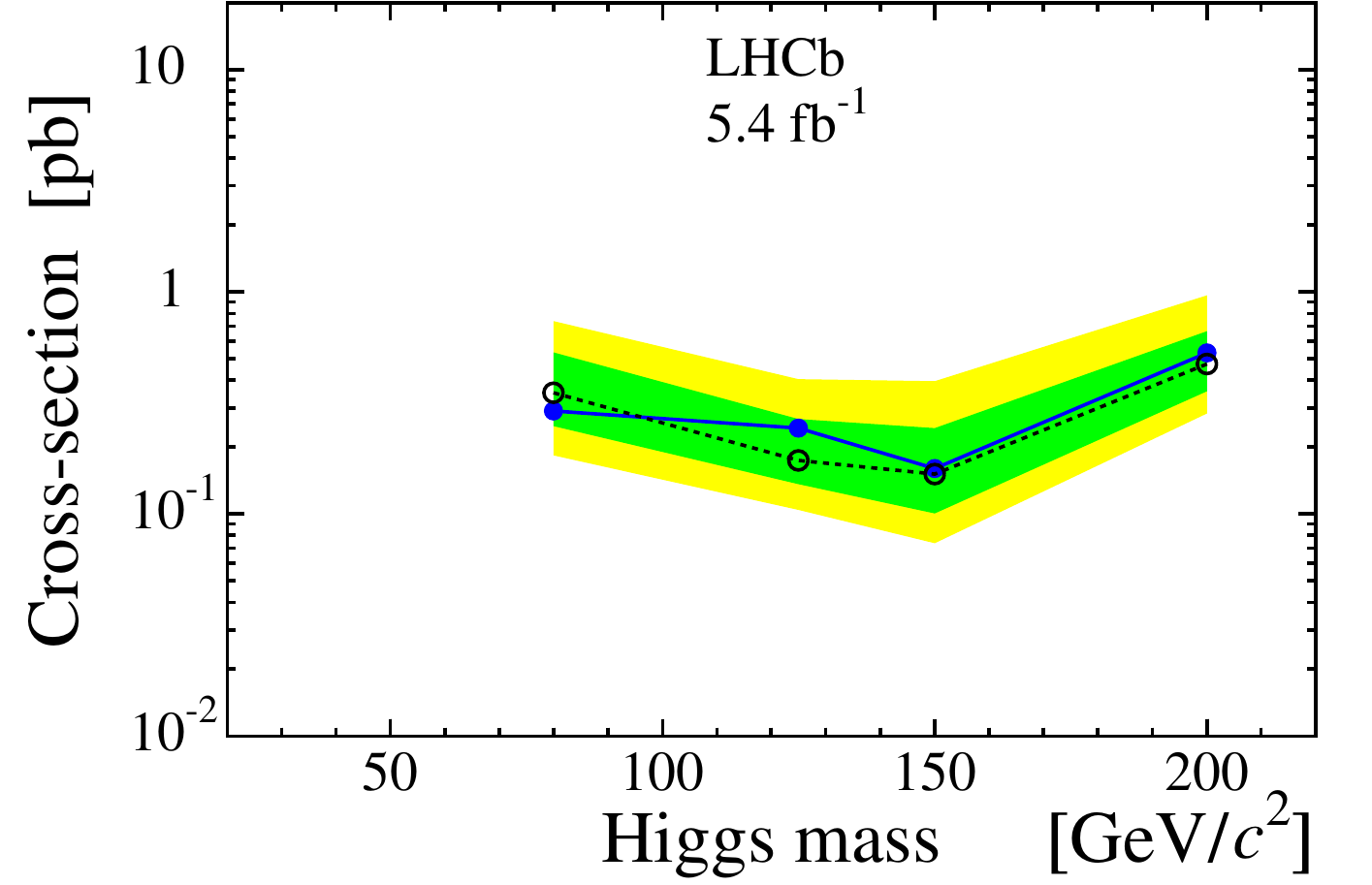}}
   {\includegraphics[width=0.5\textwidth]{ 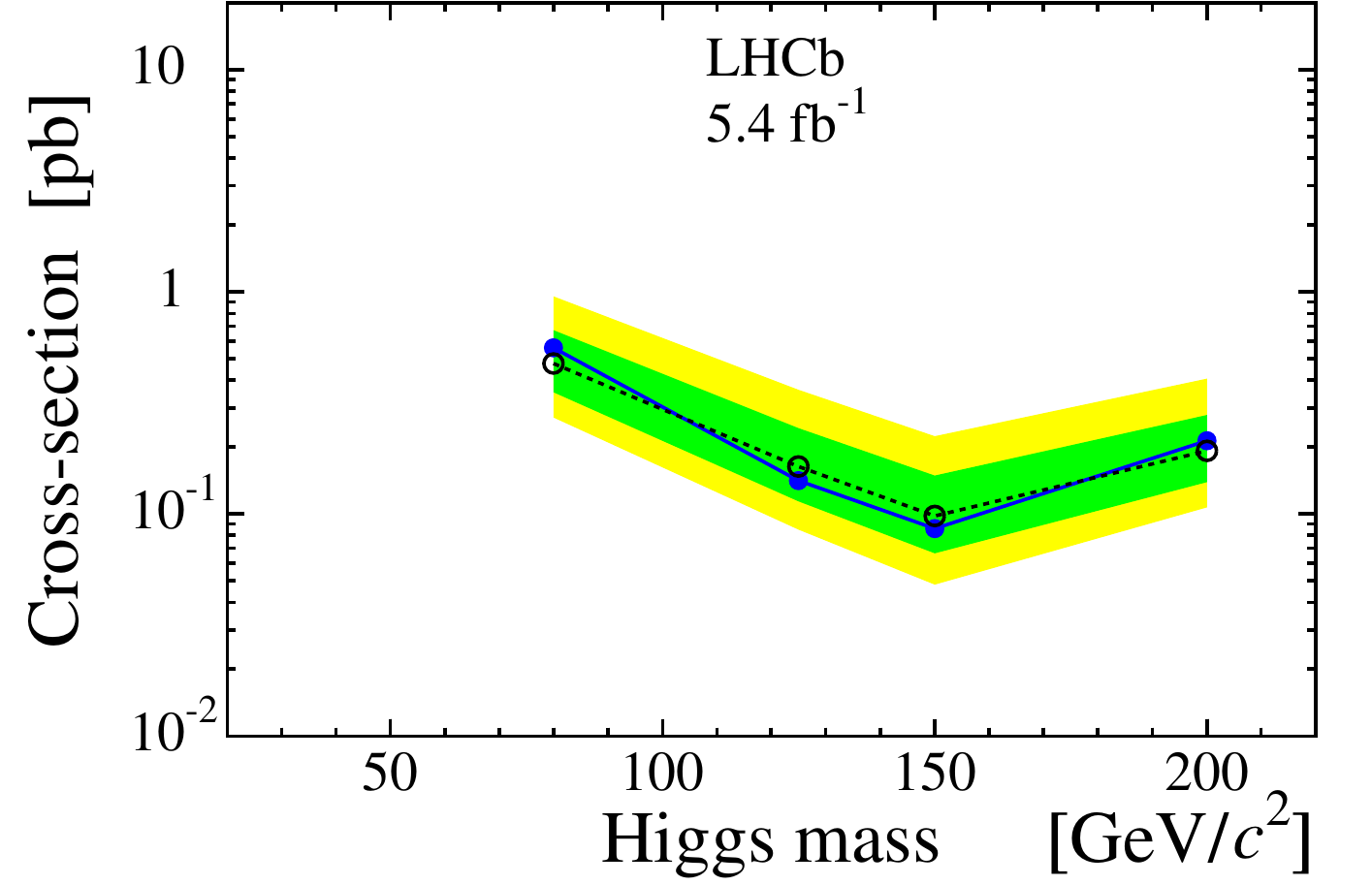}}
   \put(-270,120){(c)} \put(-35,120){(d)}
    \put(-370,105){ $m_{\khi}$=30\gevcc} \put(-135,105){ $m_{\khi}$=40\gevcc}
  }
  \mbox{\vspace{4 mm}}
\mbox{
  \centering
   {\includegraphics[width=0.5\textwidth]{ 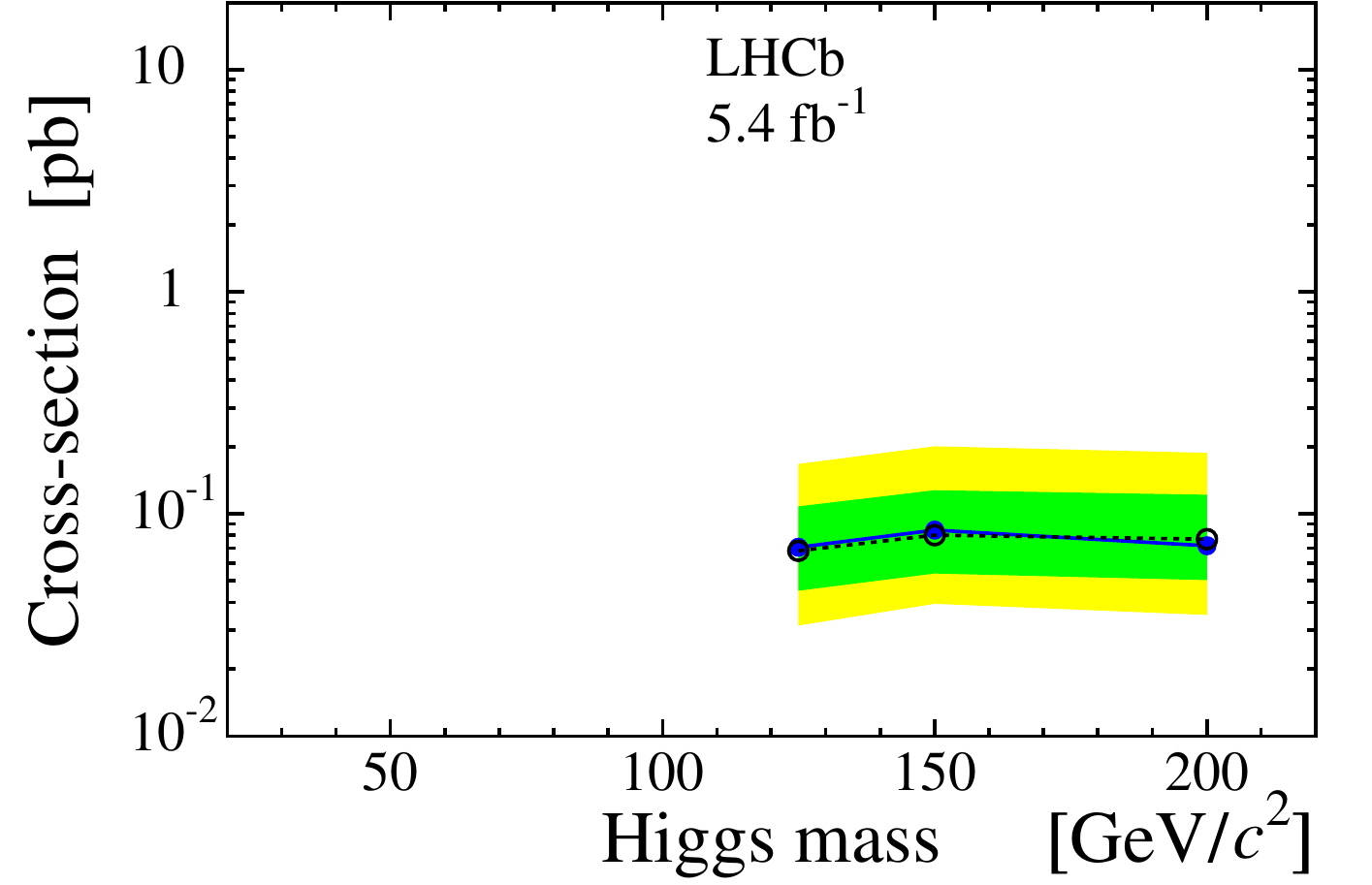}}
   {\includegraphics[width=0.5\textwidth]{ 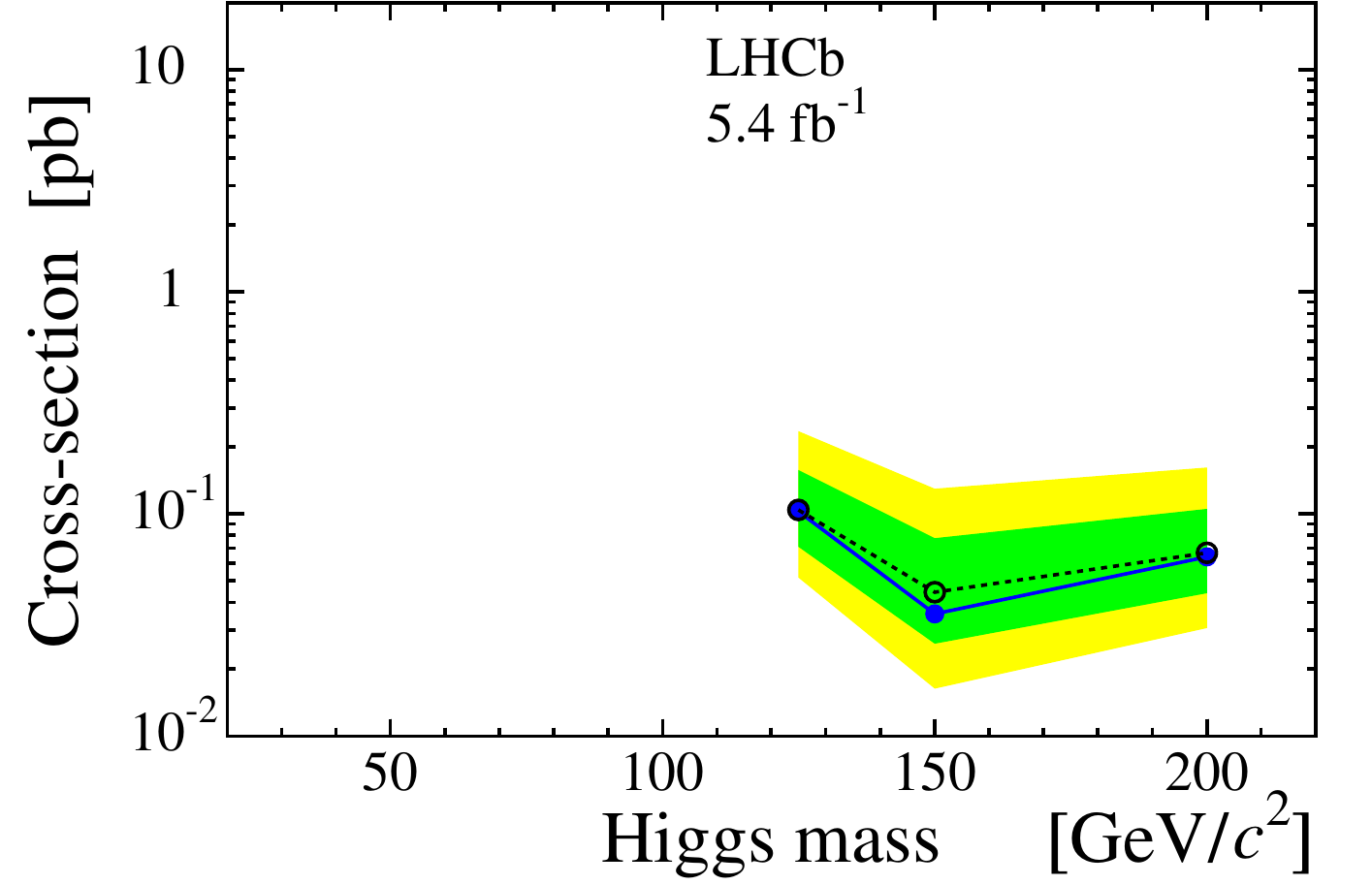}}
   \put(-270,120){(e)} \put(-35,120){(f)}
    \put(-370,105){ $m_{\khi}$=50\gevcc} \put(-135,105){ $m_{\khi}$=60\gevcc}
}
\caption{
  \small
  Expected (open dots and 1$\sigma$ and 2$\sigma$ bands) and observed (full dots)
  cross-section times branching fraction upper limits (95\% CL)
  as a function of  \mhzero and, from from (a) to (f),
  $m_{\khi}$ of 10, 20, 30, 40, 50, and 60\gevcc.
}
\label{fig:ul-3}
\end{figure}

\begin{figure}[ht]
\centering
\mbox{
\centering
 {\includegraphics[width=0.5\textwidth]{ 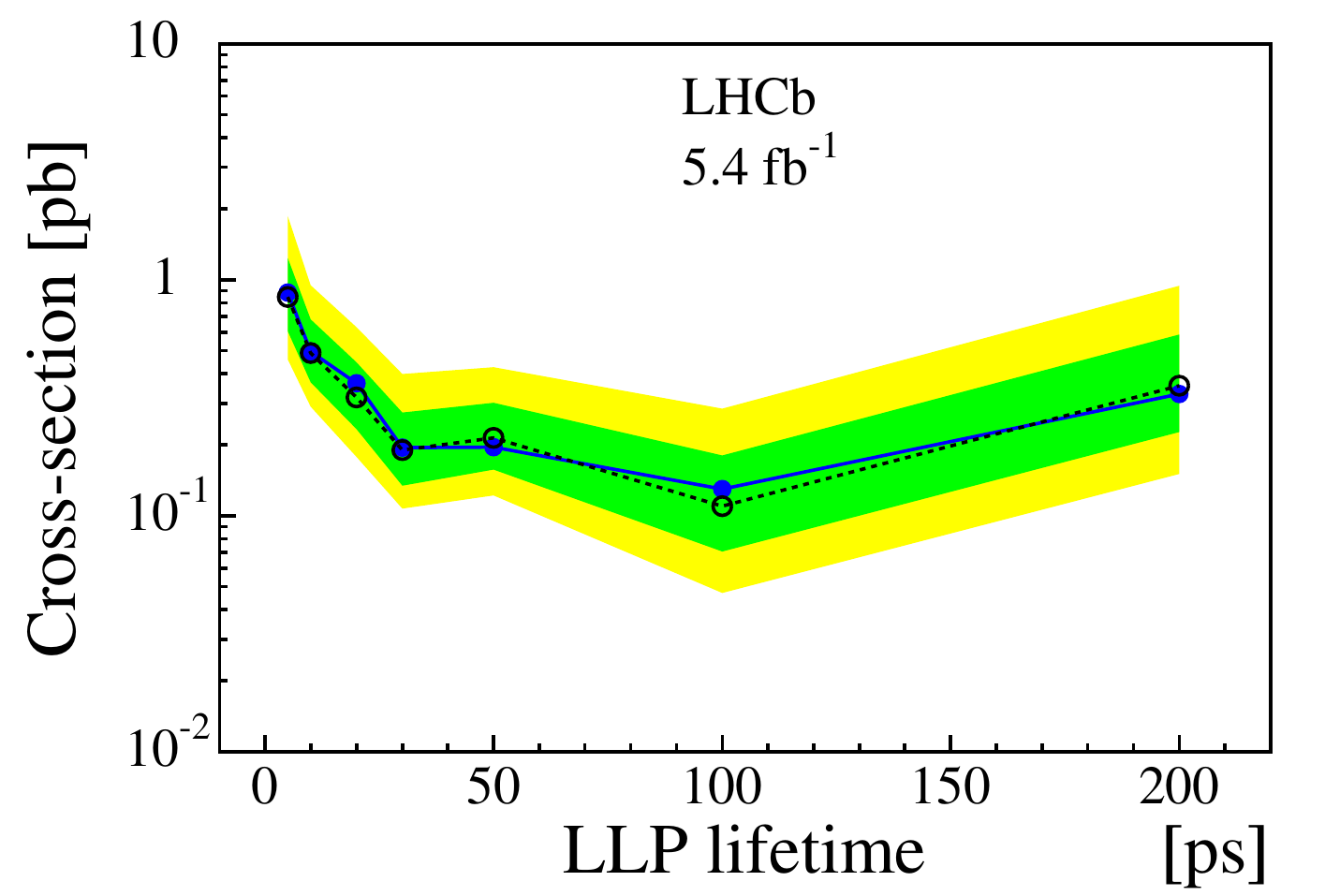}}
 {\includegraphics[width=0.5\textwidth]{ 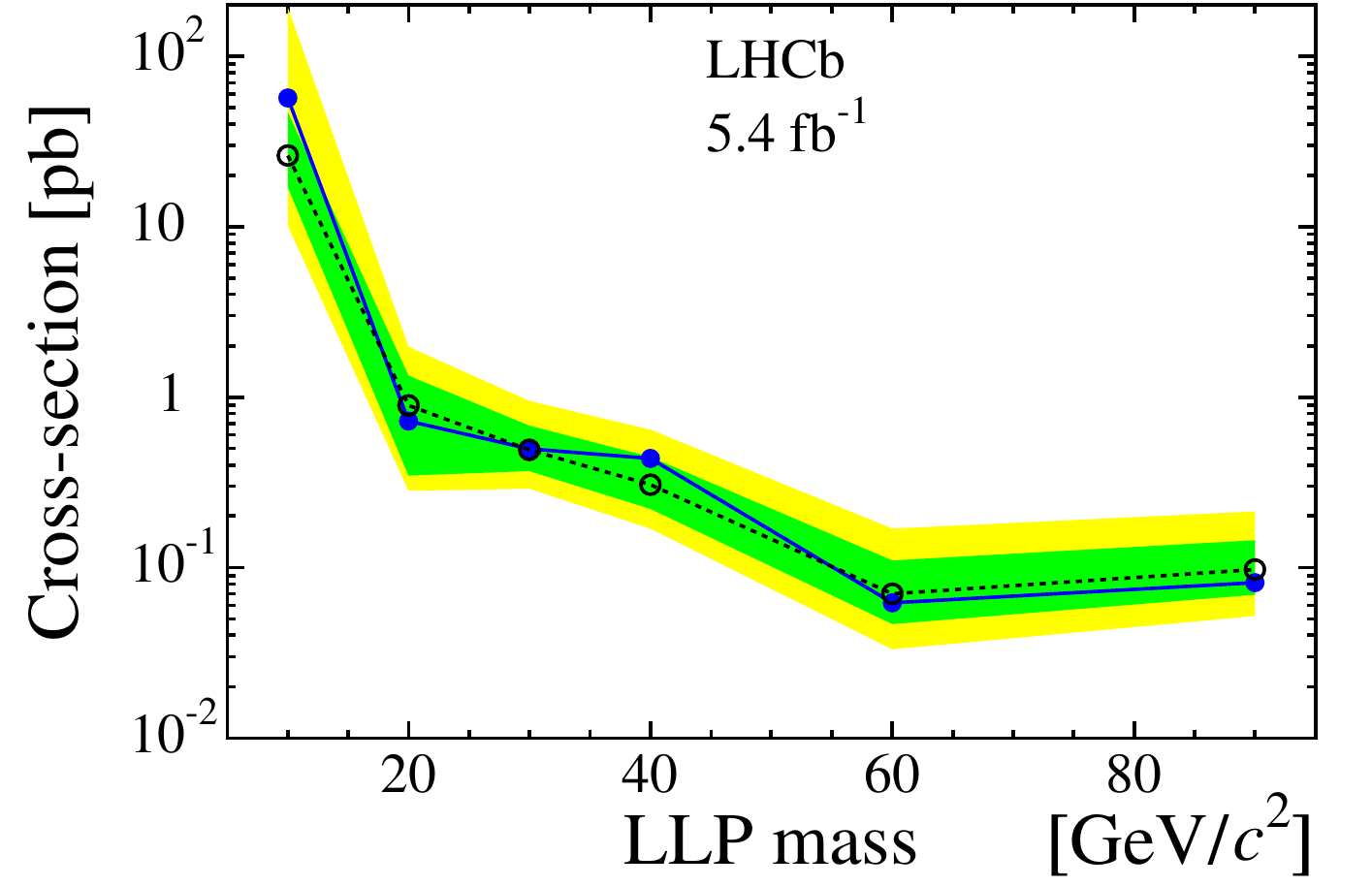}}
 \put(-270,120){(a)} \put(-35,120){(b)} 
  \put(-365,110){ $m_{\khi}$=30\gevcc} \put(-135,110){ \tauchi=10\ps}
}
\caption{
  \small
  Expected (open dots and 1$\sigma$ and 2$\sigma$ bands) and observed (full dots)
  cross-section times branching fraction upper limits (95\% CL), (a):
  as a function of  $\tau_{\khi}$  with $m_{\khi}=30\gevcc$,
  (b): as a function of $m_{\khi}$ with $\tau_{\khi}=10\ps$.
  The processes are from direct, non-resonant, LLP production.
}
\label{fig:ul-nonres}
\end{figure}

The numerical results for all the models are given in Tables~\ref{tab:ul-hX} and~\ref{tab:ul-p216}.
The figures from~\ref{fig:ul-1} to \ref{fig:ul-nonres} show the measured cross-section  times branching ratio
upper limits, for different theoretical models.
The decrease of sensitivity for relatively low LLP mass value is explained by the above-mentioned effects
on the detection efficiency.
The upper limits for the processes with $\mhzero=125\gevcc$
can be compared to the prediction of the Standard Model Higgs production cross-section
from gluon fusion of about 46\pb at $\sqs=13\tev$~\cite{PhysRevD.101.012002}.
%
\section{Conclusion}
Long-lived massive particles decaying into a muon and two quarks have been searched for
using  proton-proton collision data collected by the LHCb experiment at $\sqs =13\tev$,
corresponding to an integrated luminosity of 5.4~\invfb.
The LLP lifetime range considered is from 5 ps to 200 ps.
The background is dominated by \bbbar events and is reduced by tight selection
requirements, including a dedicated multivariate classifier.
The signal yield is determined by a fit to the LLP reconstructed mass with a signal shape
inferred from the theoretical models.

The forward acceptance of the LHCb experiment makes it complementary to other LHC experiments, while its low trigger 
\pt threshold allows exploring relatively small LLP masses.
Two types of LLP productions have been assumed. In the first
a Higgs-like particle is produced by gluon fusion and decays into two LLPs.
The analysis covers Higgs-like boson masses from 30 to 200\gevcc, and LLP mass range from 10~\gevcc
up to about one half of the mass of the parent boson.
The second mode is a direct LLP production from quark interactions,
covering the LLP mass range from 10~\gevcc up to 90~\gevcc.

The results for all theoretical models considered are compatible with the background-only hypothesis.
The upper limits at 95\% CL set on the cross-section times branching fractions are mostly of {\cal O}(0.1\pb),
but the sensitivity is limited to {\cal O}(10\pb) for the lowest LLP mass value considered of 10\gevcc.

%
\begin{footnotesize}
\begin{center}
\begin{longtable}[!t]{cccSSSS}    
 \caption {\small Upper limits at 95\% CL on the production cross-section times branching ratio
   for signal models with a resonant production via an Higgs-like boson.  
    Masses are given in \gevcc, lifetimes in \ps, cross-sections in \pb. }  
 \label{tab:ul-hX} \\
\hline
 \mhzero & m$_{\khi}$ & $\tau_{\khi}$ &  {expected UL} &  {$-\sigma$} & {$+\sigma$} & {observed UL} \\
\hline
\endfirsthead
\multicolumn{7}{c} {\tablename\ \thetable{} -- continued from previous page} \\
\hline
\mhzero & $m_{\khi}$ & $\tau_{\khi}$ & {expected UL} & {$-\sigma$} & {$+\sigma$} & {observed UL} \\
\hline
\endhead
\hline \multicolumn{7}{r}{\textit{Continued on next page}} \\
\endfoot
\hline
\endlastfoot
      30 &  10 &   10 &   40.24 &   16.52 &   31.86 &  28.34 \\
      41 &  10 &   10 &   30.89 &   11.41 &   41.72 &  22.37 \\
      41 &  20 &   10 &    7.95 &    2.24 &    3.55 &  10.07 \\
      50 &  10 &   10 &   13.72 &    4.39 &   11.90 &  17.37 \\
      50 &  20 &   10 &    2.65 &    0.82 &    1.67 &   3.46 \\
      80 &  10 &   10 &   13.72 &    4.36 &    9.37 &  14.00 \\
      80 &  20 &   10 &    0.81 &    0.26 &    0.40 &   0.64 \\
      80 &  30 &   10 &    0.36 &    0.10 &    0.18 &   0.29 \\
      80 &  40 &   10 &    0.48 &    0.12 &    0.20 &   0.56 \\
     125 &  10 &    5 &   22.06 &    6.37 &   32.95 &  21.53 \\
     125 &  10 &   10 &   17.76 &    0.26 &   15.73 &  18.08 \\
     125 &  10 &   20 &   20.64 &    7.89 &   16.67 &  12.49 \\
     125 &  10 &   30 &   17.21 &    5.37 &   13.22 &  15.40 \\
     125 &  10 &   40 &   11.10 &    4.01 &    9.27 &  20.91 \\
     125 &  10 &   50 &    8.11 &    2.63 &    6.80 &  13.42 \\
     125 &  10 &  100 &    7.89 &    2.74 &    5.99 &  16.99 \\
     125 &  10 &  200 &    7.50 &    1.13 &    8.19 &   7.82 \\
     125 &  20 &    5 &    1.38 &    0.45 &    0.60 &   0.89 \\
     125 &  20 &   10 &    0.56 &    0.15 &    0.22 &   0.47 \\
     125 &  20 &   20 &    0.49 &    0.13 &    0.22 &   0.55 \\
     125 &  20 &   30 &    0.73 &    0.21 &    0.34 &   0.87 \\
     125 &  20 &   50 &    0.91 &    0.21 &    0.32 &   1.00 \\
     125 &  20 &   80 &    1.11 &    0.20 &    0.40 &   1.11 \\
     125 &  20 &  100 &    0.82 &    0.22 &    0.38 &   0.72 \\
     125 &  20 &  200 &    0.94 &    0.38 &    0.77 &   0.98 \\
     125 &  30 &    5 &    0.55 &    0.15 &    0.23 &   0.30 \\
     125 &  30 &   10 &    0.18 &    0.04 &    0.09 &   0.25 \\
     125 &  30 &   20 &    0.20 &    0.05 &    0.08 &   0.23 \\
     125 &  30 &   30 &    0.22 &    0.06 &    0.09 &   0.19 \\
     125 &  30 &   50 &    0.34 &    0.07 &    0.11 &   0.29 \\
     125 &  30 &  100 &    0.17 &    0.06 &    0.11 &   0.17 \\
     125 &  30 &  200 &    0.48 &    0.15 &    0.17 &   0.46 \\
     125 &  40 &    5 &    0.18 &    0.06 &    0.10 &   0.15 \\
     125 &  40 &   10 &    0.17 &    0.05 &    0.08 &   0.14 \\
     125 &  40 &   20 &    0.10 &    0.02 &    0.04 &   0.11 \\
     125 &  40 &   30 &    0.16 &    0.04 &    0.06 &   0.17 \\
     125 &  40 &   50 &    0.18 &    0.05 &    0.07 &   0.14 \\
     125 &  40 &  100 &    0.20 &    0.06 &    0.11 &   0.20 \\
     125 &  40 &  200 &    0.23 &    0.07 &    0.11 &   0.20 \\
     125 &  50 &    5 &    0.28 &    0.08 &    0.13 &   0.35 \\
     125 &  50 &   10 &    0.07 &    0.02 &    0.04 &   0.07 \\
     125 &  50 &   20 &    0.05 &    0.02 &    0.03 &   0.04 \\
     125 &  50 &   30 &    0.09 &    0.02 &    0.04 &   0.10 \\
     125 &  50 &   50 &    0.08 &    0.02 &    0.03 &   0.09 \\
     125 &  50 &  100 &    0.07 &    0.03 &    0.04 &   0.05 \\
     125 &  50 &  200 &    0.18 &    0.04 &    0.08 &   0.20 \\
     125 &  60 &    5 &    0.49 &    0.13 &    0.21 &   0.49 \\
     125 &  60 &   10 &    0.11 &    0.03 &    0.05 &   0.10 \\
     125 &  60 &   20 &    0.05 &    0.02 &    0.03 &   0.06 \\
     125 &  60 &   30 &    0.04 &    0.01 &    0.03 &   0.03 \\
     125 &  60 &   50 &    0.06 &    0.02 &    0.03 &   0.06 \\
     125 &  60 &  100 &    0.07 &    0.02 &    0.03 &   0.07 \\
     125 &  60 &  200 &    0.15 &    0.02 &    0.06 &   0.14 \\
     150 &  10 &   10 &   14.85 &    5.52 &    9.07 &   9.81 \\
     150 &  20 &   10 &    0.60 &    0.18 &    0.35 &   0.41 \\
     150 &  30 &   10 &    0.15 &    0.05 &    0.09 &   0.16 \\
     150 &  40 &   10 &    0.10 &    0.03 &    0.05 &   0.09 \\
     150 &  50 &   10 &    0.08 &    0.03 &    0.05 &   0.09 \\
     150 &  60 &   10 &    0.04 &    0.02 &    0.03 &   0.04 \\
     200 &  10 &   10 &   22.48 &    6.43 &   14.67 &  28.12 \\
     200 &  20 &   10 &    2.47 &    0.61 &    0.85 &   2.71 \\
     200 &  30 &   10 &    0.48 &    0.12 &    0.19 &   0.53 \\
     200 &  40 &   10 &    0.19 &    0.05 &    0.09 &   0.22 \\
     200 &  50 &   10 &    0.08 &    0.03 &    0.05 &   0.07 \\
     200 &  60 &   10 &    0.07 &    0.02 &    0.04 &   0.06 \\
     200 &  50 &    5 &    0.12 &    0.04 &    0.06 &   0.12 \\
     200 &  50 &   20 &    0.09 &    0.02 &    0.04 &   0.09 \\
     200 &  50 &   30 &    0.12 &    0.04 &    0.07 &   0.11 \\
     200 &  50 &   50 &    0.08 &    0.03 &    0.05 &   0.07 \\
     200 &  50 &  100 &    0.15 &    0.04 &    0.08 &   0.16 \\
     200 &  50 &  200 &    0.19 &    0.06 &    0.12 &   0.20 \\
 \end{longtable}
 \end{center}
 \end{footnotesize}
%

 \begin{table}[ht]
 \centering
 \caption {\small
   Upper limits at 95\% CL on the production cross-section times branching ratio
   for signal models with a non-resonant production.  
   Masses are given in \gevcc, lifetimes in \ps, cross-sections in \pb.
 }
 \label{tab:ul-p216}
\begin{footnotesize}
 \begin{tabular}{ccSSSS}
 \hline
 $m_{\khi}$ & $\tau_{\khi}$   & {expected UL} &  {$-\sigma$} & {$+\sigma$} & {observed UL} \\
\hline
      10 &   10 &   23.67 &    6.31 &   12.90 &  28.84 \\
      20 &   10 &    0.92 &    0.24 &    0.46 &   0.82 \\
      30 &   10 &    0.49 &    0.12 &    0.19 &   0.50 \\
      40 &   10 &    0.31 &    0.09 &    0.14 &   0.44 \\
      60 &   10 &    0.08 &    0.03 &    0.04 &   0.06 \\
      90 &   10 &    0.10 &    0.03 &    0.05 &   0.08 \\
      30 &    5 &    0.86 &    0.24 &    0.40 &   0.90 \\
      30 &   20 &    0.32 &    0.08 &    0.13 &   0.36 \\
      30 &   30 &    0.19 &    0.05 &    0.09 &   0.20 \\
      30 &   50 &    0.22 &    0.06 &    0.09 &   0.20 \\
      30 &  100 &    0.11 &    0.04 &    0.07 &   0.13 \\
      30 &  200 &    0.36 &    0.12 &    0.23 &   0.33 \\
\hline
 \end{tabular}   
 \end{footnotesize}
 \end{table}
%

%


%

\section*{Acknowledgements}
%
%
\noindent We express our gratitude to our colleagues in the CERN
accelerator departments for the excellent performance of the LHC. We
thank the technical and administrative staff at the LHCb
institutes.
We acknowledge support from CERN and from the national agencies:
CAPES, CNPq, FAPERJ and FINEP (Brazil); 
MOST and NSFC (China); 
CNRS/IN2P3 (France); 
BMBF, DFG and MPG (Germany); 
INFN (Italy); 
NWO (Netherlands); 
MNiSW and NCN (Poland); 
MEN/IFA (Romania); 
MSHE (Russia); 
MICINN (Spain); 
SNSF and SER (Switzerland); 
NASU (Ukraine); 
STFC (United Kingdom); 
DOE NP and NSF (USA).
We acknowledge the computing resources that are provided by CERN, IN2P3
(France), KIT and DESY (Germany), INFN (Italy), SURF (Netherlands),
PIC (Spain), GridPP (United Kingdom), RRCKI and Yandex
LLC (Russia), CSCS (Switzerland), IFIN-HH (Romania), CBPF (Brazil),
PL-GRID (Poland) and NERSC (USA).
We are indebted to the communities behind the multiple open-source
software packages on which we depend.
Individual groups or members have received support from
ARC and ARDC (Australia);
AvH Foundation (Germany);
EPLANET, Marie Sk\l{}odowska-Curie Actions and ERC (European Union);
A*MIDEX, ANR, IPhU and Labex P2IO, and R\'{e}gion Auvergne-Rh\^{o}ne-Alpes (France);
Key Research Program of Frontier Sciences of CAS, CAS PIFI, CAS CCEPP, 
Fundamental Research Funds for the Central Universities, 
and Sci. \& Tech. Program of Guangzhou (China);
RFBR, RSF and Yandex LLC (Russia);
GVA, XuntaGal and GENCAT (Spain);
the Leverhulme Trust, the Royal Society
 and UKRI (United Kingdom).

\newpage

\newpage
\addcontentsline{toc}{section}{References}
\bibliographystyle{LHCb}
\bibliography{main,standard,LHCb-PAPER,LHCb-CONF,LHCb-DP,LHCb-TDR}

\newpage
\centerline
{\large\bf LHCb collaboration}
\begin
{flushleft}
\small
R.~Aaij$^{32}$,
A.S.W.~Abdelmotteleb$^{56}$,
C.~Abell{\'a}n~Beteta$^{50}$,
F.J.~Abudinen~Gallego$^{56}$,
T.~Ackernley$^{60}$,
B.~Adeva$^{46}$,
M.~Adinolfi$^{54}$,
H.~Afsharnia$^{9}$,
C.~Agapopoulou$^{13}$,
C.A.~Aidala$^{87}$
% ...

{\footnotesize \it

% ...
$^{9}$Universit{\'e} Clermont Auvergne, CNRS/IN2P3, LPC, Clermont-Ferrand, France\\
% ...
$^{13}$LPNHE, Sorbonne Universit{\'e}, Paris Diderot Sorbonne Paris Cit{\'e}, CNRS/IN2P3, Paris, France\\
% ...
$^{32}$Nikhef National Institute for Subatomic Physics, Amsterdam, Netherlands\\
% ...
$^{46}$Instituto Galego de F{\'\i}sica de Altas Enerx{\'\i}as (IGFAE), Universidade de Santiago de Compostela, Santiago de Compostela, Spain\\
% ...
$^{50}$Physik-Institut, Universit{\"a}t Z{\"u}rich, Z{\"u}rich, Switzerland\\
% ...
$^{54}$H.H. Wills Physics Laboratory, University of Bristol, Bristol, United Kingdom\\
% ...
$^{56}$Department of Physics, University of Warwick, Coventry, United Kingdom\\
% ...
$^{60}$Oliver Lodge Laboratory, University of Liverpool, Liverpool, United Kingdom\\
% ...
$^{68}$Syracuse University, Syracuse, NY, United States\\
% ...
$^{87}$University of Michigan, Ann Arbor, United States, associated to $^{68}$\\
\bigskip
% ...
\medskip
}
\end{flushleft}

\end{document}